\documentclass[12pt,a4paper]{article}
\nonstopmode
\usepackage{epsfig}
\usepackage[a4paper]{geometry}
\usepackage{float}
\usepackage[stable]{footmisc}
\usepackage[utf8]{inputenc}
\usepackage{a4wide}
\usepackage{amsmath,amssymb,amstext,amsthm,amssymb}
\usepackage{amsfonts}
\usepackage{framed}
\usepackage{graphicx}
\usepackage{color}
\usepackage[most]{tcolorbox}
\usepackage{bbold}
\usepackage{bbm}
\usepackage[normalem]{ulem}
\usepackage{rotating} 
\usepackage{slashed} 
\usepackage{cancel} 
\usepackage{braket} 
\usepackage{amstext}
\usepackage{array}
\usepackage{setspace}
\usepackage[urlcolor=blue,colorlinks=true,linkcolor  = black]{hyperref}
\usepackage{overpic}
\usepackage{bbm}
\usepackage{cite}
\usepackage{ulem}
\usepackage{lipsum}
\usepackage{tikz}
\usepackage{mathtools}
\usepackage{amsthm}
\usepackage{mathrsfs}
\usepackage{aas_macros}
\usepackage{enumitem}

\usepackage{hyperref} 

\newcommand{\DeclareRuneSeparators}[1]{} 

\input{arm.fd}

\usepackage{tikz-cd}
\usepackage{tikz}
\usepgflibrary{arrows} 
\usetikzlibrary{arrows,shapes,positioning} 
\usetikzlibrary{decorations.markings}
\usetikzlibrary{shapes.geometric}
\usetikzlibrary{arrows.meta,arrows}

\tikzstyle arrowstyle=[scale=1]
\tikzstyle directed=[postaction={decorate,decoration={markings,
		mark=at position .5 with {\pgftransformscale{2}\arrow[arrowstyle]{stealth}}}}]

\newcommand{\del}[0]{\partial}

\let\baraccent=\=
\renewcommand{\=}[1]{\stackrel{#1}{=}}

\DeclareSymbolFontAlphabet{\mathbb}{AMSb}

\newcommand{\be}{\begin{equation}}
	\newcommand{\ee}{\end{equation}}
\newcommand{\bea}{\begin{eqnarray}}
	\newcommand{\eea}{\end{eqnarray}}

\begin{document}
	
	\pagestyle{plain}

	\makeatletter
	\@addtoreset{equation}{section}
	\makeatother
	\renewcommand{\theequation}{\thesection.\arabic{equation}}
	\pagestyle{empty}
	{\hfill \small KCL-PH-TH/2023-49}
	\vspace{0.5cm}

	\vspace{0.5cm}
	
	\begin{center}
		
		{\LARGE \bf{Glimmers from the Axiverse}
			\\[10mm]}
	\end{center}

	\begin{center}
		\scalebox{0.95}[0.95]{{\fontsize{14}{30}\selectfont Naomi Gendler,$^{a,c}$ David J.~E.~Marsh,$^{b}$ Liam McAllister,$^{c}$ Jakob Moritz$^{c}$}} 
	\end{center}

	\begin{center}
		\vspace{0.25 cm}
		\textsl{$^{a}$Jefferson Physical Laboratory, Harvard University, Cambridge, MA 02138 USA}\\
		\textsl{$^{b}$Theoretical Particle Physics and Cosmology, King's College London, Strand, London, WC2R 2LS, United Kingdom}\\
		\textsl{$^{c}$Department of Physics, Cornell University, Ithaca, NY 14853 USA}\\

		\vspace{1cm}
		\normalsize{\bf Abstract} \\[8mm]

	\end{center}

	\begin{center}
		\begin{minipage}[h]{15.0cm}
			We study axion-photon couplings in compactifications of type IIB string theory.  We find that these couplings are systematically suppressed compared to the inverse axion periodicity, as a result of two effects.  First, couplings
			to the QED theta angle are suppressed for axion mass eigenstates that are light compared to the mass scale set by stringy instantons on the cycle supporting QED.  Second, in compactifications with many axions the intersection matrix is sparse, making kinetic mixing weak.  We study the resulting phenomenology in an ensemble of $200{,}000$ toy models constructed from the Kreuzer-Skarke database up to the maximum Hodge number $h^{1,1}=491$. We examine freeze-in production and decay of thermal axions, birefringence of the cosmic microwave background, X-ray spectrum oscillations, and constraints on the QCD axion from supernovae. We conclude that compactifications in this corner of the landscape involve many invisible axions, as well as a handful that may be detectable via photon couplings.
		\end{minipage}
	\end{center}
      \vfill
	\newpage
	\setcounter{page}{1}

	\setcounter{tocdepth}{2}
	
	\pagestyle{plain}
	\renewcommand{\thefootnote}{\arabic{footnote}}
	\setcounter{footnote}{0}
	%
	%
	
	\tableofcontents
	\newpage

	\section{Introduction}\label{sec:intro}

	Axions provide a remarkable connection between low-energy
	measurements and high-energy physics: the symmetries that protect axion masses can leave an imprint in observable phenomena.
	Axions abound in string theory, supported by the rich topologies of compactification manifolds, and may serve as an experimental window on the physics of quantum gravity~\cite{axiverse}.
	
	Ongoing searches for signatures of axions involve measurements of a vast range of phenomena in astrophysics, cosmology, and particle physics~\cite{Marsh:2015xka}.  Some searches rely on the gravitational couplings of axions~\cite{2011PhRvD..83d4026A,Mehta:2021pwf,Hlozek:2014lca,Lague:2021frh}, while others rely on axion couplings to quarks~\cite{Garconeaax4539}, leptons~\cite{2018EPJC...78..703C}, gluons~\cite{Abel:2017rtm,Aybas:2021nvn}, photons~\cite{1705.02290,Braine:2019fqb,CAPP:2020utb}, or combinations of these couplings~\cite{2008LNP...741...51R}. Gravitational constraints tend to get weaker as the axion decay constant decreases below the Planck scale, while matter couplings increase in this limit.
	The goal of the present work is to
	bring axions arising in string theory
	into contact with astrophysical constraints on the axion-photon coupling.
	
	Specifically, we will consider the axion-photon couplings in the \emph{Kreuzer-Skarke axiverse}, i.e.~in the set of low-energy effective theories of axions that arise from compactifications of type IIB string theory on orientifolds of Calabi-Yau threefold hypersurfaces \cite{Demirtas:2018akl}.  This is a setting in which the compactification geometries are comparatively well-understood thanks to
	advances in computational geometry \cite{CYTools}, and the physics of moduli stabilization has been exhaustively studied over the past two decades.  Axion physics in this setting, and in related geometries, has been studied in e.g.~\cite{Mehta:2021pwf,Broeckel:2021dpz,Demirtas:2021gsq,Gendler:2023hwg}.
	
	The axions we will consider arise from the Ramond-Ramond four-form $C_4$ reduced on four-cycles.  Their kinetic terms are dictated by classical geometric data: by the triple intersection numbers of the Calabi-Yau threefold $X$, and by the specification of a point $t_{\star}$ in the moduli space of K\"ahler parameters of the Ricci-flat metric on $X$.  The leading contributions to the axion masses result from Euclidean D3-branes wrapping holomorphic four-cycles in $X$, so the possible terms in the scalar potential can be computed by determining the topology of such four-cycles.  The magnitude of each such term is then easily evaluated given the point $t_{\star}$.  In sum, as laid out in detail in Ref.~\cite{Demirtas:2018akl}, the two-derivative effective action for all $C_4$ axions, including 
	the axion potential from Euclidean D3-branes,
	can be obtained directly from a geometric computation.  In the setting of Calabi-Yau threefold hypersurfaces in toric varieties, the necessary computation is quick and automatic using the software {\tt{CYTools}} \cite{CYTools}.  
	
	Axion couplings to the visible sector are a different matter.  At a minimum, one must specify how the visible sector is realized in the compactification, in terms of a configuration of D-branes.  Moreover, the linear combinations of axions that diagonalize the kinetic term and the mass matrix generally do not coincide with the combinations that couple simply to gauge sectors.  In particular, in a theory of $N$ axions, the QED theta angle will be a sum of kinetic-and-mass-eigenstates $\varphi^a$, $a=1,\ldots, N$,
	\begin{equation}
		\mathcal{L} \supset -   \frac{1}{4} \,g_{a\gamma\gamma}\,\varphi^a\,F_{\mu\nu} \tilde{F}^{\mu\nu}\, ,
	\end{equation} with coefficients $g_{a\gamma\gamma}$ (of mass dimension -1).
	
	The goal of the first part of this paper is to characterize the axion-photon couplings $g_{a\gamma\gamma}$ in the Kreuzer-Skarke axiverse.
	We find that structures in the compactification geometries lead to strong suppression of some of the $g_{a\gamma\gamma}$, compared to the predictions of an EFT analysis that is agnostic about the geometry.  Two effects are in play: kinetic mixing and mass hierarchies.  We find that the well-established sparseness of the triple intersection numbers of Calabi-Yau hypersurfaces causes kinetic mixing to be very weak, as already stressed in Ref.~\cite{Halverson:2019cmy}, and this \emph{kinetic isolation} effect prevents the photon from coupling democratically to all axions.
	Even stronger suppression results from mass hierarchies.
	Euclidean D3-branes wrapping the cycle that supports QED generate a potential for the QED theta angle, with an associated mass scale $m_{QED}$ that we term the \emph{light threshold}.  This is a stringy instanton effect and should be understood as a contribution from the ultraviolet completion, not from the low-energy gauge theory.  We find that axion mass eigenstates with masses $m \ll m_{QED}$ have hierarchically suppressed couplings to the photon.
	
	In the second half of the paper, we study the phenomenological consequences of the axion-photon couplings in the Kreuzer-Skarke axiverse. Constraints on axions can be separated into those that arise from vacuum processes in the local Universe (such as axion-photon conversion in magnetic fields, or axion production in stars and supernovae), and those that depend on cosmology (e.g.~by involving  the thermal and non-thermal relic densities).  Cosmological dependence is encapsulated in the vacuum misalignment angle of each axion, the reheating temperature (in the following we take this conservatively to be the temperature at which the last modulus decays, and after which uninterrupted radiation domination by the Standard Model begins), and the Hubble scale during inflation, $H_I$. We explore how the application of various limits depends on string theory inputs, in particular: the Hodge number $h^{1,1}$ (which affects the overall scale of axion decay constants \cite{Demirtas:2018akl}), the volume of the four-cycle on which the gauge group containing QED in the IR is realized (which determines the mass scale below which axion-photon couplings are suppressed), and whether or not the model is a GUT (which determines the role of the QCD axion relative to the other axions).
	
	We find that the QCD axion mass is strongly correlated with the Hodge number, and at the largest Hodge numbers the QCD axion may affect the duration of neutrino bursts of supernovae via its gluon coupling at a level that could be probed by a future galactic supernova observed with upcoming and present neutrino detectors. For all other axions, their phenomenology crucially depends on whether the model is a GUT or not, and in non-GUTs there is also dependence on the four-cycle volume. For non-GUTs, a wide range of phenomenology is open. Helioscopes and X-ray spectrum oscillations can probe the largest Hodge numbers if QED is supported on a cycle with volume $\gtrsim 30$ in string units.  Cosmic birefringence at a level consistent with current hints~\cite{Minami:2020odp} can be generated in non-GUT models with QED cycle volume $\approx 40$. In GUTs, however, the light threshold is the mass of the QCD axion, and so these phenomena are forbidden~\cite{Agrawal:2022lsp}. For GUTs and non-GUTs, the unavoidable freeze-in production and decay of axions in the mass range $1\text{ keV}\lesssim m_a\lesssim 1\text{ GeV}$ disfavors large values of the reheat temperature $T_{\rm re}\gtrsim 10^{10}\text{ GeV}$. In the Appendix, we outline other phenomenology including the dark matter composition, dark radiation production, and possibilities to redistribute the relic abundance among the different axions.
	
	\section{Axion couplings in EFT}\label{sec:EFT}
	
	We start by considering effective field theories of $N$ dimensionless axions $\theta^a$, $a=1,\ldots,N$, with integer periodicities $\theta^a\simeq \theta^a+1$, coupled to electromagnetism. The relevant part of the Lagrangian is  
	\begin{align}\label{eq:canonicalL1}
		\mathcal{L} = -\frac{1}{2}M_{\text{pl}}^2 \mathcal{K}_{ab}\partial_\mu \theta^a \partial^\mu \theta^b -V(\theta) - \mathcal{Q}^{\text{EM}}_a \theta^a  \frac{\alpha_{\text{EM}}}{4}F_{\mu\nu} \tilde{F}^{\mu\nu}\, ,
	\end{align} 
	where $F_{\mu\nu}$ is the field strength of electromagnetism, $\alpha_{\text{EM}}$ is the fine structure constant --- with value $1/137$ at low energies, and $1/127.5$ at the mass of the Z-boson --- and the axion potential generated by instantons takes the form
	\begin{equation}\label{eq:axion_potential}
		V(\theta)=\sum_{I} \Lambda_{I}^4\Bigl(1-\cos(2\pi \mathcal{Q}_{Ia} \theta^a+\delta_I)\Bigr)\, .
	\end{equation}
	The Lagrangian is parameterized in terms of the (dimensionless) kinetic matrix $\mathcal{K}_{ab}$, an integer valued \emph{electromagnetic charge vector} $\mathcal{Q}^{\text{EM}}_a$ governing the coupling between the axion sector and electromagnetism, as well as the data determining the scalar potential $V(\theta)$: the \emph{instanton mass scales} $\Lambda_I$, and the integer valued \emph{instanton charge matrix} $\mathcal{Q}_{Ia}$. We note that in this basis where $\mathcal{Q}$ takes integer values, the field space metric $\mathcal{K}_{ab}$ is in general non-diagonal.  
	
	The above data are not determined or constrained by principles of effective field theory --- since all terms obey the discrete axion shift symmetries $\theta^a\simeq \theta^a+1$ --- and must therefore be supplied by a more  fundamental computation. We will compute these data in ensembles of type IIB string compactifications, as explained in \S\ref{sec:KSaxiverse}.
	
	Given the data parameterizing \eqref{eq:canonicalL1}, one can directly compute tree-level axion-axion as well as axion-photon interactions. To this end, one expands the scalar potential 
	around a vacuum solution,
	\begin{equation}
		\theta^a=\theta^a_0+{M^a}_b \varphi^b\, ,\quad \del_{\theta^a}V|_{\theta=\theta_0}\, ,
	\end{equation}
	where the matrix $M$ is chosen to bring the kinetic terms to canonical normalization, 
	\begin{equation}
		\mathcal{L}_{\text{kinetic}}=-\frac{1}{2}\sum_a (\del \varphi^a)^2\, ,
	\end{equation}
	and the mass matrix to diagonal form, i.e.
	\begin{equation}
		V(\varphi^a)=V(\theta_0)+\frac{1}{2}\sum_a m_a^2 (\varphi^a)^2+\mathcal{O}(\varphi^3)\, .
	\end{equation}
	The photon coupling to the $a$th axion is then given by 
	\begin{equation}
		g_{a\gamma\gamma}=\alpha_{\text{EM}}\sum_b \mathcal{Q}_b^{\text{EM}}{M^b}_a\, .
	\end{equation}
	In \S\ref{ss:lightthreshold} and \S\ref{ss:kineticisolation} we will explain how two distinct effects imprint parametric hierarchies into both axion-photon as well as axion-axion interactions.

	\subsection{Axion mass hierarchies}

	As anticipated in Ref.~\cite{axiverse} and explored in detail in Refs.~\cite{Demirtas:2018akl,Mehta:2021pwf}, typical instanton mass scales are often widely hierarchical, leading to a mass spectrum of axions easily populating a range of mass scales 
	from the Planck scale down to masses much smaller than the Hubble scale $H_{0}$.
	
	We are therefore led to analyze the regime of parametric separation of instanton scales 
	\begin{equation}
		\Lambda_{I+1} \ll \Lambda_{I}\, ,\quad \forall I \, .
	\end{equation}
	One expects to find widely separated axion masses in this regime. 
	However, numerically diagonalizing the hierarchical mass matrix induced by the instantons is computationally costly and numerically unstable, though it can be done in principle as in  \cite{Mehta:2021pwf} ---  and so we will instead compute axion masses and interactions via an expansion in the small ratios
	\begin{equation}
		\epsilon_I:=\frac{\Lambda_{I+1}^4}{\Lambda_{I}^4} \ll 1\, .
		\label{eq:hierarchies}
	\end{equation}
	We will see momentarily that various coupling constants are systematically suppressed by these ratios.
	
	We first truncate the scalar potential to the terms generated by the $N$ most relevant instantons, and treat the remaining terms as a small --- generally CP-breaking ---  
	perturbation:
	\begin{equation}
		V(\theta)=V_0(\theta)+\delta V_{\cancel{CP}}(\theta)\, ,
	\end{equation}
	with
	\begin{align}\label{eq:potential_split}
		V_0(\theta)&=\sum_{a=1}^{N} \Lambda_a^4\left(1-\cos\left(2\pi \sum_b Q_{ab}\theta^b\right)\right)\, ,\\
		\delta V_{\cancel{CP}}(\theta)&=\sum_{\alpha=1}^{\delta N} \hat{\Lambda}_{\alpha}^4\left(1-\cos\left(2\pi \sum_b \hat{Q}_{\alpha b}\theta^b+\delta_{\alpha}\right)\right)\, .
	\end{align}
	We define the rows of the \emph{reduced charge matrix} $Q_{ab}$ as the charges of the most relevant instantons: these are obtained by deleting from $\mathcal{Q}_{Ia}$ all rows that are contained in the linear span of the preceding rows.\footnote{We caution the reader that as a result, the ordering of the $\Lambda_I$ in \eqref{eq:axion_potential} does not coincide with that of the $\Lambda_a$ in \eqref{eq:potential_split}.  That is, for any given $k \in \{1,\ldots N\}$, in general $\Lambda_{a=k} \neq \Lambda_{I=k}$, although $\{\Lambda_a, a \in \{1,\ldots, N\}\} \subset \{ \Lambda_I, I \in \mathbb{N}\}.$} The deleted rows are collected in $\hat{Q}_{\alpha a}$. We have furthermore used up our freedom to shift the $N$ axions by constants, in order to set to zero all the additive phases that would otherwise appear in $V_0(\theta)$, and so the leading order potential $V_0(\theta)$ is manifestly CP-preserving.
	
	As the reduced charge matrix $Q_{ab}$ is full rank, it is always possible to work in a (not necessarily integer) axion basis in which it becomes the identity. In such a basis it is clear that the $N$ axions couple to each other at most via the CP-breaking potential $\delta V_{\text{CP-breaking}}(\theta)$, and through kinetic mixing. Indeed, one cannot in general simultaneously diagonalize the reduced charge matrix $Q_{ab}$ as well as the kinetic term $K_{ab}$, and for now we will assume that the kinetic term $K_{ab}$ is indeed generic.
	
	However, there is a unique axion basis, which we will denote $\phi^a={\Lambda^a}_b \theta^b$, in which the kinetic term is canonical, modulo the overall factor of $M_{\text{pl}}^2$, and the reduced charge matrix is lower triangular, i.e.
	\begin{equation}
		\sum_{cd}{{\Lambda^{-1}}^c}_a\,\mathcal{K}_{cd}\, {{\Lambda^{-1}}^d}_b=\delta_{ab}\, ,\quad q_{ab}:=\sum_c Q_{ac}\, {{\Lambda^{-1}}^c}_b=0\,\quad \forall b>a\, .
	\end{equation}
	The newly obtained $q_{ab}$ will in general not take integer values.
	
	One obtains the basis $\phi^a$ by applying the Gram-Schmidt process to the rows of the charge matrix, and therefore we will refer to it as the \emph{Gram-Schmidt basis}.  
	Crucially, in units $M_{\text{pl}}=1$ the \emph{Gram-Schmidt basis is equal to the mass and kinetic eigenbasis}, to leading order in the ratios of scales $\epsilon_I$. One shows this via successively integrating out axions as in \cite{Demirtas:2021gsq}: first, one approximates the potential by the leading instanton term
	\begin{equation}
		V(\phi)\approx \Lambda_1^4\Bigl(1-\cos\bigl(2\pi q_{11} \phi^1\bigr)\Bigr)\, .
	\end{equation}
	This potential depends only on $\phi^1$, and thus $\phi^1$ is a mass and kinetic eigenstate in this approximation. Integrating out\footnote{Although in the present computation we integrate out the heavy axions in order to obtain an approximation to the mass matrix, we stress that in our subsequent treatment of axiverse phenomenology we will keep all $h^{1,1}$ axions in the EFT. In other words, we fix the heavy fields successively to their vacuum values in $V$ in order to approximate the mass hierarchy, but we in fact retain the resulting mass terms and kinetic terms and thus all heavy field dynamics in the Lagrangian.} 
	the heaviest axion $\phi^1$ at tree level amounts to setting $\phi^1=0$, and the low energy effective theory of $N-1$ axions is parameterized by the axions $\phi^{2,\ldots,N}$, and the scalar potential is parameterized by the sub-matrix $q_{ab}$, $a,b=2,\ldots,N$, which is again lower triangular. Thus, also $\phi^2$ is a mass and kinetic eigenstate and, by induction, so are all the $\phi^a$.
	
	The axion masses are easily computed as
	\begin{equation}
		m_a^2=\frac{\Lambda_a^4}{f_a^2}\, ,
	\end{equation}
	where by analogy to the single axion case, we define \emph{axion decay constants}
	\begin{equation}\label{eq:fdef}
		f_a:=\frac{M_{\text{pl}} }{2\pi q_{aa}}\, .
	\end{equation}
	Denoting as above the exact mass eigenstates by $\varphi^a$, we may write the scalar potential as
	\begin{equation}
		V_0(\varphi)=\sum_{a=1}^{N} \Lambda_a^4\left(1-\cos\left(\sum_b \Theta_{ab}\frac{\varphi^b}{f_b}\right)\right)\, ,
	\end{equation}
	in terms of mixing angles $\Theta_{ab}$. In the limit $\epsilon_a\rightarrow 0$ the mixing angles are simply proportional to entries of the reduced charge matrix $q_{ab}$, and thus also lower triangular, while to leading order in the $\epsilon_a$ one finds
	\begin{equation}\label{eq:mixing_angles}
		\Theta_{ab}\simeq \begin{cases}
			\frac{q_{ab}}{q_{bb}} & b\leq a\,,\\
			-\frac{\Lambda_b^4}{\Lambda_a^4}\frac{q_{ba}}{q_{aa}} & b>a\,.
		\end{cases}
	\end{equation}
	In particular, the 
	argument of the cosine
	associated with an instanton with scale $\Lambda_a^4$ has an unsuppressed dependence only on the axions $\varphi^1,\ldots \varphi^a$, while its dependence on the lighter axions $\varphi^{a+1},\ldots,\varphi^N$ is suppressed by the small ratios of scales $\epsilon_a$.
	
	\subsection{Axion couplings and the light threshold}\label{ss:lightthreshold}
	Expanding the electromagnetic charge vector $\mathcal{Q}_a^{\text{EM}}$ in terms of the rows of the reduced instanton charge matrix --- $\mathcal{Q}_a^{\text{EM}}=n^b_{\text{EM}} Q_{ba}$ --- one expresses the axion photon couplings $g_{a\gamma\gamma}$,  
	\begin{equation}\label{eq:axion_photon_vertex}
		\mathcal{L}\supset -\sum_{a=1}^N g_{a\gamma\gamma} \varphi^a\, \frac{1}{4}F_{\mu\nu}\tilde{F}^{\mu\nu}\, ,
	\end{equation}
	in terms of the mixing angles:
	\begin{equation}\label{eq:axion_photon_couplings}
		g_{a\gamma\gamma}=\frac{\alpha_{\rm EM}}{2\pi f_a} \sum_b n^b_{\text{EM}}\Theta_{ba}\, .
	\end{equation}
	The partial decay width of a non-relativistic axion $\varphi^a$ to two photons --- which will be crucial for the phenomenological analyses in \S\ref{sec:cosmology} and \S\ref{sec:axion-photon-physics} --- is given in terms of these as
	\begin{equation}\label{eq:decay_width_axion_phothon}
		\Gamma_{a\rightarrow \gamma\gamma}=\frac{m_a^3g_{a\gamma\gamma}^2}{64\pi}\, .
	\end{equation}
	From \eqref{eq:axion_photon_couplings} we learn the following important lesson: if $\mathcal{Q}_a^{\text{EM}}$ is contained in the linear span of the $k<N$ leading instantons, i.e.
	\begin{equation}
		\mathcal{Q}_a^{\text{EM}}\in \text{span}_{\mathbb{Q}}\langle Q_{1a},\ldots, Q_{ka}\rangle\, ,
	\end{equation}
	then all axion photon couplings $g_{a\gamma\gamma}$ with $a>k$ receive a suppression factor, 
	\begin{equation}
		g_{a\gamma\gamma}=\mathcal{O}\left( \frac{\Lambda_a^4}{\Lambda_k^4}\right)\, ,
	\end{equation}
	and are then typically negligibly small. In particular, if the EM charge vector $\mathcal{Q}_a^{\text{EM}}$ itself contributes one of the leading instanton terms in the scalar potential $V_0(\theta)$ --- i.e.~if $Q_{a_{\text{EM}}b}=\mathcal{Q}_b^{\text{EM}}$ for some index $a_{\text{EM}}$ ---  then the corresponding instanton introduces an important mass scale $m_{a_{EM}} \equiv m_{\text{QED}}$ that we will refer to as the \emph{light threshold}.
	The coupling of electromagnetism to all axions $\varphi^a$ with mass below the light threshold
	is suppressed by the ratio of mass scales,
	\begin{equation}
		f_a \,g_{a\gamma\gamma} = \mathcal{O}\left(\frac{m^2_a}{m^2_{a_{\text{EM}}}}\right)\qquad \text{for}~~ m_a<m_{a_{\text{EM}}}\, .
	\end{equation}
	In our ensemble of string compactifications there always exists an instanton associated to QED (specifically, a Euclidean D3-brane wrapping the QED cycle), an effect that is absent from a field theory perspective. The light threshold is thus a stringy generalization of the idea analyzed for GUTs in \cite{Agrawal:2022lsp} and alluded to in \cite{axiverse}.
	
	We will later argue that a combination of this suppression in terms of the light threshold ---  combined with a form of kinetic decoupling --- dramatically suppresses axion-photon couplings in the string axiverse, independent of whether or not the Standard Model is completed into a GUT in the UV: \emph{in typical models with many axions, photons couple significantly only to the electroweak axions and at most a few other axions.} 
	
	Let us also point out that axion-flavor changing self interactions are similarly suppressed by ratios of mass scales. Specifically, we will be interested in processes where a non-relativistic axion mass eigenstate decays to lighter axions. The relevant interactions are the cubic and quartic axion-axion interactions,
	\begin{equation}\label{eq:cubic_quartic_int}
		\mathcal{L}\supset -\frac{\chi_{abc}}{3!}\varphi^a \varphi^b \varphi^c-\frac{\lambda_{abcd}}{4!}\varphi^a \varphi^b \varphi^c \varphi^d\, .
	\end{equation}
	First, we assume that the CP-breaking interactions encoded in $\delta V_{\text{CP-breaking}}$ are negligible (i.e. we neglect the cubic couplings). Then, the leading decay channel of the $a$-th axion to other axions is the one-to-three axion decay parameterized by the dimensionless quartic coupling
	\begin{equation}
		\mathcal{L}\supset -\frac{\lambda_{abcd}}{4!}\varphi^a\varphi^b\varphi^c\varphi^d\supset \frac{1}{3!}\frac{\Lambda_{b}^4}{f_a f_{b}^3}\Theta_{ba}\varphi_a \varphi_{b}^3\equiv -\frac{1}{3!}\lambda_{abbb}\varphi_a (\varphi_{b})^3 \, ,
	\end{equation}
	with $b>a$ on the r.h. side. Indeed, this flavor changing quartic coupling is suppressed in comparison with quartic axion self-interactions $\lambda_{aaaa}=\Lambda_a^4/f_a^4$.
	
	In the limit $m_b\ll m_a$ the corresponding flavor changing partial decay width for a non-relativistic axion $\varphi^a$ is given by
	\begin{equation}\label{eq:quartic_decay_rate}
		\Gamma^{(4)}_{a\rightarrow bbb}= \frac{\lambda_{abbb}^2 m_a}{128\pi^3}\simeq \frac{\Lambda_a^2\Lambda_{b}^8}{128\pi^3 f_a^3 f_{b}^6}\Theta_{ba}^2\, ,
	\end{equation}
	For generic field space metric $K_{ab}$ in \eqref{eq:canonicalL1} the leading decay is to the neighboring axion $\varphi^{b=a+1}$, but in general the decay to even lighter axions $\varphi^{b>a+1}$ may dominate over this if the lower triangular components of the matrix of mixing angles are suppressed for other reasons, such as kinetic isolation as discussed below. For simplicity we only include nearest-neighbor decays in \S\ref{sec:decaying_DM_indirect}.
	
	We note, finally, that one-to-two axion decays via cubic interactions break CP, and are therefore at most induced by the terms in $\delta V_{\text{CP-breaking}}$. Namely, the CP-breaking phases $\delta_\alpha$ associated with the instantons of charges $\hat{Q}_{\alpha a}$ shift the expectation values of the phases of the leading cosines in the scalar potential to
	\begin{equation}\label{eq:CP_breaking_spurion}
		\zeta_a:=\left\langle 2\pi \sum_{b}q_{ab}\phi^b \right\rangle \simeq -\sum_{\alpha}\frac{\hat{\Lambda}^4_\alpha}{\Lambda_a^4}\frac{\hat{q}_{\alpha a}}{q_{aa}}\sin(\delta_{\alpha})\, ,
	\end{equation}
	By construction of the reduced instanton charge matrix, and the Gram-Schmidt basis, we have $\hat{q}_{\alpha a}=0$ for all $(\alpha,a)$ such that $\hat{\Lambda}^4_\alpha > \Lambda_a^4$, and thus the CP-breaking spurions $\zeta_a$ are small when the instanton scales are hierarchical.
	
	While the $\zeta_a$ can induce non-vanishing cubic interactions
	\begin{equation}
		\mathcal{L}\supset -\frac{1}{2}\chi_{abb}\varphi_a \varphi_b^2\simeq -\frac{1}{2}\frac{\Lambda_b^4}{f_a f_b^2} \Theta_{ba} \zeta_b\, ,
	\end{equation}
	the corresponding cubic decay width is 
	\begin{equation}\label{eq:cubic_decay_rate}
		\Gamma^{(3)}_{a\rightarrow bb}= \frac{\chi_{abb}^2}{16\pi m_a}\simeq \frac{\Lambda_{b}^8}{16\pi \Lambda_a^2 f_a f_{b}^4}\Theta_{ba}^2 \zeta_b^2\, ,
	\end{equation}
	and is thus suppressed by $\zeta_b$. Comparing to the quartic decay width one finds
	\begin{equation}
		\frac{	\Gamma^{(3)}_{a\rightarrow bb}}{	\Gamma^{(4)}_{a\rightarrow bbb}}\sim \frac{f_a^2f_b^2}{\Lambda_a^4}\zeta_b^2\, ,
	\end{equation}
	and the hierarchical enhancement of the cubic decay rate over the quartic one by $f_a^2f_b^2/\Lambda_a^4$ is typically overcome by the smallness of the CP-breaking spurions $\zeta_a$.

	\subsection{Kinetic isolation}\label{ss:kineticisolation}
	We will now discuss abstractly how axion interactions are affected, quantitatively, when kinetic mixing is weak. For this, as before, let $q_{ab}$ be the reduced instanton charge matrix. Then  we may use the rows of $q_{ab}$ as a basis of instanton charges, and in this basis one has by construction $q_{ab}=\delta_{ab}$.
	
	We now \emph{assume} that in this basis the off-diagonal terms in the kinetic term $\mathcal{K}_{ab}$ are suppressed in comparison to the diagonal terms. We write
	\begin{equation}
		\mathcal{K}_{ab} = (2\pi)^2 \text{diag}\Bigl(f_{(0)1}^2,\ldots,f_{(0)N}^2\Bigr)_{ab}+\delta \mathcal{K}_{ab}\, .
	\end{equation}
	In the limit $\delta \mathcal{K}\rightarrow 0$ the mass eigenstates are given by $\varphi^a_{(0)} := f_{(0)a} \phi^a$ (no sum over $a$). In particular, the field space metric is near-canonical,
	\begin{equation}
		\mathcal{L}_{\text{kinetic}}=-\frac{1}{2}\tilde{\mathcal{K}}_{ab}\del \varphi^a_{(0)} \del \varphi^b_{(0)}\, ,\quad \tilde{\mathcal{K}}_{ab} := \frac{\mathcal{K}_{ab}}{(2\pi)^2f_{(0)a} f_{(0)b}}=\delta_{ab}+\xi_{ab}\, ,
	\end{equation}
	for some small symmetric matrix $\xi_{ab}$. Defining the lower triangular matrix
	\begin{equation}
		\varepsilon_{ab}:=\begin{cases}
			\xi_{ab}& b<a\\
			0 & a\leq b
		\end{cases}\, ,
	\end{equation}
	which we will refer to as \emph{kinetic isolation factors},
	then in the Gram-Schmidt basis one has
	\begin{equation}
		q_{ab}= \delta_{ab}-\varepsilon_{ab}+\mathcal{O}(\varepsilon^2)\, ,
	\end{equation}
	and therefore the off-diagonal mixing angles $\Theta_{ab}$ are of order $\varepsilon_{ab}$. In \S\ref{sec:string}, as in \cite{Halverson:2019kna}, we will see that in string compactifications most off-diagonal terms in the basis of the leading instantons are indeed very small. This leads to a 
	significant suppression in axion-photon interactions, even for axions above the light threshold: from \eqref{eq:axion_photon_couplings} we find that $g_{a\gamma\gamma}$ is of order $\varepsilon$ for these axions.

	\section{Axion couplings in type IIB string theory}\label{sec:string}

	In \S\ref{sec:EFT} we laid out a procedure for translating the Lagrangian \eqref{eq:canonicalL1}, expressed in the variables that are natural in string theory, to a form in which the physical axion-photon couplings can be read off. The only input from string theory used in \S\ref{sec:EFT} was the general expectation that the instanton scales in \eqref{eq:axion_potential} should be hierarchical, as in \eqref{eq:hierarchies}, which allowed for efficient computation of the approximate mass eigenbasis. 
	
	We will now describe the method used for constructing the ensemble of axion theories resulting from explicit string theory compactifications that we will study. Then, in \S\ref{sec:distributions} we will present the resulting distributions of axion-photon couplings and axion masses that arise in this ensemble.

	\subsection{The Kreuzer-Skarke axiverse}\label{sec:KSaxiverse}
	
	As a first step in constructing an ensemble of axion theories,
	we consider a Calabi-Yau threefold, $X$, and suppose that $\sigma: X \to X$ is a holomorphic involution whose fixed loci are points and four-cycles in $X$.
	Compactifying type IIB string theory on the corresponding O3/O7 orientifold gives rise to $h^{1,1}$ axions\footnote{Because $\sigma^2$ is the identity, the action of $\sigma$ on $H^4(X)$ can be diagonalized to give eigenspaces $H^{4}_{\pm}$ with eigenvalues $+1$ and $-1$, and with corresponding dimensions $h^{1,1}_{\pm}$.  In this work we will consider involutions for which $h^{1,1}_-=0$, and so $h^{1,1}_+ = h^{1,1}$.  More general orientifolds with $h^{1,1}_- >0$ lead to axions from reducing the two-forms $C_2$ and $B_2$, but we leave the physics of the resulting two-form axiverse as a question for the future.} from the Kaluza-Klein zero modes of the Ramond-Ramond four-form $C_4$.
	
	Denoting by $[D_a]$ the homology class of a divisor $D_a$, we take $\{[D_a]\}$, $a=1,\ldots,h^{1,1}$ to be a basis of $H_4(X,\mathbb{Z})$, and let $J$ be the K\"ahler form on $X$, parameterized in terms of K\"ahler parameters $t_a$  
	\begin{equation}
		J=\sum_{a=1}^{h^{1,1}} t_a [D_a]\, .
	\end{equation}
	Then, the low energy effective theory contains $h^{1,1}$ closed string axions defined as
	\begin{equation}\label{eq:defax}
		\theta^a := \int_{D_a} C_4\,.
	\end{equation}
	Defining the four-cycle volumes
	\begin{equation}
		\tau^a := \frac{1}{2}\int_{D_a}J\wedge J\, ,
	\end{equation} 
	the holomorphic coordinates on the moduli space of complexified K\"ahler parameters are
	\begin{equation}
		T^a := \tau^a + i \theta^a\equiv \int_{D_a}\left(\frac{1}{2}J\wedge J+i C_4\right)\, .
	\end{equation}
	The $\tau^a$ are \emph{saxions}, i.e.~real scalar fields that each combine with an axion to form a complex scalar in a chiral multiplet of $\mathcal{N}=1$ supersymmetry.
	
	The overall volume of $X$ and the four-cycle volumes $\tau^a$ are computed in terms of the K\"ahler parameters $t_a$ and the triple intersection form $\kappa^{abc}:=\int_X [D_a]\wedge [D_b]\wedge [D_c]$,
	\begin{equation}
		\mathcal{V}  =\int_X \frac{1}{3!}J\wedge J\wedge J= \frac{1}{3!}\kappa^{abc}t_a t_b t_c\, , \quad \tau^a=\frac{1}{2}\kappa^{abc}t_bt_c\, .
	\end{equation}
	Dimensional reduction of the ten-dimensional gravitational action yields
	an axion kinetic term of the form \eqref{eq:canonicalL1}, with
	\begin{equation}
		\mathcal{L}_{\text{kin}} = -\frac{M_{\text{pl}}^2}{2} \mathcal{K}_{ab} \partial_{\mu} \theta^a \partial^{\mu} \theta^b \,,
	\end{equation}
	where $\mathcal{K}_{ab}$ is the K\"ahler metric on K\"ahler moduli space.
	The inverse metric $\mathcal{K}^{ab}$ has a simple expression in terms of the $t_a$, $\tau^a$, and $\kappa^{abc}$:
	\begin{equation}\label{eq:inverse_metric_CY}
		\mathcal{K}^{ab} = 2\mathcal{V} \left(c^{ab}+\frac{\tau^a\tau^b}{\mathcal{V}}\right)\, ,
	\end{equation}
	where $c^{ab}:=- \kappa^{abc}t_c$.
	It is convenient to express $\mathcal{K}_{ab}$ in terms of a K\"ahler potential $\mathscr{K}$,
	\begin{equation}\label{eq:kis}
		\mathscr{K} = - 2\,\text{log}\mathcal{V}\,,
	\end{equation}
	obeying
	\begin{equation}\label{eq:kisfromv}
		\mathcal{K}_{ab} = 2\frac{\partial}{\partial T^a\phantom{\bigr|}} \frac{\partial}{\partial \overline{T^b}} \mathscr{K} = \frac{1}{2}\frac{\partial}{\partial \tau^a} \frac{\partial}{\partial \tau^b} \mathscr{K}\,.
	\end{equation}
	To ensure control of the $\alpha'$ expansion, we will only consider the subregion of K\"ahler moduli space in
	which all effective curves and all effective divisors have volume $\ge 1$ in units of the string length $\ell_s = 2\pi\sqrt{\alpha'}$.  Following \cite{Demirtas:2018akl} we refer to this region as the \emph{stretched K\"ahler cone}.
	Within the stretched K\"ahler cone, and for small values of the string coupling $g_s$, the corrections to \eqref{eq:kis} in the $g_s$ and $\alpha'$ expansions are nonzero, but have small impact on the phenomena studied here.
	
	The axions $\theta_a$ enjoy perturbative shift symmetries  $\theta^a \to \theta^a + const.$ that descend from ten-dimensional gauge invariance, and are broken only by nonperturbative effects.  
	In particular, because fundamental strings do not carry Ramond-Ramond charge, no potential for the $\theta_a$ can arise at any order in the string loop expansion.  Only D-branes --- either ordinary D-branes filling spacetime, or Eulidean D-branes --- can break the shift symmetries: consider a Euclidean D3-brane wrapping a four-cycle in a class $[\Sigma] \in H_4(X,\mathbb{Z})$. We can express it in our basis $\{[D_a]\}$ as
	$[\Sigma] = \sum_a Q_{a} [D_a]$ in terms of an instanton charge $Q_{a}$.
	The complexified action $S$ of the Euclidean D3-brane obeys
	\begin{equation}
		\frac{S}{2\pi} = \text{Vol}(\Sigma) + i \int_{\Sigma} C_4 = \text{Vol}(\Sigma) + i Q_{a} \theta^a\,.
	\end{equation}
	If $\Sigma$ is holomorphic\footnote{Non-holomorphic cycles  $\Sigma$ can produce  Euclidean D3-brane contributions to the K\"ahler potential, but such effects are typically subleading. For analyses of the exceptional circumstances in which non-holomorphic cycles support effects that dominate the breaking of axion symmetries, see \cite{Demirtas:2018akl,Demirtas:2019lfi}.}
	(i.e., an effective divisor) then $\text{Vol}(\Sigma)=Q_{a} \tau^a$.  If in addition the Euclidean D3-brane worldvolume theory on $\Sigma$ supports exactly two fermionic zero modes\footnote{If $\Sigma$ is smooth and one neglects the effects of flux in the compactification and on the Euclidean D3-brane worldvolume, then a necessary condition for the existence of precisely two fermionic zero modes is that $\Sigma$ is rigid --- i.e.~has no normal bundle deformations --- is irreducible, and is mapped to itself by the orientifold involution. For the more general case where $\Sigma$ is singular, see e.g.~\cite{Gendler:2022qof}.}, then there is a resulting contribution to the superpotential,
	\begin{equation}
		W \supset \mathcal{A} e^{-S} = \mathcal{A}_{\Sigma}\,\text{exp}\Bigl(-2\pi Q_a \tau^a - 2\pi i Q_a \theta^a\Bigr)\,.
		\label{eq:nonpertW}
	\end{equation}
	Here $\mathcal{A}_{\Sigma}$ is a Pfaffian prefactor that we will treat as a constant and set to unity.\footnote{See \cite{Alexandrov:2022mmy,Kim:2023cbh} for recent progress in computing the Pfaffians.}

	\subsection{Generating models} \label{ss:models}

	We define a model as a compactification of type IIB string theory in which we have specified:
	\begin{enumerate}
		\item A Calabi-Yau threefold $X$, on which we compactify.
		\item A set of assumptions about the UV completion of the Standard Model.
		\item A point in moduli space at which we compute the axion effective theory.
	\end{enumerate}
	We will now explain how we determine each of these items in turn.
	
	\paragraph{Choosing Calabi-Yau threefolds:} We will compactify type IIB string theory on (orientifolds of) Calabi-Yau threefold hypersurfaces in four-dimensional toric varieties. The Kreuzer-Skarke database \cite{Kreuzer:2000xy} is a list of $473{,}800{,}776$ four-dimensional reflexive polytopes, of which fine, regular, and star triangulations (FRSTs) determine toric varieties, from which smooth Calabi-Yau threefolds are constructed as anti-canonical hypersurfaces.
	
	Our ensemble consists of Calabi-Yau threefolds with $h^{1,1} = 50, \, 100, \, 200, \, \text{and } 491$.
	We sample from $131$ polytopes ($50$, $50$, $30$, and $1$ at $h^{1,1} = 50, \, 100, \, 200,$ and $491$, respectively). Given a polytope with $h^{1,1}<491$, we generate a set of threefolds using the \texttt{random\_triangulations\_fast} method in \texttt{CYTools}.\footnote{For the unique polytope with $h^{1,1}=491$, we adapt this method.} For polytopes with $h^{1,1} < 491$, we sampled $10$ triangulations per polytope, while at $h^{1,1} = 491$ we sampled $100$ triangulations.  
	
	\paragraph{Assumptions about the UV completion of the Standard Model:}
	In this work, we assume that the Standard Model is realized via a configuration of D7-branes in the compactification manifold. As a general toy model, we assume that QCD is realized on some prime toric divisor in $X$, while QED is realized on an intersecting divisor.\footnote{Note that while the integer charge vector governing the axion coupling to a simple gauge group factor is topological, it jumps under Higgsing. For example, given the axion couplings $(\vec{Q}^\text{ W},\vec{Q}^\text{ Y})$ to the electroweak gauge group $SU(2)\times U(1)_Y$, the coupling to electromagnetism after electroweak symmetry breaking is given by $\vec{Q}^{\text{ EM}}=4\vec{Q}^{\text{ W}}+\vec{Q}^\text{ Y}$. We do not expect that a realization of the QED charge as this linear combination of prime toric divisors versus as a single prime toric divisor will make a difference in our conclusions.}

	We will not explicitly engineer the matter content of the Standard Model in our ensemble. We will, however, make the following assumptions about the spectrum of states above the mass of the Z-boson:
	\begin{itemize}
		\item There exist no vectorlike pairs charged under QCD. This assumption ensures that the strong CP problem is most easily solved by the QCD axion (see \cite{Demirtas:2021gsq} for a discussion).
		\item The Standard Model is completed into a supersymmetric theory at an energy scale $M_{\text{SUSY}}$ defined below.
		\item The fine structure constant of QED in the UV is no less than $1/127.5$, but we impose no further constraints on this coupling.
	\end{itemize}

	\paragraph{Picking points in moduli space:}
	
	The data characterizing the axion effective theory \eqref{eq:canonicalL1} is computed at a particular point in moduli space. In a fully realistic setup, one would explicitly stabilize the K\"ahler moduli, and compute the EFT data in terms of those stabilized values.  But, while the structure of the $4d$ $\mathcal{N}=1$ effective supergravity theories that arise from type IIB flux compactifications in principle allows for a rich landscape of isolated minima for the K\"ahler moduli --- arising for example via a competition between tree level terms and the leading perturbative corrections --- not enough is currently known about the subleading corrections, preventing us from explicitly enumerating such vacuum solutions.\footnote{See however \cite{Berg:2004ek,Berg:2007wt,Shiu:2008ry,Frey:2013bha,Martucci:2016pzt,Gao:2022uop,Kim:2023sfs,Kim:2023eut,Cho:2023mhw}.} On the other hand, we find it plausible that the structure of perturbative corrections, some of which are determined by the choice of flux background \cite{Berg:2004ek,Berg:2007wt,Shiu:2008ry,Frey:2013bha,Martucci:2016pzt,Gao:2022uop,Kim:2023sfs,Kim:2023eut,Cho:2023mhw}, is generic enough to densely populate K\"ahler moduli space. Thus, in this work we will remain agnostic about the precise moduli stabilization mechanism, and instead assume that moduli may get stabilized perturbatively at essentially any point in moduli space. After integrating out the ``saxions'' we are then left with an effective theory of light axions, whose masses are generated non-perturbatively, and our scheme is then to compute axion data at \textit{generic} points in moduli space.
		
	A word of caution is in order: in explicit moduli stabilization schemes, such as KKLT \cite{2003PhRvD..68d6005K} or the Large Volume Scenario (LVS) \cite{2005JHEP...03..007B}, axion physics may look different from what we find here. For example, one key feature of the KKLT setup is that many divisor volumes must be commensurate, so that instanton corrections to the superpotential equitably compete with the flux-induced superpotential. Likewise, in LVS the divisor volumes must be commensurate in order to compete with the perturbative K\"ahler potential correction. This property invariably changes the hierarchical structure of the instanton scales that are observed in the current work. As we are focused on obtaining large-scale statistics across the landscape, we will leave studying axion physics in setups where  moduli stabilization is explicit and fully understood for future work.\footnote{See e.g. \cite{2006JHEP...05..078C, 2012JHEP...10..146C, Broeckel:2021dpz} for work along these lines.}
	
	The most important data that determines axion physics is the spectrum of volumes of divisors, determining the axion potential and the kinetic terms. As explained in \cite{Demirtas:2021gsq}, these are only weakly sensitive to the precise angle in the K\"ahler cone for threefolds with large $h^{1,1}$.
	In this work, we therefore calculate axion EFTs at single points in moduli space, with the understanding that our conclusions will be representative of a uniform sample across the moduli space through our sampling of Calabi-Yau threefolds.
	
	To pick this representative point, we must specify an angle and a radial coordinate within the K\"ahler cone. We choose the angular coordinate as the one defined by the tip of the stretched K\"ahler cone, which is defined as the point closest to the origin such that all 2-cycles have volumes bigger than $1$ in string units \cite{Demirtas:2018akl}.  
	The radial coordinate is then determined by (a) the flux superpotential, and (b) the assumptions on the matter content, as follows:
	\begin{itemize}
		\item We set the flux superpotential $W_0 = 1$ in the axion EFT. This, in turn, sets the scale of supersymmetry breaking to be $M_{\text{SUSY}} \simeq  \frac{M_{\text{pl}}}{\mathcal{V}}$.
		\item We dilate the overall volume of $X$ so that the gauge coupling of QCD matches the observed value in the infrared. We set the volume of $D_{\text{QCD}}$ to be $40$, which is the value that would give the correct IR gauge coupling in the case that the Standard Model is non-supersymmetric to high scales.\footnote{For an arbitrary scale of supersymmetry breaking, one would impose the matching condition
			\begin{align}
				e^{-2\pi \text{vol}(D_{\text{QCD}})} = \left(\frac{m_Z}{M_{\text{SUSY}}} \right)^3 \times \frac{\mathcal{V} \times m_Z^4}{M_{\text{pl}}^3 M_{\text{SUSY}}} e^{-2\pi/\alpha_{\text{QCD}}(m_Z)}
				\label{matching}
			\end{align}
			where $m_Z$ is the mass of the Z-boson and $\alpha_{\text{QCD}}(m_Z)$ is the observed QCD fine structure constant at the scale of the Z-boson.}
	\end{itemize}
	Imposing that $\text{vol}(D_{QCD})=40$ along the ray of the tip of the stretched K\"ahler cone uniquely determines a point in moduli space, and consequently the volumes of all submanifolds. In order to ensure that the $\alpha'$ expansion and instanton series are under control, we further impose that at this location in moduli space, all divisors have volumes bigger than $1$ in string units. Concretely, we calculate all divisor volumes at this point: if the minimal divisor volume is less than $1$, we discard the model.

	To summarize, the pipeline to start with the Kreuzer-Skarke database and arrive at an axion EFT ready for analysis is as follows:
	\begin{enumerate}
		\item Select a polytope from the Kreuzer-Skarke database.
		\item Obtain an FRST of this polytope and construct the associated Calabi-Yau.
		\item Choose, at random, a triple of intersecting divisors to host the Standard Model.
		\item Pick moduli corresponding to the tip of the stretched K\"ahler cone.
		\item Perform a homogeneous dilatation of the Calabi-Yau until $\text{vol}(D_{\text{QCD}})=40$, so that the gauge coupling of QCD in the UV runs to the observed IR value.
		\item Discard the model if any divisors have volumes less than $1$ in string units.
	\end{enumerate}

	\subsection{Kinetic isolation in string theory}
	
	We now return to address kinetic mixing in axiverse models that arise from Calabi-Yau compactifications in the Kreuzer-Skarke database. This discussion will be largely a review of results obtained in \cite{Halverson:2019kna}.
	
	We begin by inspecting the inverse field space metric $K^{ab}$, as given in \eqref{eq:inverse_metric_CY}. Importantly, terms in the matrix $c^{ab}$ with $a\neq b$ are non-vanishing if and only if the basis divisors $D_a$ and $D_b$ intersect, in which case $c^{ab}\equiv \text{Vol}(D_a\cap D_b)$. The diagonal term $c^{aa}$ is interpreted as the volume of the anti-canonical divisor in $D_a$. Thus, the $c^{ab}$ arise from \emph{local} intersections in the Calabi-Yau threefold, while the second term $\frac{\tau^a\tau^b}{\mathcal{V}}$ is non-local, and thus suppressed by the overall volume. In particular, given a pair of small intersecting divisors $D_a$ and $D_b$, i.e.~a pair with $D_a\cap D_b\neq \emptyset$ and $\text{Vol}(D_{a,b})\ll \sqrt{\mathcal{V}}$, the local term dominates over the non-local term. 
	
\begin{figure}
		\centering
		\includegraphics[keepaspectratio,width=6cm,angle=-90,clip,trim=2.3cm 2.3cm 4.2cm 2.3cm]{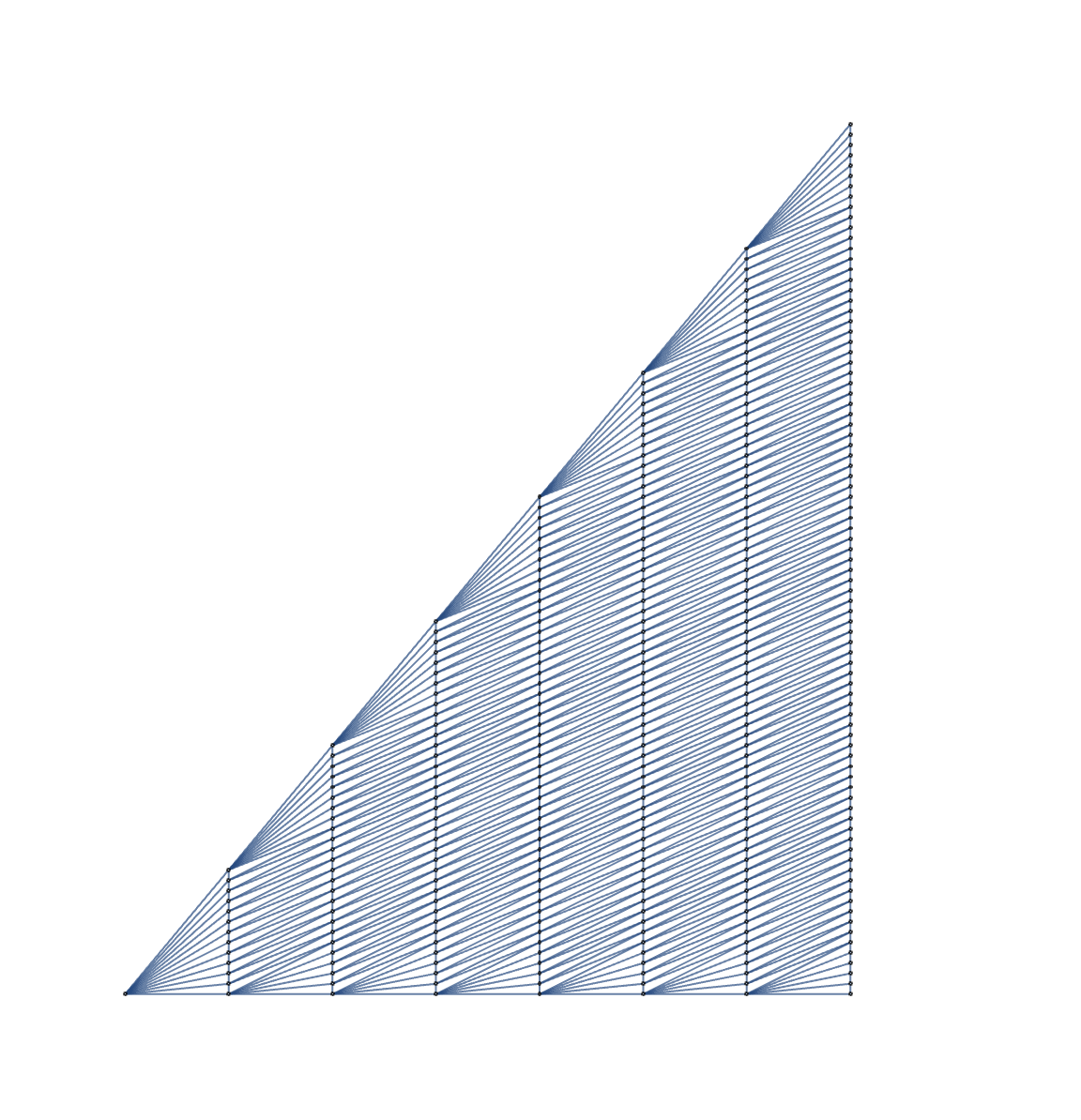}
		\caption{The Delaunay triangulation of the largest two-face of the reflexive polytope $\Delta^\circ_{h^{1,1}=491}$ that yields the largest Hodge number $h^{1,1}=491$ in the Kreuzer-Skarke database. Each point represents a prime toric divisor $D_a$, and each edge between a pair of points indicates that $D_a\cap D_b\neq \emptyset$.}
		\label{fig:two_face_triangulation_491}
	\end{figure}

	Following our discussions in \S\ref{sec:EFT} and \S\ref{sec:KSaxiverse}, the relevant basis of divisors is the one specified by the rows of the reduced instanton charge matrix, i.e.~a certain subset of prime toric divisors. A crucial property of the Calabi-Yau threefold hypersurfaces constructed from the Kreuzer-Skarke database is that the graph of prime toric divisor intersections is sparse at large $h^{1,1}$.  This can be seen as follows. 
	Each of the Calabi-Yau threefolds in question is a hypersurface in a toric fourfold whose toric fan is induced by a suitable triangulation of a four-dimensional reflexive polytope $\Delta^\circ$ (see e.g.~\cite{Demirtas:2018akl}). A pair of prime toric divisors $D_{a,b}$, associated with integer points $p_{a,b}\in \Delta^\circ$, intersects in a non-trivial curve if and only if there exists a two-face of $\Delta^\circ$ containing both $p_a$ and $p_{b}$, and an edge of the two-face triangulation that connects them. At large $h^{1,1}$ most divisors arise from points interior to two-faces of $\Delta^\circ$, and therefore one can think of the relevant portion of the intersection graph as a two-dimensional planar graph, see Figure \ref{fig:two_face_triangulation_491}
	for an example. Clearly, most divisors will intersect only with a few of their neighbors in the two-face of $\Delta^\circ$ and thus one expects the number of non-vanishing off-diagonal entries in the matrix $c^{ab}$ to be of $\mathcal{O}(h^{1,1})$ rather than $\mathcal{O}((h^{1,1})^2)$.
	
	Combining the light threshold effect explained in \S\ref{sec:EFT} with the kinetic isolation explained above,
	one arrives at the following conclusion: \emph{a mixing angle $\Theta_{ab}$, as defined in \eqref{eq:mixing_angles},
		is unsuppressed only if $b\geq a$ --- i.e.~if $\text{Vol}(D_a)<\text{Vol}(D_b)$ --- and $D_a\cap D_b\neq \emptyset$}.

	This is of particular importance for the computation of axion-photon couplings in the Kreuzer-Skarke axiverse. Given the typical distribution of prime toric divisor volumes over a few orders of magnitude at large $h^{1,1}$, and demanding that the Standard Model is not too weakly coupled in the UV, implies that the simple factors of the Standard Model gauge group (or more precisely the possibly grand-unified UV gauge group) must be realized on some of the smallest four-cycles, or else it is hard or impossible to match with the observed IR gauge couplings. 
	
	The suppression of axion-photon couplings in ratios of hierarchical mass scales then implies that only the heaviest axions can couple significantly to electromagnetism. More precisely, assuming that $U(1)_{\text{EM}}$ is embedded in some product UV gauge group $U(1)_{\text{EM}}\hookrightarrow G=G_1\times \cdots \times G_n$, with non-trivial embedding into all $n$ simple factors $G_i$, realized by seven-branes wrapping intersecting four-cycles $\Sigma_1,\ldots,\Sigma_n$, then all axions lifted by Euclidean D3-branes wrapped on cycles with volumes $\leq \max_i(\text{Vol}(\Sigma_i))$, and that also intersect with at least one of the $\Sigma_i$, can couple to electromagnetism without mass or mixing suppression. Typically, the only axions that satisfy these constraints are precisely the ones lifted by Euclidean D3-branes wrapped on the $\Sigma_i$. In a simple model where no unification occurs, i.e.~$G=SU(2)\times SU(3)\times U(1)_Y$, electromagnetism couples significantly only to the axion lifted by the Euclidean D3-brane wrapped on the $SU(2)$-cycle and the one lifted by the Euclidean D3-brane wrapped on the $U(1)$-hypercharge cycle. We will refer to this pair of axions as the \emph{electroweak axions}.

	\section{Axion-photon couplings in the Kreuzer-Skarke axiverse}\label{sec:distributions}
	
	We now examine the couplings of axions in the ensemble constructed in \S\ref{sec:string}.
	The results presented in this section\footnote{Later, in \S\ref{sec:cosmology}, we will present an analysis of cosmological axion dynamics, including production and decay of axion DM, and compare to current limits.  The analysis of \S\ref{sec:cosmology} necessarily involves cosmological assumptions.} do not rely on any cosmological assumptions or modeling, and involve only the microphysical choices --- e.g., the modeling of QED, and the sampling of geometries --- that led to the set of Lagrangians found in \S\ref{sec:string}.  One such choice bears special mention:
	unless stated otherwise, models are required to have
	$\alpha_{\rm EM}(\Lambda_{\rm UV})>\alpha_{\rm EM}(M_Z)$,
	which implies that the volume of the EM divisor must obey $\text{Vol}<127.5$.
	
	\begin{figure}
		\centering
		\includegraphics[width=0.6\textwidth]{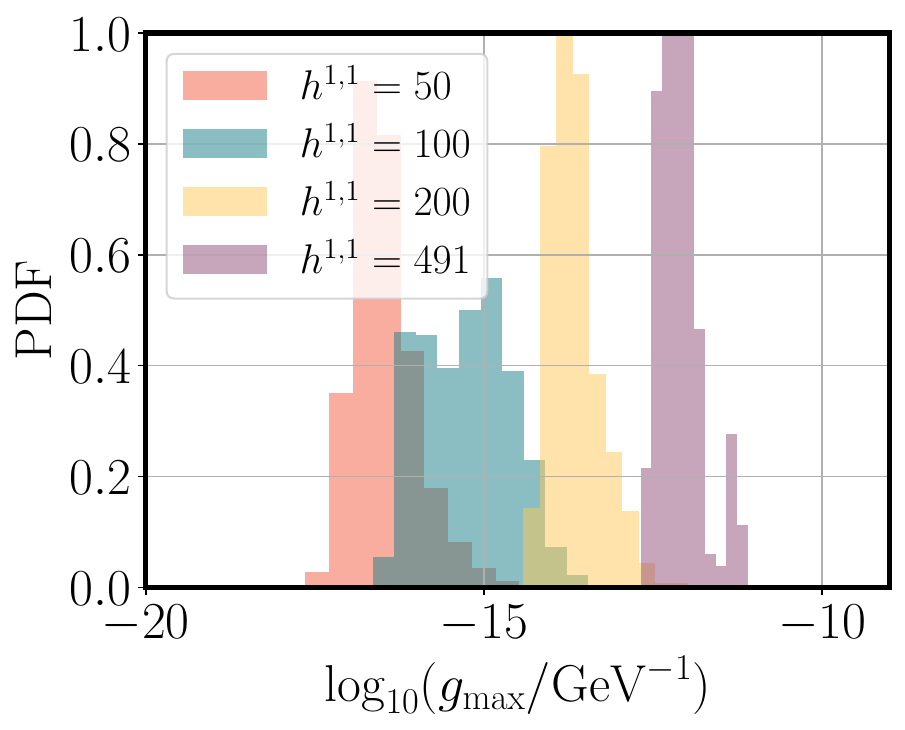}
		\caption{Maximum axion-photon coupling in each model, for reference values of $h^{1,1}$.}
		\label{fig:gMAX}
	\end{figure}

	\begin{figure}
		\includegraphics[width=0.5\textwidth]{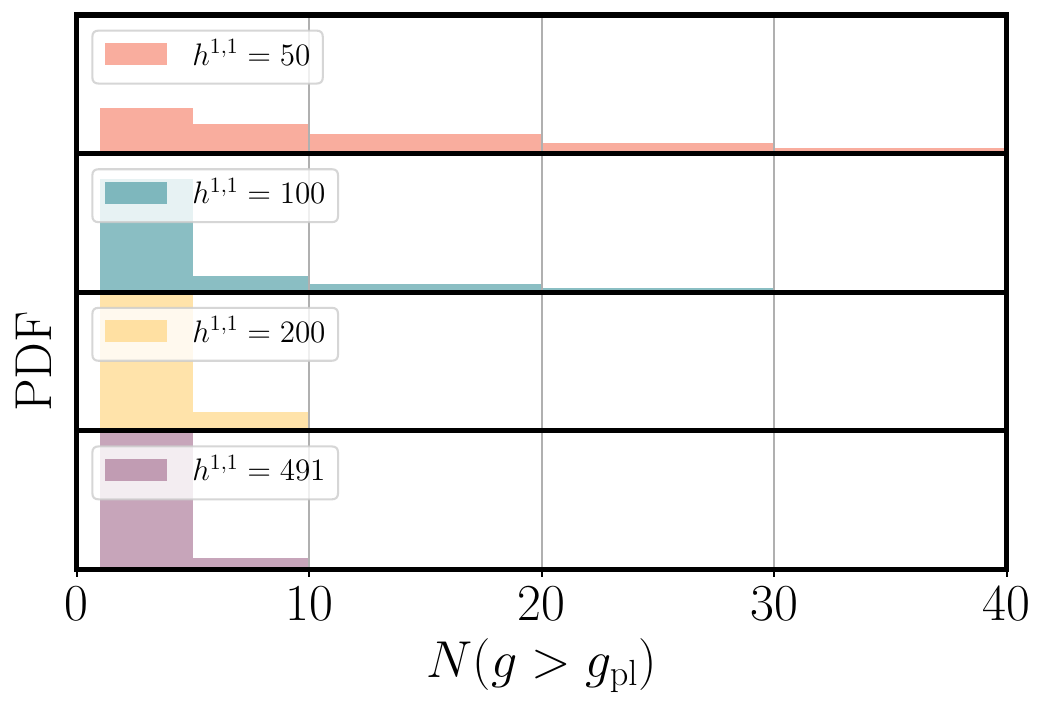}
		\includegraphics[width=0.5\textwidth]{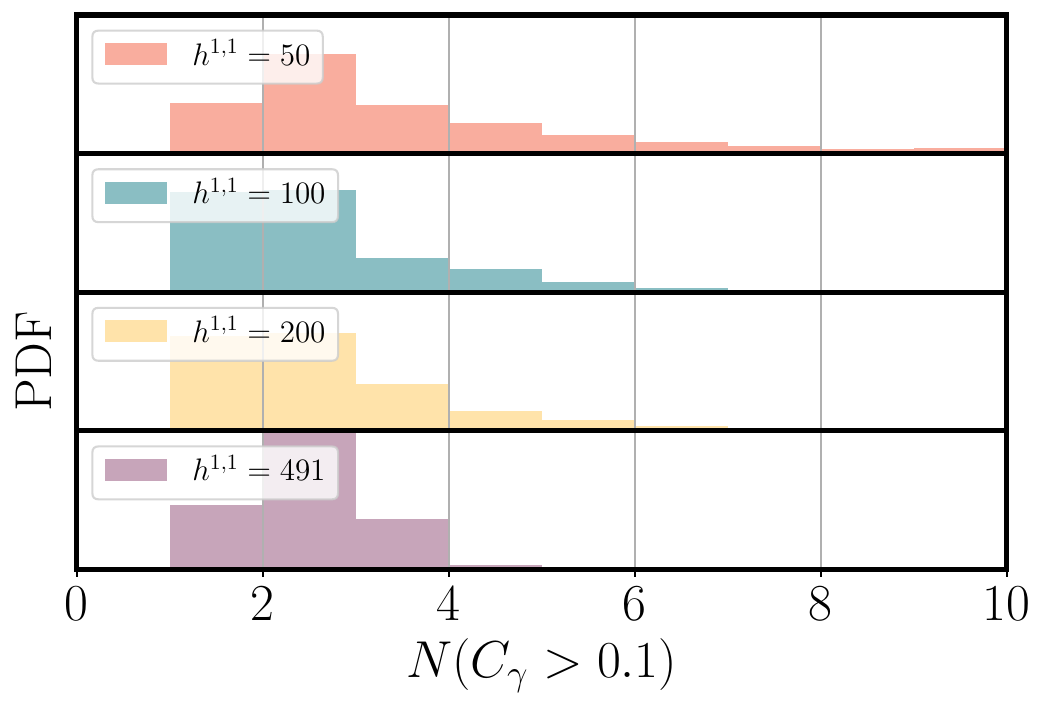}
		\caption{\emph{Left panel:} Number of axions per model that couple to photons with unsuppressed (i.e. $g>g_{\text{pl}}$) couplings, shown for each value of $h^{1,1}$ in our ensemble. \emph{Right panel:} Number of axions per model with dimensionless coupling $C^a_\gamma$  larger than $0.1$, shown for each value of $h^{1,1}$ in our ensemble.}
		\label{fig:num_axions}
	\end{figure}

	\begin{figure}
		\centering
		\includegraphics[width=0.45\textwidth]{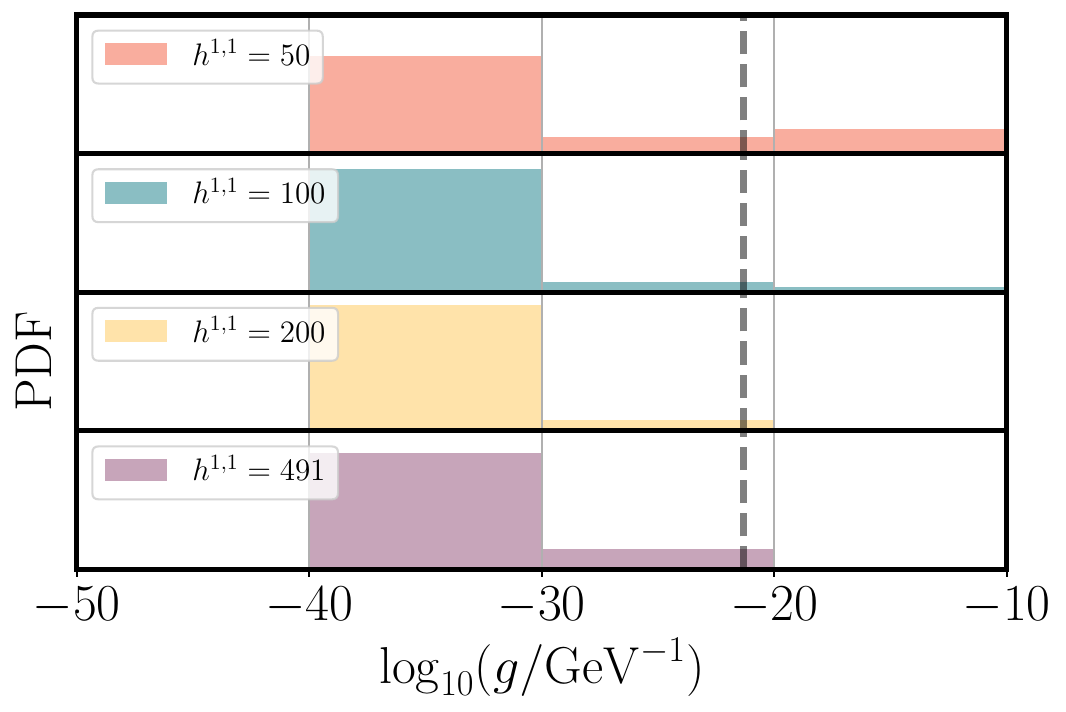}
		\includegraphics[width=0.45\textwidth]{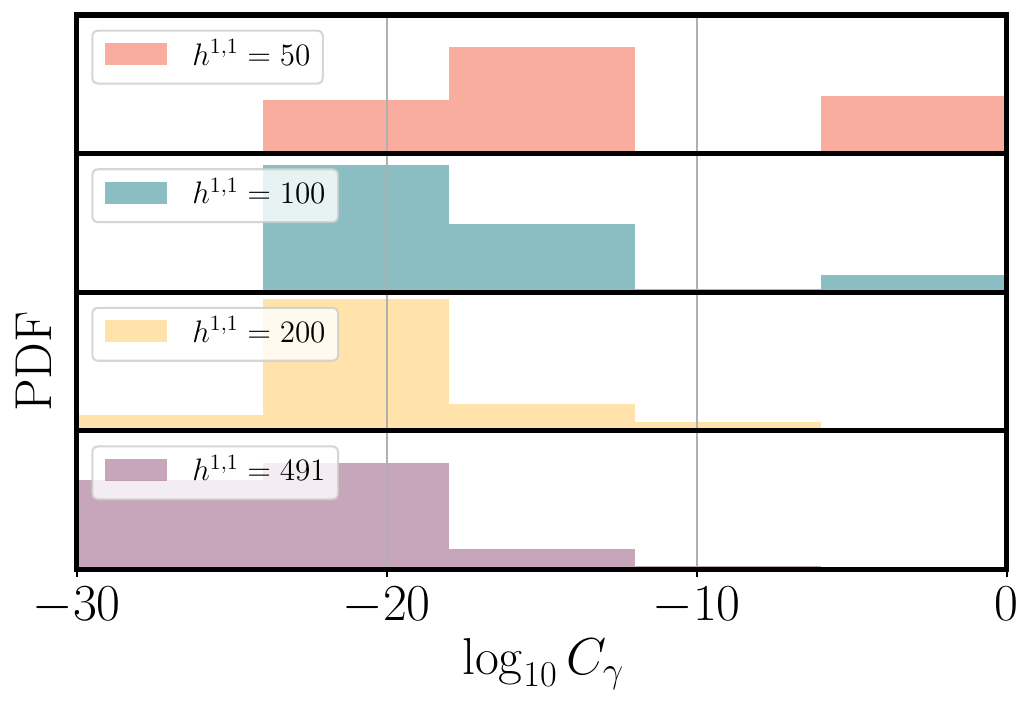}
		\includegraphics[width=0.45\textwidth]{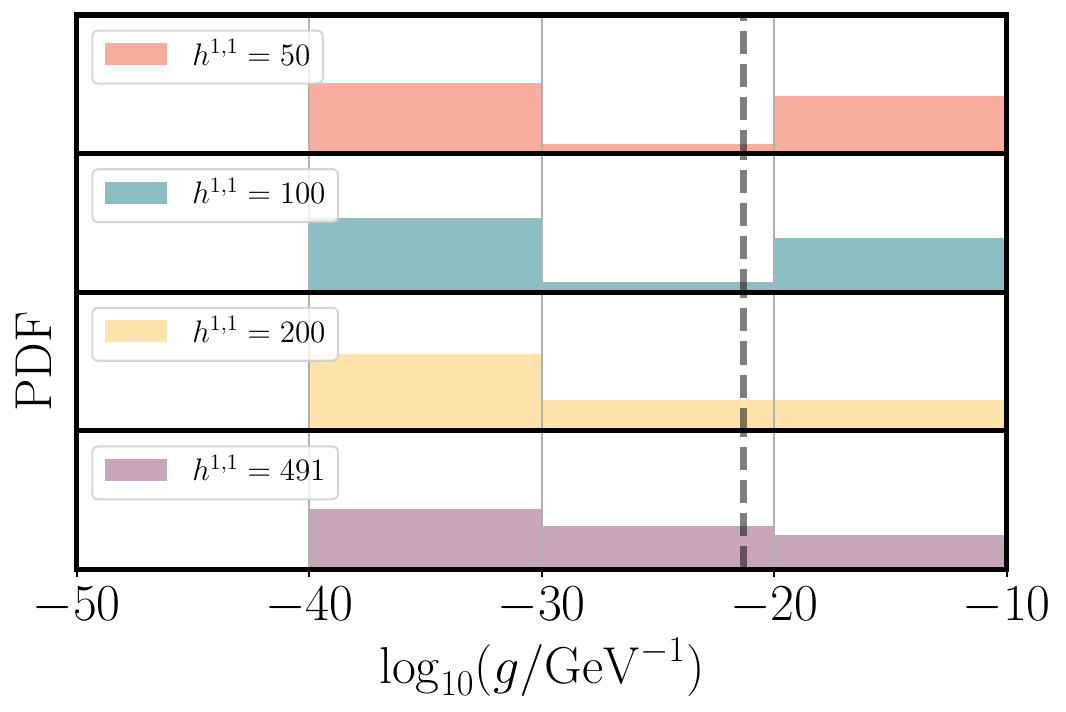}
		\includegraphics[width=0.45\textwidth]{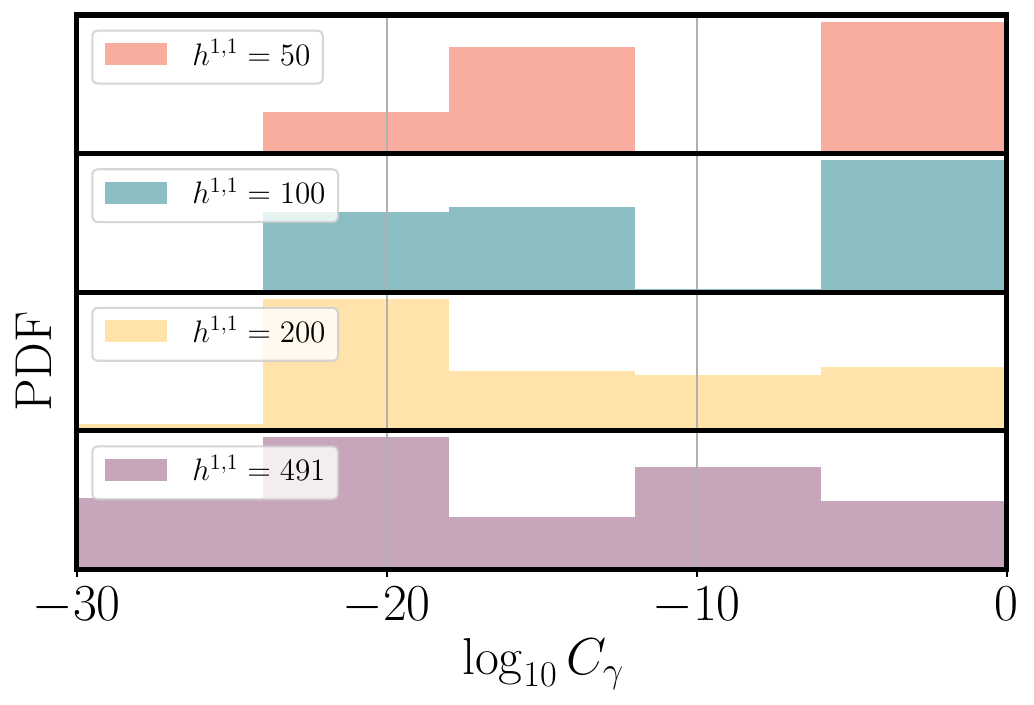}
		\caption{\emph{Left Panel:} Distribution of the dimensionful axion-photon coupling $g$. The vertical dashed line represents a Planck-suppressed value $g_{\text{pl}}:=\alpha/(2\pi M_{\text{pl}})$. \emph{Right panel:} Distribution of the dimensionless axion-photon coupling $C^a_\gamma := 2\pi g_{a\gamma\gamma} f_{a}/\alpha$. In each panel, the left portion of the distribution has extremely small couplings, as a consequence of the light threshold discussed in \S\ref{ss:lightthreshold}.  However, even the right portion of each distribution displays considerable suppression from kinetic isolation (\S\ref{ss:kineticisolation}): in a single-axion theory one expects the quantity $C^a_\gamma$ shown in the right panel to be of order unity. Thus, most of the Kreuzer-Skarke axiverse is invisible. The top row shows distributions cut for QED cycle volume $\text{Vol}<127.5$, while the bottom row has no such cut. } 
		\label{fig:gAll}
	\end{figure}
	
	There are two general trends in axion-photon couplings that are in tension with each other, revealed by scanning over the Kreuzer-Skarke database: first of all, as previously observed in \cite{Halverson:2019cmy,Demirtas:2021gsq}, axion periodicities typically decrease as a function of $h^{1,1}$, which drives axion-photon couplings up. Therefore, Calabi-Yaus with larger $h^{1,1}$ are more constrained by electromagnetic observations. At the same time, couplings are lowered by the two effects explained in \S\ref{sec:EFT} --- namely, kinetic isolation and the light threshold --- whose impact becomes stronger at larger $h^{1,1}$.
	
	These features can be clearly seen in Figs.~\ref{fig:gMAX} and \ref{fig:num_axions}. In Fig.~\ref{fig:gMAX}, we see that the maximum effective axion-photon couplings shift up as $h^{1,1}$ increases. On the other hand, defining $g_{\text{pl}}=\alpha/(2\pi M_{\text{pl}})$ as the threshold of Planck-suppressed couplings, in Fig.~\ref{fig:num_axions} we see that the actual number of axion species that couples with $g_{a\gamma\gamma} > g_{\text{pl}}$ to photons in each of the models we consider remains very low, even as $h^{1,1}$ increases.

	Axions that are suppressed by kinetic isolation, but not by the light threshold, generally maintain $g_{a\gamma\gamma}>g_{\text{pl}}$, i.e.~the kinetic isolation factors  $\varepsilon$ defined in \S\ref{ss:kineticisolation} obey $\varepsilon>f/M_{\text{pl}}$.  Such axions may be visible in some circumstances.
	However, axions below the light threshold have $g_{a\gamma\gamma}\ll g_{\text{pl}}$, and cannot be detected via their axion-photon couplings.
	
	Another useful representation of these suppression effects in the axiverse uses the dimensionless axion-photon coupling, 
	\begin{equation}
		C^a_\gamma := 2\pi \frac{g_{a\gamma\gamma} f_a}{\alpha}\,,
	\end{equation}
	with the axion decay constant $f_a$ defined in \eqref{eq:fdef},
	as shown in Fig.~\ref{fig:gAll} (right panel). We notice two populations: the right population, which is only suppressed by the kinetic isolation effects, and the left population, which is suppressed by virtue of being below the light threshold. Here we can observe the trend that kinetic isolation becomes more severe at large $h^{1,1}$.
	Because the $f_a$ decrease as $h^{1,1}$ increases, $C^a_\gamma$ and $g$ show different trends with $h^{1,1}$.
	In particular, the maximum value of the dimensionful coupling $g_{a\gamma\gamma}$ rises with $h^{1,1}$: the handful of axions with $g_{a\gamma\gamma} > g_{\text{pl}}$ become more strongly coupled to electromagnetism as the overall scale of the cycle volumes increases.  We will revisit this finding in Fig.~\ref{fig:mass_coupling_EM} and the surrounding discussion.

\begin{figure}
		\includegraphics[width=0.5\textwidth]{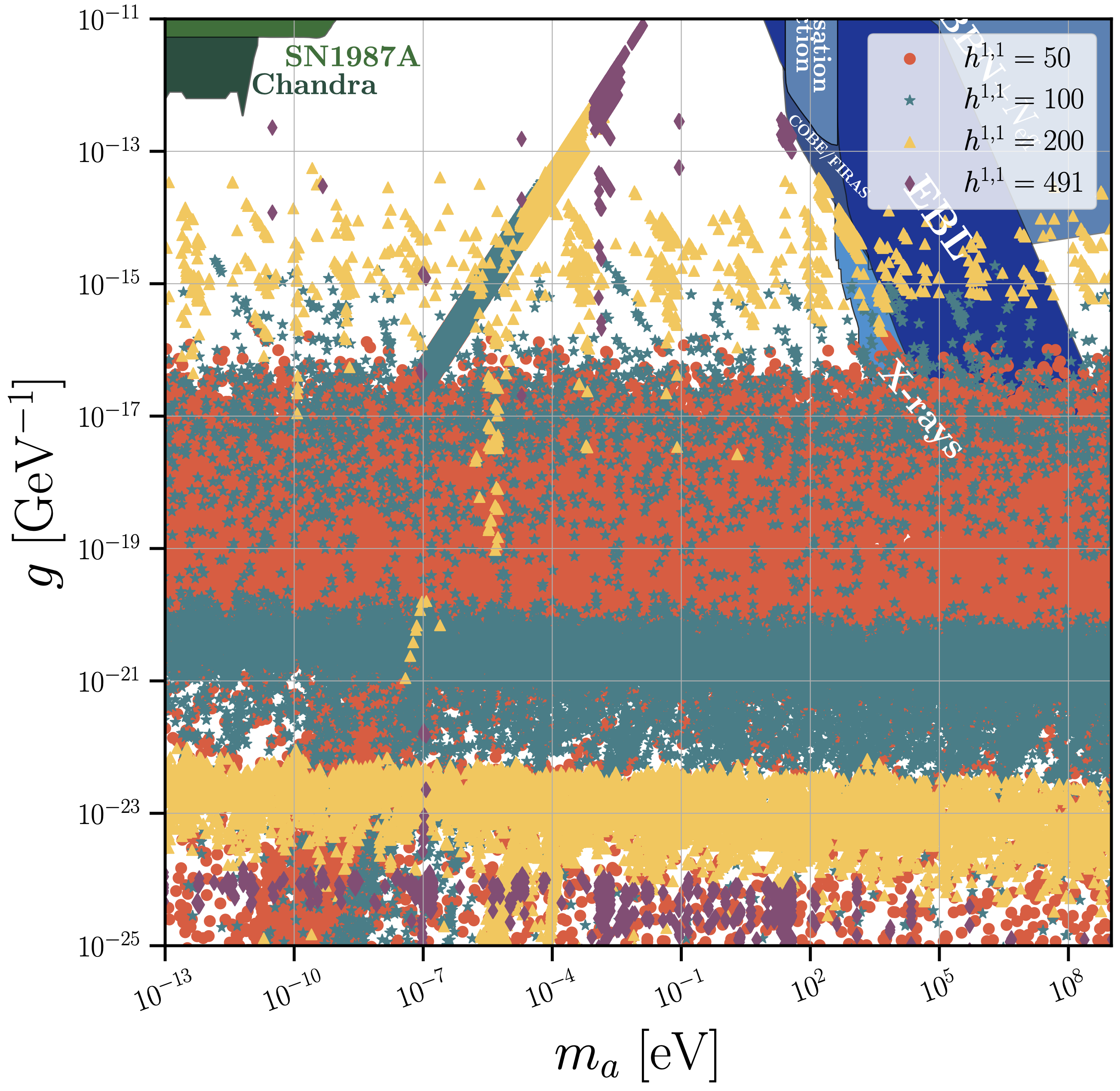}
		\includegraphics[width=0.5\textwidth]{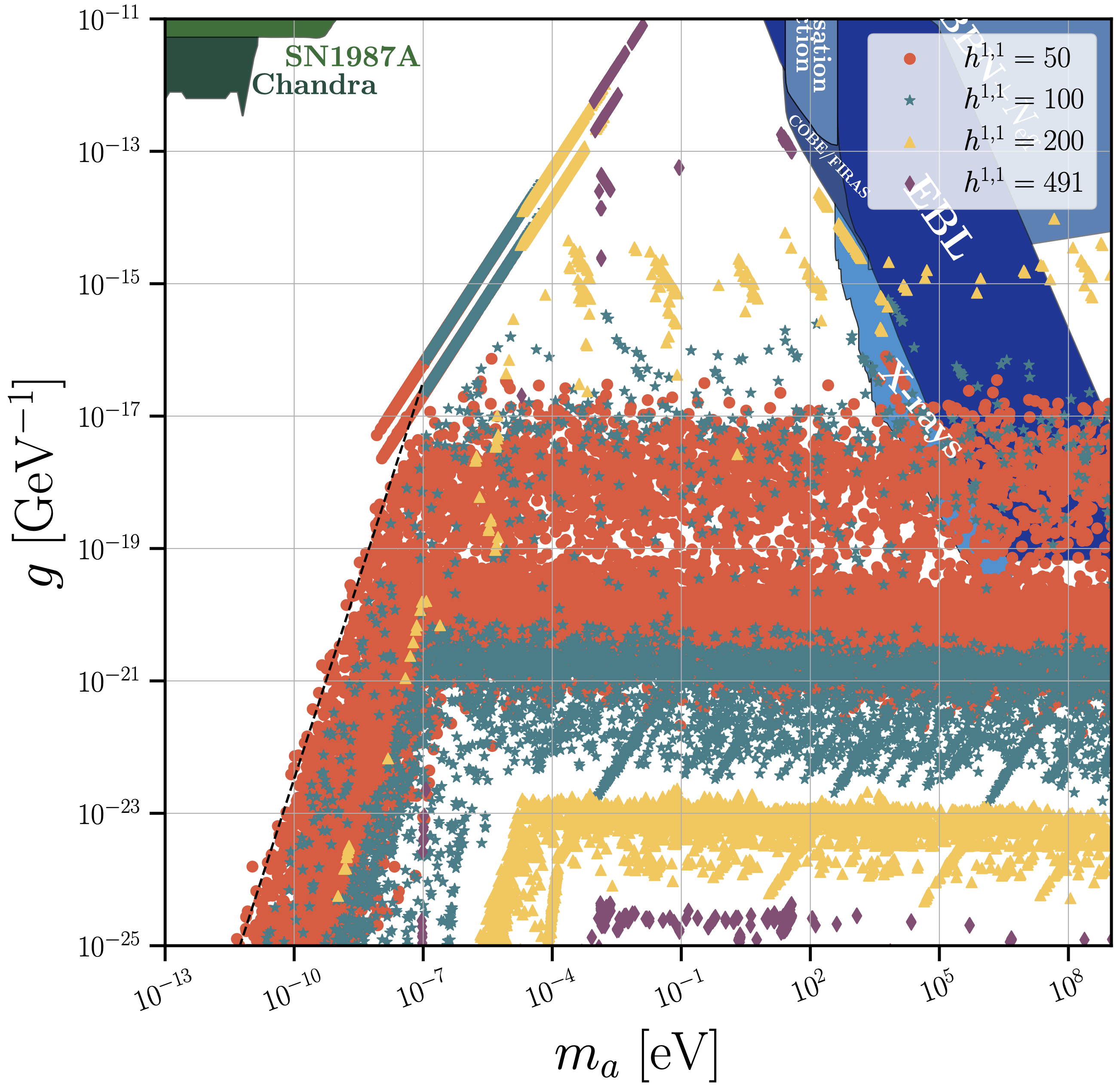}	
		\caption{Distribution of axion mass and coupling for different $h^{1,1}$. \emph{Left panel}: Non-GUT realizations of electromagnetism in the UV. \emph{Right panel}: GUTs. In the right panel, the dashed black line has a slope $g\propto m^{2}$, consistent with the results of Ref.~\cite{Agrawal:2022lsp} for the suppression of low-mass axion couplings below the QCD line. Constraints on the mass and coupling that do not rely on dark matter direct detection (which involves additional assumptions not covered in this work) are also shown, taken from Ref.~\cite{AxionLimits} (where a full list of references can be found). }
		\label{fig:mass_coupling_main}
	\end{figure}

	An important piece of information is to know where the axions in our ensemble fall in the standard exclusion plots for axion-photon couplings as a function of axion mass. These exclusion plots are shown in Figs.~\ref{fig:mass_coupling_main}, along with the axion parameters obtained in the scan, color-coded by $h^{1,1}$.  
	In the left panel, we allow electromagnetism to be on any divisor intersecting QCD, with no restriction on the divisor volume.   In the right panel, we show only models that allow a simple gauge group GUT (i.e.~with electromagnetism on the same cycle as QCD). In GUTs, we observe results consistent with Ref.~\cite{Agrawal:2022lsp}, with no models above the QCD line, and those axions lighter than the QCD axion remaining below a limiting distribution with $g\propto m^2$. Fig.~\ref{fig:mass_coupling_main} makes it clear that the axiverse at low $h^{1,1}$ is unaffected by the astrophysical limits (in green, top left) on light axions. Future searches for axions with DM direct detection, or improved astrophysical searches (above the QCD line, not shown) will probe non-GUT models. On the other hand, cosmological limits (blue, right) can probe the axiverse across all $h^{1,1}$ almost independently of the UV completion of the Standard Model. The consequences of this are explored in \S\ref{sec:decaying_DM_indirect}.
	
	The mass-mixing suppressed population is at such small values of $g$ that it  does not appear in Fig.~\ref{fig:mass_coupling_main}.
	Instead we observe the splitting between the kinetic-mixing suppressed population and the handful of axions with $g_{a\gamma\gamma}\sim \alpha / (2\pi f_a)$ or equivalently $C^a_{\gamma} \sim 1$, i.e.~with an unsuppressed coupling to electromagnetism. The separation between these populations grows with $h^{1,1}$.
	
\begin{figure}
		\centering
		\includegraphics[width=0.6\textwidth]{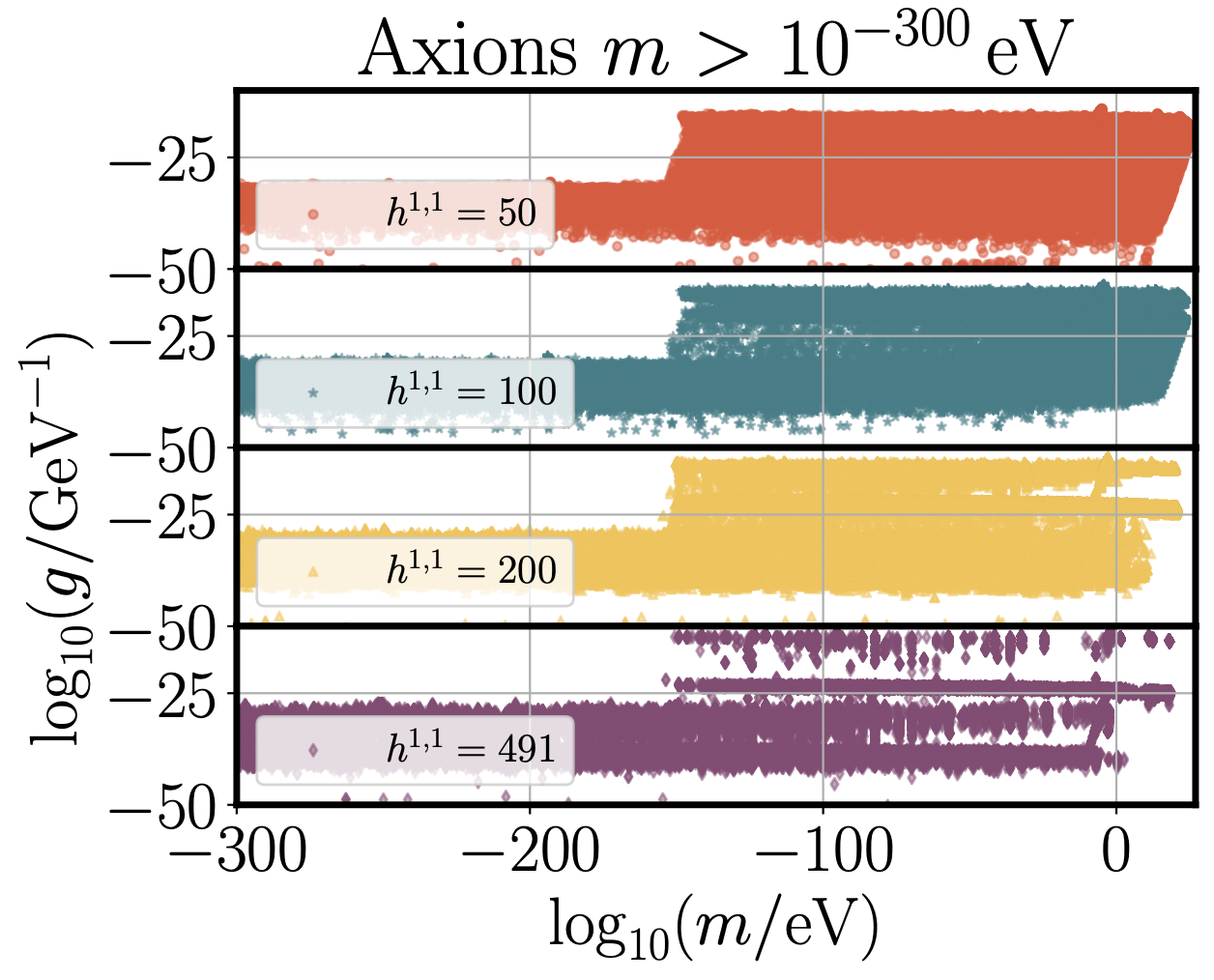}
		\caption{Distribution of axion masses and axion-photon couplings for all axions with $m>10^{-300}\text{ eV}$ (the minimum mass resolution of our computations).  The step occurs at the mass scale $m_{\text{QED}}$, i.e.~the light threshold.}
		\label{fig:mAll_gAll}
	\end{figure}

\begin{figure}
		\centering
		\includegraphics[width=0.6\textwidth]{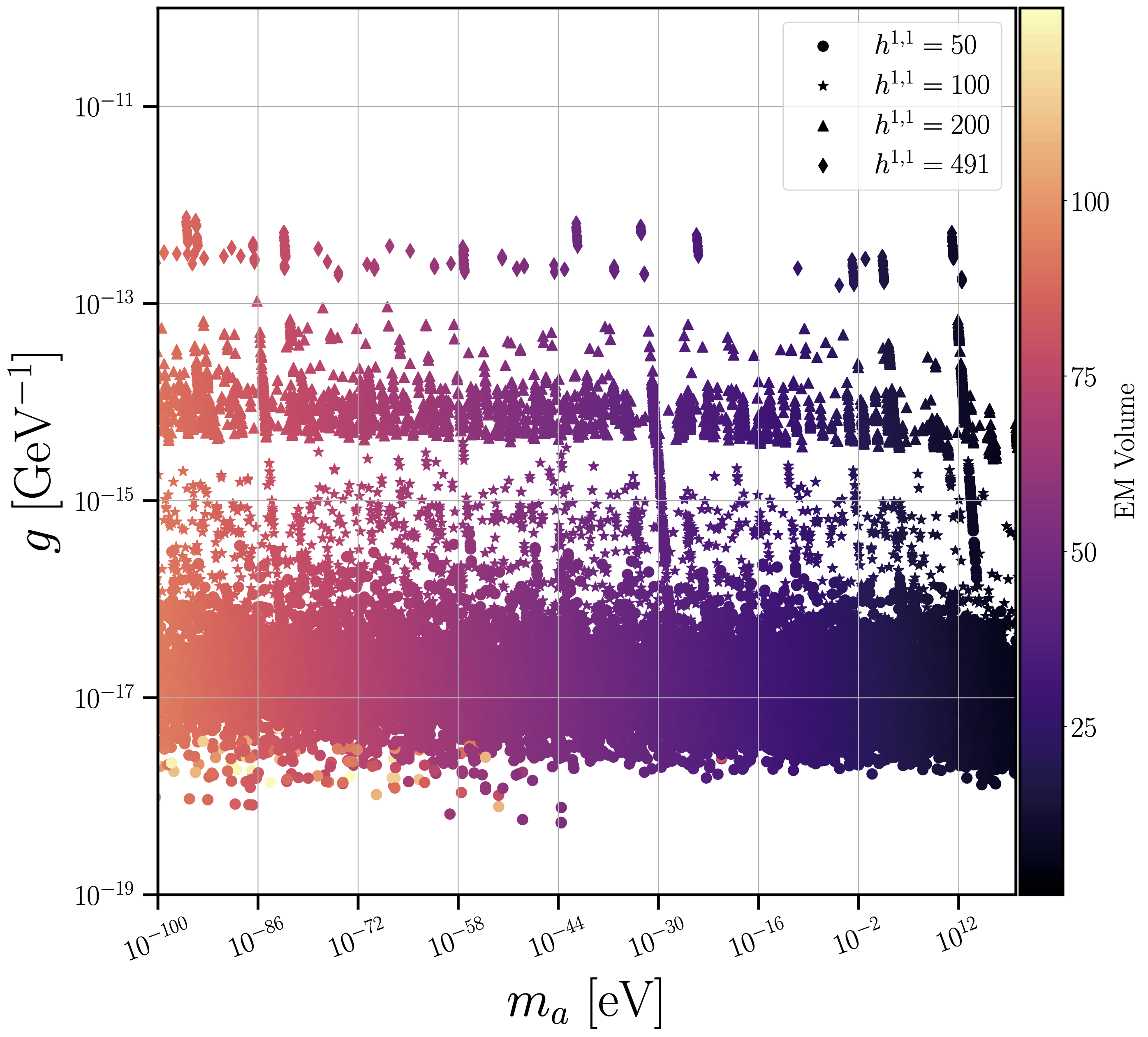}
		\caption{Distribution of axion mass and coupling for different $h^{1,1}$, selecting only those axions for which $C^a_\gamma>0.9$, displaying the dependence on the volume of the divisor selected to host EM. }
		\label{fig:mass_coupling_EM}
	\end{figure}

	For clarity, the same data is displayed for each $h^{1,1}$ in Fig.~\ref{fig:mAll_gAll}. At small $m$, we see that the entire population of axions are below the light threshold and so have suppressed couplings, while at large $m$ the couplings cover a wide range of values.
	Apart from this, when the QED cycle is allowed to have any volume, the distribution of $g$ is largely independent of $m$ and there are visible and invisible axions of all masses.
	The distinct step in the distributions in Fig.~\ref{fig:mAll_gAll} corresponds to the light threshold.
	
	We have seen in \S\ref{sec:EFT} that an axion only couples non-trivially to photons when its mass is larger than the mass set by the instanton associated to the QED cycle in the compactification. Therefore, the axion-photon couplings depend crucially on the volume of the QED cycle. We show this dependence  Fig.~\ref{fig:mass_coupling_EM}. To illustrate the effect clearly, we select only the axions in each model with unsuppressed coupling to EM, defined arbitrarily as $C^a_\gamma>0.9$, and show the distribution extending to very low values of $m\ll H_0$. We observe that axions of extremely low mass can have unsuppressed couplings to EM even for moderately large divisor volumes, here capped at 127.5. Demanding a particular mass range of ultralight axion with unsuppressed EM coupling requires a non-GUT (because such an axion lies above the QCD line, which in our ensemble is not possible in GUTs, cf.~Fig.~\ref{fig:mass_coupling_main}), and a particular value for the EM divisor volume. This result will be of key importance applied to helioscopes, X-ray spectrum oscillations, and birefringence. 
	
	The resulting picture of the axiverse that emerges is one where:
	\begin{enumerate}[label=(\roman*)]
		\item  the distribution of suppressed axion-photon couplings is almost universal and independent of $h^{1,1}$ (see Fig.~\ref{fig:gAll}).
		\item the high end of the axion-photon coupling distribution increases with $h^{1,1}$ (see Fig.~\ref{fig:gMAX}).
		\item  the entire distribution is suppressed due to kinetic isolation, an effect that grows stronger with $h^{1,1}$ (see Fig.~\ref{fig:num_axions}). 
	\end{enumerate}
	We find that models with large numbers of axions are constrained by current and future helioscope bounds, but that only a handful of axions in each model are detectable with these experiments.
	
	\section{Axion cosmology}\label{sec:cosmology}

	We now consider the cosmological dynamics of axions. The possibilities for axion cosmology are vast, and we restrict ourselves here to just two phenomena: production of axion DM by the Primakoff process and vacuum realignment, and the subsequent decay of that DM by axion-photon and axion-axion interactions. The Appendix contains some comments on other phenomena, the detailed study of which we leave for future works.
	
	\subsection{Freeze-in, freeze-out, shake it all about}\label{sec:hokey_kokey}
	
	The axion field can be modeled as a classical condensate plus quantum fluctuations:
	\be
	\varphi = \varphi_{\rm cond}+\hat{\varphi}\, .
	\ee
	The condensate is the vacuum expectation value (one-point function), $\langle\varphi\rangle= \varphi_{\rm cond}$, and we drop the subscript ``cond'' from hereon. The particles are described by the two-point function, $\langle\hat{\varphi}\hat{\varphi}\rangle$, which is related to the standard Boltzmann distribution function in the WKB approximation (see e.g.~Ref.~\cite{Drewes:2012qw}). The distribution function in turn defines the number density, $n_i$, which obeys the collisional Boltzmann equation, while the condensate obeys the Klein-Gordon equation, which itself obeys a Boltzmann equation in the appropriate limit. We ignore interactions between the condensate and the particles (for more on the out-of-equilibrium quantum dynamics of axions see e.g.~Refs.~\cite{Berges:2014xea,Cao:2022kjn}).
	
	The equations of motion giving rise to the relic density are, somewhat schematically:
	\begin{align}
		\partial_t n_a+3H n_a = C[n]\, , \\
		\Box \varphi_a-\Gamma_a\partial_t\varphi_a - \partial_a V = 0\, ,
	\end{align}
	where $C[n]$ is the collision operator for all interactions of axion $i$ with the Standard Model and other axions, and $\Gamma_a$ is the total width of axion $a$, which phenomenologically encodes the possible decay of the condensate (see e.g.~Ref.~\cite{1997PhRvD..56.3258K}). In the following we outline approximate solutions to these equations that allow estimation of the axion relic density.
	
	Axion particles described by $n_a$ can behave as dark matter (DM) or dark radiation (DR) depending on their temperature, while the condensate can behave as DM or dark energy (DE) depending on the value of $\dot{\varphi}$. We parameterize the DM density 
	using the standard relic density parameter, $\Omega_a h^2$ for axion species $a$, defined from the energy density as:
	\be\label{eq:relic_density_parameters}
	\Omega_a h^2 = \frac{\rho_a}{3 M_H^2 M_{\text{pl}}^2} = \frac{\rho_a}{8.1\times 10^{-11}\text{ eV}^4}\, .
	\ee
	We also use the fractional abundance, $\xi_a = \Omega_a h^2/0.12$ where $0.12$ is the measured CDM density today. For unstable axions, i.e.~those with $\Gamma_a>H_0$, following Ref.~\cite{Balazs:2022tjl} we define $\xi$ as the DM density the axions would have today if they were stable, which is a useful way to compare to observational constraints. Thus, it is possible to have $\xi>1$ and not be excluded by the relic density if decays occur sufficiently early.
	
	We consider axion particle production via the Primakoff process: axion-photon conversion mediated by free charged particles. This mechanism leads to either freeze-in or freeze-out abundance of DM and DR, depending on the strength of the coupling and the reheat temperature. Another important process is the axion decay and inverse decay to photons: axions decay when this process comes into equilibrium. Although production via freeze-in is not technically a thermal process, since the axions do not come into thermal equilibrium, we refer to both freeze-in and freeze-out as ``thermal'', since in these processes axions are unavoidably produced by the thermal bath of Standard Model particles (i.e.~can arise from a vacuum initial state).\footnote{We only consider production by the axion-photon coupling. In principle all the axions also have an axion-gluon coupling that we can compute, and a thermal abundance produced from this coupling as well (see Ref.~\cite{Arias:2023wyg} for a recent discusssion). For the values of $f_a$ in our ensemble, production of the QCD axion by either coupling is small. 
		Production of other axions by the gluon coupling is further suppressed by mixing effects.}

	We consider the initial population of axion particles created during reheating to be zero, thus making our limits conservative. As described in Refs.~\cite{Balazs:2022tjl,Langhoff:2022bij}, freeze-in and freeze-out provide an \emph{irreducible} contribution to $\xi$, which is minimized when the reheat temperature is at its lowest observationally allowed value, $T_{\rm re}\approx 5\text{ MeV}$. Freeze-out occurs when the Primakoff process brings axions into thermal equilibrium for temperatures $T<T_{\rm re}$. The Primakoff process then goes out of equilibrium at the freeze-out temperature defined as usual by:
	\be
	H(T_{\rm fo})=\Gamma_{\rm Prim}(T_{\rm fo})\, .
	\ee
	For the Hubble rate, we use the effective number of relativistic degrees of freedom for the energy density, $g_{\star,\rho}(T)$, from Ref.~\cite{Saikawa:2018rcs}. For the Primakoff rate we use the expression given in Ref~\cite{Cadamuro:2011fd} with the effective number of charged degrees of freedom, $g_{\star,Q}(T)$, supplied to us by the authors of Ref.~\cite{Depta:2020wmr}. We note that for string axions the axion-photon coupling we compute is present in the UV, and so unlike field theory axions~\cite{Arias-Aragon:2020shv}, this coupling, and the axion production rate, does not substantially change going through the electroweak phase transition.
	
	After freeze-out, the abundance can be estimated following standard arguments~\cite{1990eaun.book.....K}, and is given by:
	\be
	n_a \approx n_{a,\rm eq}(T_{\rm fo})\frac{T_0^3 g_{\star,S}(T_0)}{T_{\rm fo}^3 g_{\star,S}(T_{\rm fo})}\, ,
	\ee
	where $T_0$ is the temperature today, $n_{\rm eq}$ is the equilibrium abundance, and $g_{\star, S}$ is the effective number of relativistic degrees of freedom for the entropy.
	
	Freeze-in is defined as axion production where thermal equilibrium cannot be established at temperatures $T<T_{\rm re}$, i.e.~$T_{\rm fo}>T_{\rm re}$. In this case, the approximate abundance can be estimated from the equilibrium abundance at $T_{\rm re}$ suppressed by the ratio of the Primakoff rate to the Hubble rate at reheating, and redshifted to today as usual, i.e.~\cite{Hall:2009bx,Jaeckel:2014qea}:
	\be
	n_a \propto n_{a,\rm eq}(T_{\rm re})\frac{\Gamma_{\rm Prim}(T_{\rm re})}{H(T_{\rm re})}  \frac{T_0^3 g_{\star,S}(T_0)}{T_{\rm re}^3 g_{\star,S}(T_{\rm re})}\, .
	\label{eqn:jaeckel_freeze-in}
	\ee
	A slightly better approximation to the Primakoff freeze-in abundance can be found by integrating directly the simplified Boltzmann equation:
	\be
	\frac{dn_a}{dt} +3H n_a = n_{a,\mathrm{eq} \, a}\Gamma_{\rm Prim}\, , \label{eqn:approx_boltzmann}
	\ee
	which treats the axions as massless with $T\gg m_a$. The analytic fit can be found in Ref.~\cite{Balazs:2022tjl} where it was shown that this result, along with Boltzmann suppression for $T<m_a$, agrees  well with the result (including also axion decays and inverse decays, and the muon-induced Primakoff effect) computed using \textsc{micromegas}~\cite{Belanger:2001fz}. In this work, we found that the general form of Eq.~\eqref{eqn:jaeckel_freeze-in} is most easily adaptable to a wide range of masses and axion-photon couplings, and matches the fits in Ref.~\cite{Balazs:2022tjl} by fixing the constant of proportionality in Eq.~\eqref{eqn:jaeckel_freeze-in} to 0.16.
	
	We assume that freeze-in produces axions with an effective freeze-in temperature $T_{\rm fi}$ obeying $T_{\rm fi} \approx T_{\rm re}$ (i.e.~we assume kinetic equilibrium, which is implicit in Eq.~\ref{eqn:approx_boltzmann}), and then redshift $T_{\rm fi}$ to calculate the present-day temperature of this population as if it decoupled at $T_{\rm re}$ (i.e.~not sharing in entropy production). After freeze-in or freeze-out production, we track the temperature of the produced axions by assuming $T\propto a^{-1}$ for $T>m$ and $T\propto a^{-2}$ for $T<m$. Freeze-in and freeze-out produce relativistic axions as DR and contribute to $N_{\rm eff}$ if $T>m$ at the present day. We do not otherwise separate between hot, warm, and cold DM, and leave a detailed study of the DM composition to a future work.
	
	A complication in multi-axion theories compared to those with a single axion is the difference between the interaction and mass eigenbasis.
	Naively, one might expect to produce $N$ axion degrees of freedom from the Primakoff interaction, for $N$ mass eigenstates. However, the direction in field space coupled to $F\tilde{F}$ is only one direction in flavor space, so on the other hand one might expect to produce just one thermalized degree of freedom. As we now argue, the actual situation is a mix between the two.
	
	Firstly, we note that the Boltzmann equations are not field redefinition invariant, and do not necessarily respect the symmetries of the relativistic Lagrangian. A full treatment of axion production in a multi-field model is beyond the scope of this work, and we content ourselves with approximations that we now justify.
	
	In the massless limit and considering only the EM interaction, there is a conserved charge corresponding to the freedom to rotate the axion fields. In this limit, only the single degree of freedom coupled to EM will be produced, and the Boltzmann equation should be solved in the EM basis to guarantee the conservation of charge. Mass terms break the rotational symmetry and charge conservation, since in the EM basis there are mass mixing interactions. We should thus ask what the scale of symmetry breaking is. If mass mixing strongly breaks the flavor symmetry, then we should consider Primakoff production of individual mass eigenstates, while if mass mixing is only a soft breaking, then these axions are produced as a single degree of freedom in the flavor basis, which we term \emph{collective Primakoff production}.
	
	We make the split between production of mass eigenstates and collective Primakoff production by comparing the mass mixing timescale $\Gamma_{\rm mix}$ to the Primakoff rate, $\Gamma_{\rm Prim}(T_f)$ at freeze-out or freeze-in (whichever is the lower temperature). If the mixing timescale is much faster than the Primakoff production rate, then the Primakoff interaction will ``see'' individual mass eigenstates. If instead the mixing timescale is much slower, then mixing only happens after the Primakoff interaction has frozen out, leading to the single flavor eigenstate being produced.  That is:
	\begin{align}
		\frac{\Gamma_{\rm mix}}{\Gamma_{\rm Prim}}>1 &\Rightarrow \text{mass eigenstates produced,} \label{eqn:mass_Primakoff} \\
		\frac{\Gamma_{\rm mix}}{\Gamma_{\rm Prim}}<1 &\Rightarrow \text{flavor eigenstate produced.}\label{eqn:collective_Primakoff}
	\end{align}
	The conditions Eqs.~\eqref{eqn:mass_Primakoff} and \eqref{eqn:collective_Primakoff} should be checked for all $N$ axion directions. The most naive expectation is that $\Gamma_{\rm mix}=m_i$ for axion $i$. In fact, in analogy to neutrino mixing and photon/dark-photon kinetic mixing, the axion polarization vector rotates at a rate given by Stodolsky's equation~\cite{Stodolsky:1986dx,Jaeckel:2008fi}. Ignoring damping due to axion-axion scattering, assuming an $\mathcal{O}(1)$ mixing angle between the EM basis and the mass eigenbasis, and in the hierarchical approximation for axion massses, the rate of rotation into a given mass eigendirection is approximated by:
	\be
	\Gamma_{{\rm mix},a} \approx \frac{m_a^2}{\omega_a}\approx \frac{m_a^2}{\sqrt{m_a^2+T^2}}\, .\label{eqn:mixing_timescale}
	\ee
	
	The mixing rate, Eq.~\eqref{eqn:mixing_timescale}, scales as $m$ for $m\gg T$ and is suppressed for $m\ll T$. This fits an intuition that cold axion DM is produced in the mass eigenbasis. On the other hand, we find that \emph{Primakoff production produces only one degree of freedom of relativistic axions}. The Primakoff interaction typically freezes out at temperatures far above the top quark mass, implying that axion DR will only ever contribute of order $\Delta N_{\rm eff}=0.027$ regardless of how many axions there are. This is explored in more detail in Appendix~\ref{appendix:additonal_pheno}.
	
	The condensate leads to production of axion DM via vacuum realignment~\cite{1983PhLB..120..133A,1983PhLB..120..137D,1983PhLB..120..127P}. Misalignment-produced DM arises due to coherent motion of the axion field, damped by the expansion of the Universe. Working in the mass eigenbasis introduced in \S\ref{sec:EFT}, an axion field $\varphi^a$ of mass $m_a$ starts oscillating  at a temperature $T_{\rm osc}$ defined by
	\be
	3H(T_{\rm osc}) = m_a\, ,
	\ee
	from some initial expectation value $\langle \varphi^a \rangle$, naturally expressed in units of the axion decay constants $f_a$ via dimensionless angles 
	\begin{equation}
		\vartheta^a:=\frac{\langle \varphi^a \rangle}{f_a}\, .
	\end{equation}
	The oscillation of the axion fields around the minimum of the potential are well approximated by the quadratic approximation if $\vartheta^a\ll 1$. For simplicity, even for $\vartheta^a=\mathcal{O}(1)$ we will work in the quadratic approximation, accepting an $\mathcal{O}(1)$ error in the resulting DM abundance. 
	
	In this approximation, the energy density at some lower temperature $T<T_{\rm osc}$ given by
	\be
	\rho(T) = \frac{1}{2}\Lambda_a^4 \vartheta_a^2 \frac{g_{\star,S}(T)}{g_{\star,S}(T_{\rm osc})}\left(\frac{T}{T_{\rm osc}}\right)^3\, .
	\label{eqn:misalignment_density}
	\ee
	Generically, one expects $\theta^a=\mathcal{O}(1)$ for all axions of masses $m_a\lesssim H_I$, where $H_I$ denotes the Hubble scale during inflation, and $\vartheta^a\approx 0$ otherwise. Theta angles much smaller than this --- or very close to $\pi$ --- amount to a fine tuning of initial conditions. Unnaturally small $\vartheta^a$ would yield smaller DM abundances. We will also discuss cosmic birefringence induced by axion misalignment in \S\ref{sec:birefringence}.

	Finally, we note that in the initial field displacements $\vartheta^a$ we also include inhomogeneous inflationary fluctuations $\delta \vartheta^a\sim \frac{H_I}{2\pi f_a}$, which satisfy $\delta \vartheta^a\ll 1$ under the standard assumption that the Hubble scale during inflation lies below the cutoff of four-dimensional effective field theory, i.e.~$H_I\ll m_{KK}\lesssim f_a$.\footnote{We thank Andrew Long for a useful discussion on this point.}

	\subsection{Decays}\label{sec:decay_problem}
	
	The axions produced as described above can subsequently undergo one-particle to many-particle decays, either by the axion-photon vertex \eqref{eq:axion_photon_vertex}, or by axion flavor changing interactions \eqref{eq:cubic_quartic_int}, as discussed in \S\ref{sec:EFT}.
	
	In order to compute the total decay width $\Gamma_a$ for an axion $\varphi^a$ we compute the sum of the partial decay widths to two photons \eqref{eq:decay_width_axion_phothon}, two lighter axions \eqref{eq:cubic_decay_rate} and three lighter axions \eqref{eq:quartic_decay_rate}, and denote the branching ratios for process $X$ as $\text{BR}_X = \Gamma_X/\Gamma_a$.
	
	If all decay constants $f_a$ were the same, and likewise all instanton scales $\Lambda_a$ were of the same order $\Lambda_a\simeq \Lambda$, the mixing angles $\Theta_{ab}$ in \eqref{eq:mixing_angles} were all $\mathcal{O}(1)$, and furthermore the CP-breaking spurions $\zeta_a$ in \eqref{eq:CP_breaking_spurion} were $\mathcal{O}(1)$, one would conclude that
	\begin{equation}
		\text{BR}_{a\rightarrow bb}\simeq 1\, ,\quad \text{BR}_{a\rightarrow \gamma\gamma}\simeq \frac{\alpha_{\text{EM}}}{64\pi^2}\sim 10^{-5}\, ,\quad \text{BR}_{a\rightarrow bbb}\simeq \frac{\Lambda^4}{8\pi^2f^4}\ll 1\, ,
	\end{equation}
	and thus cubic decays would dominate the total decay width. In contrast, in our ensemble of string compactifications \S\ref{sec:string}, we find that the hierarchical suppression effects discussed in \S\ref{sec:EFT} render cubic decays negligible, and leaving quartic decays and anomalous decay to two photons as the leading \emph{known} decay channels in our models. However, our string models are incomplete and in particular do not take into account the possibly generic presence of dark sector D-branes, which could host dark $U(1)$'s and/or confining gauge groups. Thus, our computation of partial decay widths generally yields a lower bound for the total axion decay widths.

	In general an axion $\varphi^a$ decays when $\Gamma_a=H(T_{\rm dec})$. We will assume that the relic abundance of axions is created before decays and inverse decays lead to re-equilibration, i.e.~$T_{\rm dec}<T_{\rm osc,fi,fo}$.
	In particular, we will neglect all populations of axions
	created at $T_{\rm dec}>T_{\rm osc,fi,fo}$. 
	When the time scale of axion decays is sufficiently separated from axion production, then the decay products can have phenomenological consequences, which we now outline.

	We assign to each axion $\varphi^a$ a parameter $\xi_a$ described below \eqref{eq:relic_density_parameters}, which measures the relic density the axion would have if it were stable. As long as the axion remains sub-dominant to radiation, and decays before matter-radiation equality at $T\sim 1\text{ eV}$, then $\xi_i>1$ can be consistent with observations. Constraints on such a scenario can be readily obtained from studies of decaying DM in the literature. Constraints depend on the branching ratios to photons versus axions, and on whether or not the decay products have time to thermalize. The phenomenology of decaying axion DM is discussed in \S\ref{sec:decaying_DM_indirect}.

	
	\section{Axiverse photon physics}\label{sec:axion-photon-physics}
	
	This section outlines some of the phenomenology that
	we regard as the most
	promising to probe the Kreuzer-Skarke axiverse.
	We do not discuss black hole superradiance, as this was covered in detail in Ref.~\cite{Mehta:2021pwf}. Some additional phenomenology is treated briefly in Appendix~\ref{appendix:additonal_pheno}. As above, unless stated otherwise, models are required to have the volume of the EM divisor obeying $\text{Vol}<127.5$, so that $\alpha_{\text{EM}}(\Lambda_{\text{UV}})>\alpha_{\text{EM}}(M_Z)$.
	
	\subsection{The QCD axion}\label{sec:qcd_axion_details}
	
	\begin{figure}
		\centering
		\includegraphics[width=0.6\textwidth]{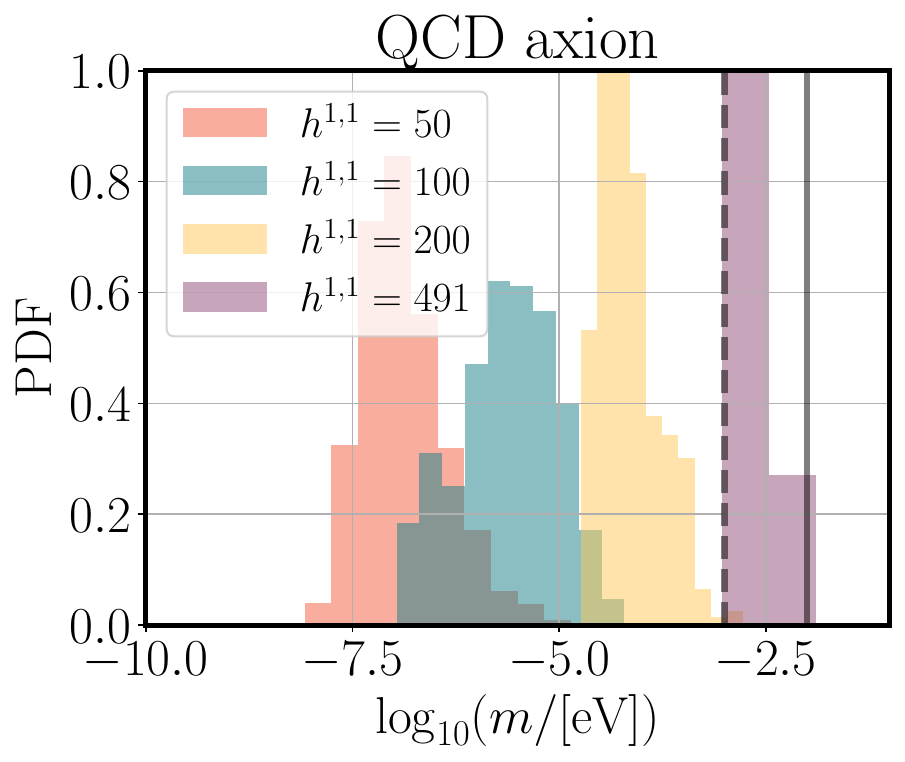}
		\caption{Distribution of QCD axion mass for different $h^{1,1}$. The vertical solid line shows the upper limit on the mass $m<10\text{ meV}$~\cite{Lella:2023bfb}, which applies assuming that no flavor-specific fine tuning of the axion couplings to the $u$ and $d$ quarks leads to cancellations in the axion-nucleon coupling (see Ref.~\cite{diCortona:2015ldu}). The dashed line shows a forecast for the limit from a future galactic supernova neutrino burst measured by Hyper Kamiokande with percent level determination of the luminosity~\cite{GalloRosso:2017hbp}.}
		\label{fig:mQCD}
	\end{figure}
	
	In our model the QCD axion coupling to EM is given by the UV contribution from D7-branes (which gives $C^a_\gamma=\pm 1$ in a GUT, or $|C^a_\gamma|<1$ from kinetic isolation or the light threshold) plus the additional contribution arising from mixing of the QCD axion with the pion in the IR, mediated by the axion-gluon coupling. This means that our models of the QCD axion populate a narrow region of the QCD axion model band~\cite{DiLuzio:2016sbl,DiLuzio:2020wdo,Plakkot:2021xyx}. However, we have not included the possibility of additional charged fermions interacting with the QCD axion (or indeed with any axions in our constructions). Our values for the coupling are thus \emph{lower bounds} (barring accidental cancellations for particular anomaly ratios), and in particular the QCD axion in the axiverse can populate the entire range of the model band.
	
\begin{figure}
		\centering
		\includegraphics[width=0.6\textwidth]{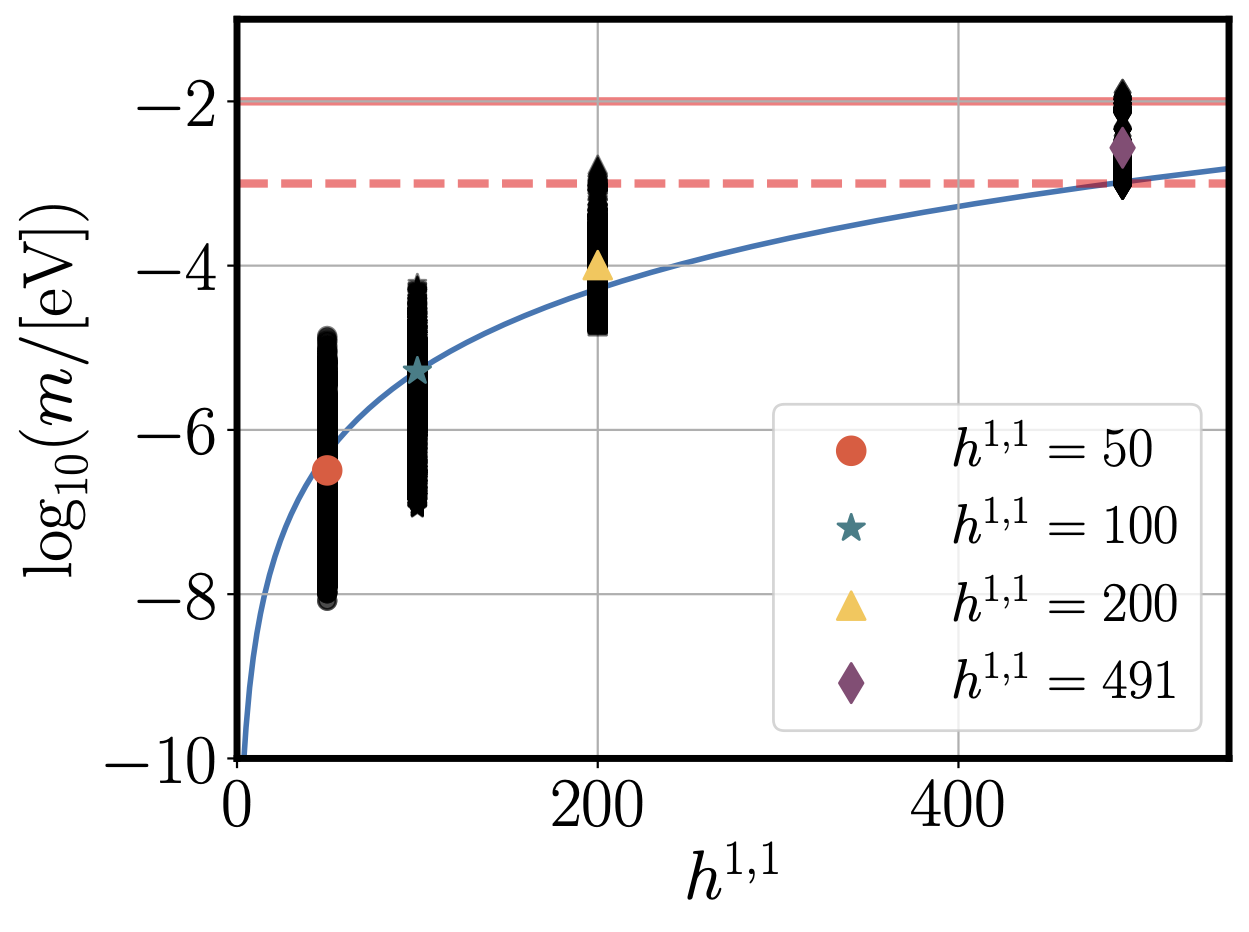}
		\caption{Distribution of QCD axion mass for different $h^{1,1}$. The solid line shows the expected trend $\langle m\rangle \sim (h^{1,1})^{10/3}$~\cite{Demirtas:2018akl}, normalized to the mean of the distribution at $h^{1,1}=100$, which is a reasonable fit to the data at other $h^{1,1}$. The horizontal lines show the upper limit on the mass $m<10\text{ meV}$ from the SN1987A neutrino burst~\cite{Lella:2023bfb} (solid), and a projection for the limit from a hypothetical future galactic
			supernova measured by Hyper Kamiokande with percent-level precision~\cite{GalloRosso:2017hbp} (dashed). At $h^{1,1}=491$ we show the results of 100 triangulations, with the predicted mean and 1 dex error.}
		\label{fig:mQCD_Hodge}
	\end{figure}

	What our constructions predict for the QCD axion without model ambiguity is the decay constant, and thus the axion mass, which we compute at zero temperature using the next to leading order chiral perturbation theory result from Ref.~\cite{diCortona:2015ldu} (for the realignment relic density we use a simple $m(T)\sim T^{-4}$ model for $T>\Lambda_{\rm QCD}\sim 200\text{ MeV}$). Fig.~\ref{fig:mQCD} shows the resulting distribution of the QCD axion mass, which follows a peaked distribution in log-space spanning around two orders of magnitude for a given $h^{1,1}$. The mean value of $m$ increases with $h^{1,1}$. The increase in $m$ with $h^{1,1}$ follows from the trend in $f$ as a function $h^{1,1}$ observed in our previous studies~\cite{Demirtas:2018akl,Mehta:2021pwf,Demirtas:2021gsq}:
	namely, $f$ decreases with increasing $h^{1,1}$, as a consequences of the narrowing K\"{a}hler cones and increasing Calabi-Yau volumes at large $h^{1,1}$.
	The predicted scaling is $\mathcal{V}\sim (h^{1,1})^5$, $\langle f \rangle \sim \mathcal{V}^{-2/3}\Rightarrow \langle m\rangle \sim (h^{1,1})^{10/3}$~\cite{Demirtas:2018akl}, which is illustrated in Fig.~\ref{fig:mQCD_Hodge}. The trend describes the data well, with a $\mathcal{O}(1)$ dex spread at each $h^{1,1}$. At $h^{1,1}=491$, the distribution is very close to the observational upper limit. The QCD axion mass therefore serves as a powerful tool for probing $h^{1,1}$ of a Calabi-Yau:
	experimental upper bounds on the QCD axion mass can be used to set upper bounds on $h^{1,1}$, and disfavor large Hodge numbers, within the context of the compactifications considered here.
	
	A powerful constraint on the QCD axion, which is independent of the DM relic abundance, arises from the duration of the neutrino burst from the galactic supernova SN1987A~\cite{Raffelt:1987yt,Burrows:1988ah}, reviewed in Ref.~\cite{2008LNP...741...51R}. In order that axion emission does not cause the neutrino burst to be shorter than its observed duration, the luminosity in axions, $L_a$, in the supernova core, at the `post bounce time' of $t\approx 1\text{ s}$, should be lower than the luminosity in neutrinos, $L_\nu$. By numerically modelling  both emission channels, and the supernova evolution including the nuclear equation of state, neutrino transport and general relativistic effects (see e.g.~Ref~\cite{Fischer:2016cyd}), Ref.~\cite{Chang:2018rso}, in combination with the Particle Data Group~\cite{Workman:2022ynf}, constrain the QCD axion mass to be $m_a\leq 20 \text{ meV}$, corresponding to $f_a\geq 4\times 10^8\text{ GeV}$. Ref.~\cite{Lella:2023bfb} have improved the limit to $m_a\leq 10 \text{ meV}$ by including axion emission by pion Compton processes~\cite{Lella:2022uwi} in addition to nuclear Bremsstrahlung. This limit is indicated on Figs.~\ref{fig:mQCD} and~\ref{fig:mQCD_Hodge} with solid lines.
	
	The limits on the QCD axion mass apply for the KSVZ axion model, and can be rescaled by $(C_{\rm KSVZ}/C_N)$ when nuclear processes dominate (when including pion Compton processes the rescaling is slightly different), where $C_N$ is the axion-nucleon coupling constant,
	\begin{equation}
		C_N^2 = Y_n C_n^2+Y_p C_p^2\, ,
	\end{equation}
	with $Y_i$ the neutron/proton abundance (in this case, that estimated in SN1987A) and $C_i$ the axion-proton/neutron coupling. In the Kreuzer-Skarke axiverse,
	theories of $h^{1,1} \gtrsim 15$ axions realize the Peccei-Quinn solution to the strong CP problem \cite{Demirtas:2021gsq}, and in such cases the dominant source of mass for the QCD axion comes from its coupling to gluons.  We therefore do not include the effects of axion-fermion couplings. Including only the axion-gluon coupling, our fiducial QCD axion model thus resembles the KSVZ axion with $C_n=-0.02$ and $C_p=-0.47$~\cite{diCortona:2015ldu}. Tree-level axion-fermion couplings alter the value of $C$ away from its reference KSVZ value, with the up and down quark couplings contributing with opposite sign to those from the other quarks and from the gluons. It is thus possible that the bare quark couplings could be tuned to avoid the SN1987A bound on the mass by causing cancellations that set $C_n\approx C_p\approx 0$. We take the bound  $m\lesssim 10 \text{ meV} $ as indicative of current bounds in models without fine tuning (and consistent with our calculations being correct up to $\mathcal{O}(1)$ constants).
	
	The neutrino burst of a  future galactic supernova could be observed in much greater detail by Hyper Kamiokande and other modern neutrino observatories. Ref.~\cite{GalloRosso:2017hbp} estimate that the supernova binding energy could be measured to around 2.5\% accuracy with Hyper Kamiokande in the event of a supernova within 10 kpc of Earth. Given that the axion luminosity obeys $L_a\sim m_{\rm QCD}^2$, such an event could improve the upper limit on the QCD axion mass by almost an order of magnitude, which we indicate on Figs.~\ref{fig:mQCD} and~\ref{fig:mQCD_Hodge} as a forecast limit at $m\lesssim 1 \text{ meV} $ with a dashed line. A future measurement indicating significant axion emission could provide positive evidence for a high mass QCD axion and thus favor large $h^{1,1}$ in the landscape, while excluding such a high mass axion would reduce the number of observationally viable Calabi-Yau compactifications substantially.
	
	Searches for the QCD axion and constraints on its couplings that are independent of the DM abundance are comparatively  
	model-independent.  However, the future program of axion DM direct detection has the potential to probe the entire mass range allowed by existing limits from SN1987A and black hole superradiance~\cite{Adams:2022pbo}.
	Due to the model dependence of relying on the relic abundance, a null result by a direct detection experiment cannot exclude a value of $h^{1,1}$. On the other hand, any measurement of the QCD axion mass could favor certain ranges of $h^{1,1}$. For example, a detection by ADMX-G2 at $m_a=10^{-5}\text{ eV}$ corresponds to the mean of the distribution at $h^{1,1}=100$, would exclude $h^{1,1}\gtrsim 200$ and is in the tail of the distribution at $h^{1,1}=50$.
	
	\subsection{Helioscopes and X-ray spectrum oscillations}\label{sec:chandra}
	
	Axion-photon conversion in magnetic fields can lead to constraints on the axion-photon coupling that are independent of the DM density and of cosmology via a number of astrophysical processes, of which we focus on two. In both cases, we ignore the possible complications in deriving a robust limit caused by mass-mixing among axions~\cite{Chadha-Day:2021uyt} and thus make only indicative statements about the utility of each probe. A complete statistical analysis including such effects will be the subject of future work.
	\begin{figure}
		\includegraphics[width=0.5\textwidth]{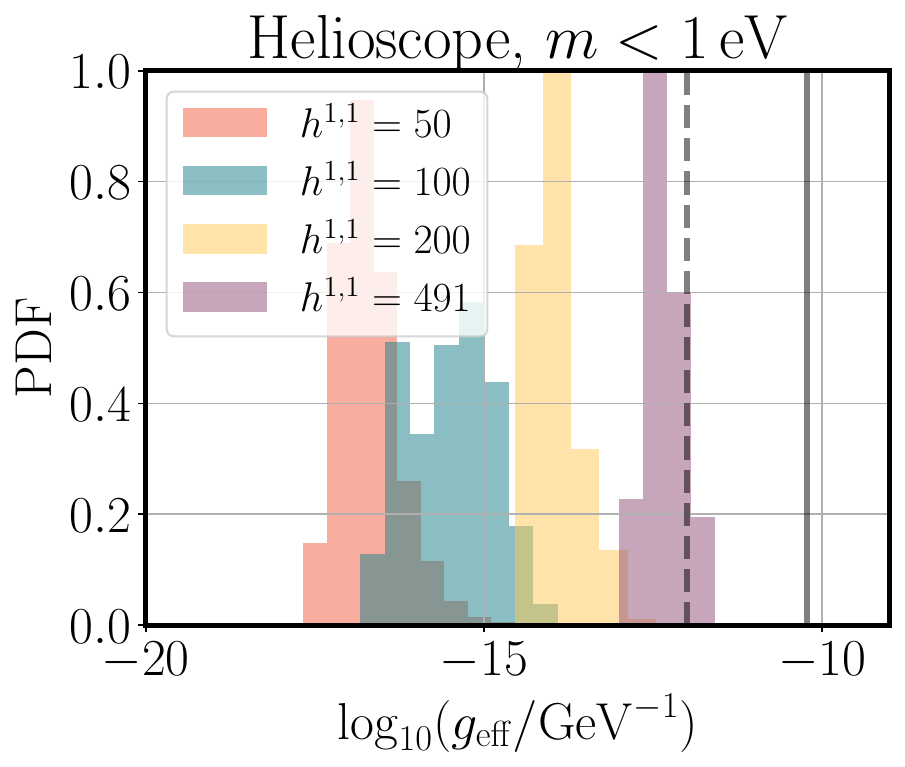}
		\includegraphics[width=0.5\textwidth]{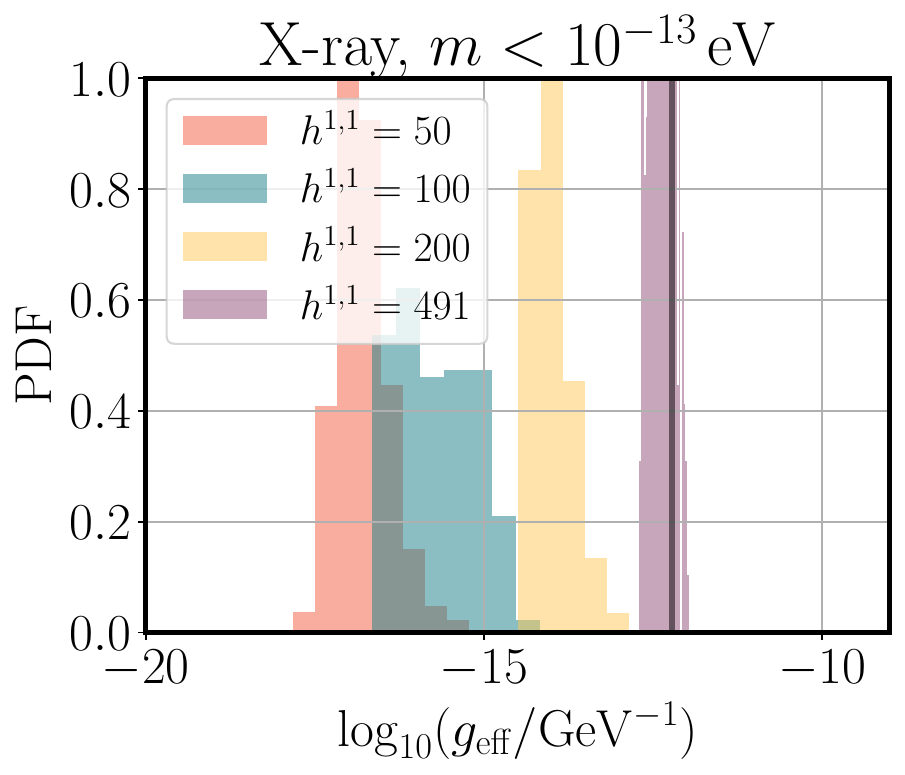}
		\caption{Distribution of the effective axion-photon coupling for helioscopes (left) and X-ray spectrum observations (right) in all models. In the left panel, the vertical lines indicate the current CAST limit (solid) and future IAXO limit (dashed). In the right panel, the vertical line indicates the current Chandra limit.}
		\label{fig:g_effective}
	\end{figure}
	
	An axion \emph{helioscope}~\cite{1983PhRvL..51.1415S} searches for axions produced in the Sun by the Primakoff process, and subsequently converted into X-rays in a laboratory magnetic field. The strongest current limits are set by the CERN Axion Solar Telescope (CAST)~\cite{2014PhRvL.112i1302A,1705.02290}, with future improvements planned by the International Axion Observatory (IAXO)~\cite{1904.09155}. Very roughly, such limits apply to all axions below a mass threshold of around 1 eV, and the effective coupling is approximately the Pythagorean sum of all couplings below this threshold (e.g.~Refs.~\cite{Chadha-Day:2021uyt,Demirtas:2021gsq}).
	
	Axion-photon conversion can also occur in the magnetic fields of galaxy clusters~\cite{Raffelt:1987im}. X-ray photons produced in active galactic nuclei could convert to axions along the line of sight, and if axions exist this would lead to oscillations in the predicted X-ray spectrum. The non-observation of such oscillations can be used to set limits on the axion-photon coupling (see e.g.~Refs.~\cite{Berg:2016ese,Conlon:2017qcw,Reynes:2021bpe,Matthews:2022gqi,Marsh:2021ajy} and references therein).  This effect occurs for all axions with masses less than the typical photon plasma frequency in the observed galaxy cluster, which is very roughly $m_a\lesssim 10^{-13}\text{ eV}$. Observations by the Chandra X-ray satellite place the strongest limits on the axion-photon coupling in this mass range that are independent of the DM density.

	Fig.~\ref{fig:g_effective} shows the effective axion-photon coupling for helioscopes and X-ray spectrum oscillations for different values of $h^{1,1}$. The distribution at $h^{1,1}=491$ is extremely narrow since, after restricting only to models with $\text{Vol}>127.5$ for the EM divisor, our ensemble contained only a small number of models. As we observed in Fig.~\ref{fig:mass_coupling_main}, detection by helioscopes or X-ray spectrum oscillations is generally possible only for 
	models with $(m,g)$ above the QCD line. These are necessarily non-GUT models. We observe the trend of increasing $g$ with $h^{1,1}$ discussed in \S\ref{sec:distributions}, and driven by the trend in the K\"{a}hler metric eigenvalues resulting from the narrowing of the K\"{a}hler cone at large $h^{1,1}$. Compared to Refs.~\cite{Halverson:2019cmy,Demirtas:2021gsq} our results in this regard include the effects of the light threshold, while being specific to type IIB string theory.  
	
	\begin{figure}
		\centering
		\includegraphics[width=0.6\textwidth]{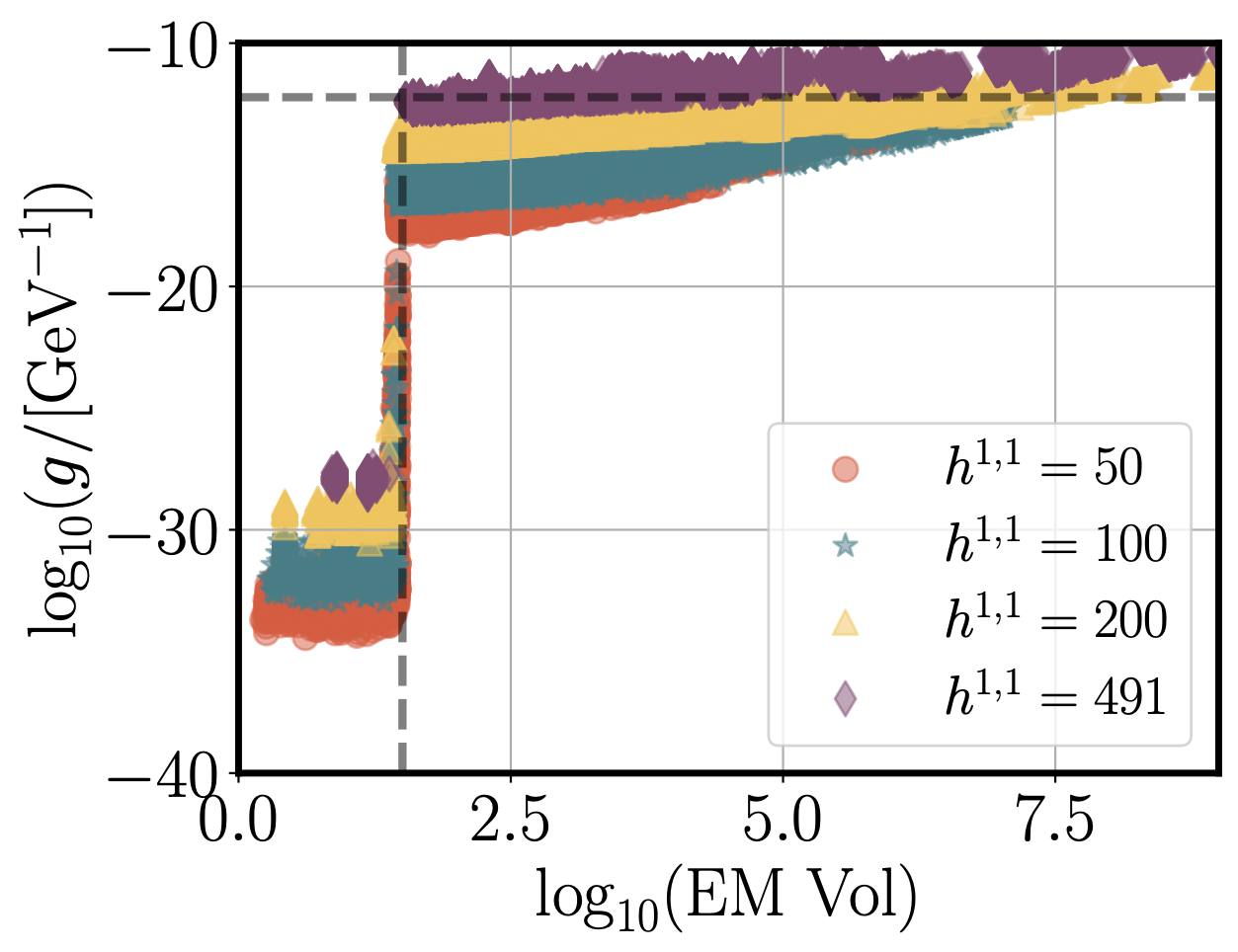}
		\caption{Effect of the EM divisor volume on the effective coupling for X-ray spectrum oscillations. The vertical dashed line is $\text{Vol}=30$, the threshold above which the mass scale generated by Euclidean D3-branes falls below the typical plasma frequency in galaxy clusters. The horizontal dashed line is the existing limit on this coupling from Chandra.}
		\label{fig:gChandra_EM}
	\end{figure}
	Fig.~\ref{fig:gChandra_EM} explores the effect of the realization of EM in the UV on the effective axion-photon coupling relevant for X-ray spectrum oscillations. We display the effective coupling as a function of EM divisor volume for non-GUT models. We observe that if the divisor volume is $\text{Vol}\lesssim 30$ then the coupling is strongly suppressed. This is because the mass generated by Euclidean D3-branes on a cycle with $\text{Vol}\lesssim 30$ is larger than the typical plasma frequency in galaxy clusters:
	thus, all axions that are
	above the light threshold,
	and so could have non-negligible couplings to the photon, 
	are too heavy to
	mix with the X-ray photons. We thus conclude that \emph{constraints on axions from X-ray spectrum oscillations apply to non-GUT models with EM divisor volumes larger than 30 in string units}. The corresponding limit on divisor volumes from helioscopes is fairly similar, because of the exponential dependence of masses on divisor volumes. However, because CAST and IAXO can probe beyond the QCD line, the helioscope limits can also apply to GUTs. 
	
	Fig.~\ref{fig:gChandra_EM} demonstrates that existing limits from Chandra already exclude models with $h^{1,1}=491$ if the EM divisor volume is very large. However, referring back to Fig.~\ref{fig:g_effective}, right panel, we see that once the EM divisor volume is restricted to the phenomenologically motivated range $\text{Vol}\leq 127.5$, then existing limits are less powerful.

	\subsection{Cosmic birefringence}\label{sec:birefringence}
	
	Recently, analysis of the large scale polarization anisotropies in the CMB as measured by the \emph{Planck} satellite showed evidence for an intrinsic rotation $\beta$ of the polarization angle between the surface of last scattering and today by $\beta=(6.1\pm 2.4)\times 10^{-3}$ (radians, 68\% C.L.)~\cite{Minami:2020odp}.  Subsequent studies found that the exact value of $\beta$ depends on the galactic mask used in the analysis to separate cosmic rotation from local effects, including rotation by dust and calibration of the polarization angle itself, with the preferred galactic model and mask giving $\beta=(5.2\pm 1.7)\times 10^{-3} $~\cite{Diego-Palazuelos:2022dsq}. The trend with mask size is consistent with a signal of cosmic origin, but the inferred value of $\beta$ and the statistical significance of the result are still under investigation, and will require confirmation by CMB measurements using different methods of polarization calibration and foreground modelling. For the purposes of this section we suppose that what is seen is indeed rotation of cosmic origin, and we examine whether axion-photon couplings in string theory can account for the observations.
	
	Rotation of the angle of polarization of the CMB is known as \emph{cosmic birefringence}, and can be caused by the axion-photon coupling (see e.g.~Refs.~\cite{axiverse,Carroll:1989vb,Pospelov:2008gg,Caldwell:2011pu,Zhao:2014yna,Fedderke:2019ajk,Fujita:2020ecn,Takahashi:2020tqv}). An all-sky signal such as that hinted at in Ref.~\cite{Minami:2020odp} requires that the axion mass is $H_0\lesssim m_a\lesssim H(z_{\rm dec})$ where $H(z_{\rm dec})\approx 10^{-28}\text{ eV}$ is the Hubble scale at CMB decoupling, and $H_0\approx 10^{-33}\text{ eV}$ is the Hubble scale today. 
	
	In Ref.~\cite{Mehta:2021pwf} we already identified that axions in this mass range exist in the Kreuzer-Skarke axiverse, with the possibility to produce the observed birefringent signal. Our present work can elaborate on this possibility further. By reference to Fig.~\ref{fig:mass_coupling_main} we find that birefringence will only be possible in non-GUTs, and by reference to Fig.~\ref{fig:mass_coupling_EM} we observe that an unsuppressed coupling to EM of an axion in the birefringence window requires EM realized on a divisor volume of order 40-50.\footnote{Ref.~\cite{Gasparotto:2023psh} recently investigated birefringence in the axiverse using the $m$ and $f$ distributions we found in Ref.~\cite{Mehta:2021pwf}. Ref.~\cite{Gasparotto:2023psh} did not include the extra suppression effects we have identified in the present work, nor was it possible to relate the presence of birefringence to aspects of QED in the UV.}   
	\begin{figure}
		\centering
		\includegraphics[width=0.6\textwidth]{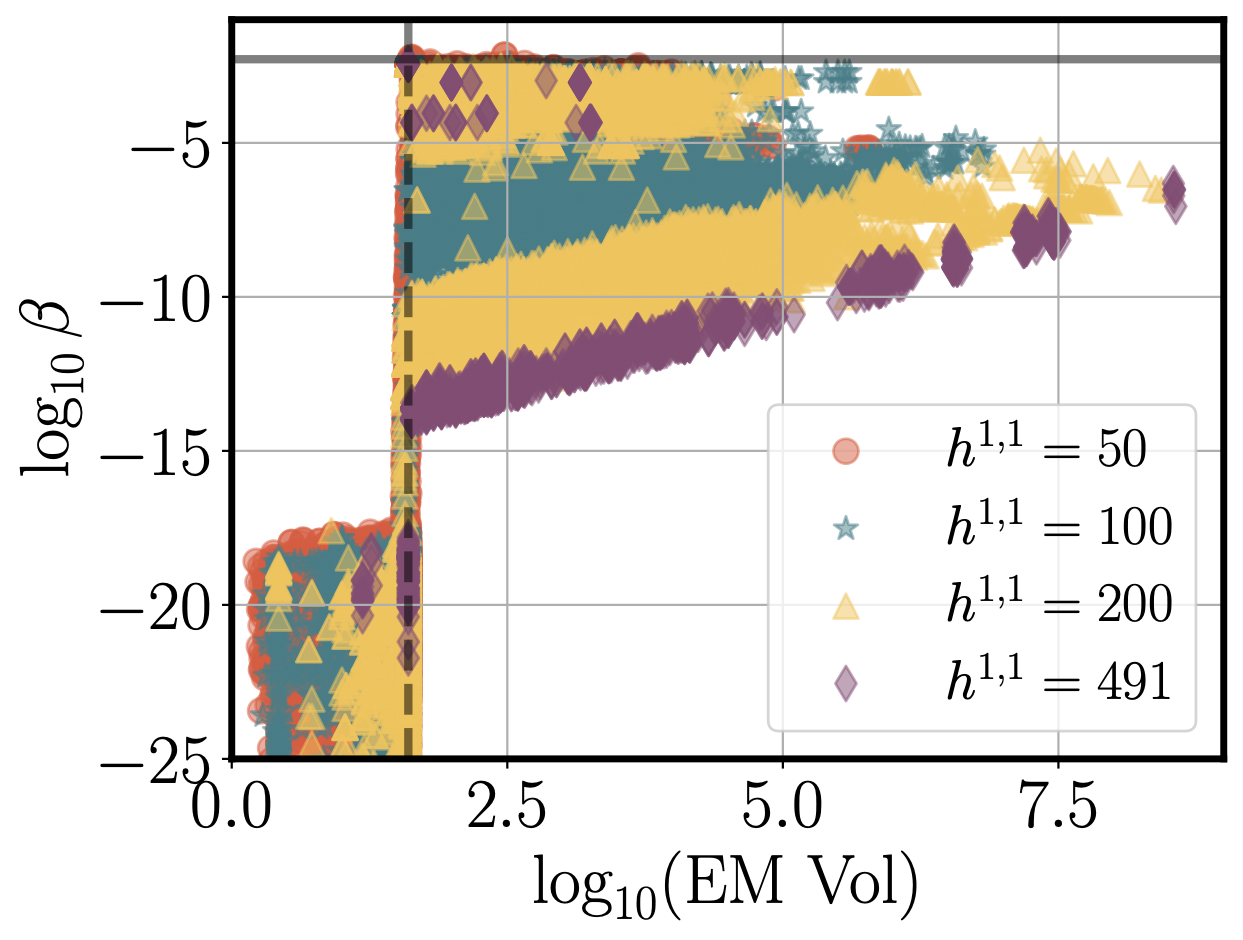}
		\caption{Effect of the EM divisor volume on the maximum birefringence angle, i.e.~obtained with all contributing axions having initial misalignment $\approx \pi$.  The vertical dashed line is $\text{Vol}=40$,
			which corresponds to the light threshold above which the mass scale generated by Euclidean D3-branes falls below $H(z_{\rm dec})\approx 10^{-28}\text{ eV}$.
			The horizontal line is the observed value, $\beta=5.2\times 10^{-3}$~\cite{Diego-Palazuelos:2022dsq}.  We conclude that with suitable initial conditions, the observed birefringence can be accommodated in models with $\text{Vol} \approx 40$.    }
		\label{fig:birefringence_EM}
	\end{figure}
	
	The rotation angle contribution of a single axion is given by integrating the motion of the axion along the line of sight from the CMB to the observer: 
	\be
	\beta_a = g_{a\gamma\gamma}\cdot\Bigl[ \varphi_a(z=0) - \varphi_a(z=z_{\rm dec})\Bigr] \approx \frac{\alpha_{EM}}{2\pi} C^a_{\gamma}\vartheta^a  \, ,
	\ee
	where the last approximation holds for the mass range stated above, and given the initial conditions on the condensate outlined in \S\ref{sec:hokey_kokey}. 
	For reference, in a theory of a single axion with decay constant $f$ and axion-photon coupling $g_{a\gamma\gamma} \approx \alpha/(2\pi f)$, and with initial misalignment $\langle\varphi\rangle \approx f$,
	one finds $\beta \approx 10^{-3}$ for a single axion. As discussed in Ref.~\cite{Mehta:2021pwf}, the rotation angle can be increased when there are $N$ contributing axions: as $N$ for aligned initial displacements from the vacuum, or as $\sqrt{N}$ for random displacements.\footnote{Ref.~\cite{Eskilt:2023nxm} studied axion-like fields as an Early Dark Energy component and placed bounds on $g$ from birefringence. However, they considered the potential $V\propto(\cos \theta)^3$ motivated by the $H_0$ tension. This potential does not arise in our models and so the bounds do not apply.}
	
	We show the total \emph{maximum} birefringence angle as a function of EM divisor volume in Fig.~\ref{fig:birefringence_EM}: this is computed by displacing all axions by $\pi$ from the vacuum, and adding (linearly) the contributions to the angle from all axions in the birefringent window.  
	We observe that if the divisor volume is $\text{Vol}\lesssim 40$ then the birefringence angle is strongly suppressed. This is because the mass generated by Euclidean D3-branes on a cycle with $\text{Vol}\lesssim 40$ is larger than $H_{\rm CMB} \approx 10^{-28}\text{ eV}$, and thus all axions in the birefringent window will be below the light threshold and so have couplings to EM that are hierarchically suppressed. If the EM divisor volume is $\text{Vol}\approx 40$, then a misalignment angle of $\pi$ leads to exactly the observed birefringence angle, since $\beta_{\rm obs}\approx \alpha$. Larger EM divisor volumes lead to an average $\beta$ slightly below the observed value due to kinetic isolation. We conclude that \emph{explaining the birefringence hint within the Kreuzer-Skarke axiverse requires non-GUT models with EM realized on divisors of volume $\approx 40$ in string units}. It is interesting to note that $1/\alpha=40$ near the GUT scale if renormalization group running includes only Standard Model degrees of freedom (c.f. a GUT in the MSSM, which gives $1/\alpha=25$).
	
	\subsection{Decaying dark matter and constraints on reheating}\label{sec:decaying_DM_indirect}
	
	In this section, we study constraints on Calabi-Yau threefolds in our ensemble originating from considerations of dark matter and reheating. This investigation serves as a proof of concept for ruling in and ruling out certain geometries based on cosmological observations, rather than as a conclusive analysis of the constraints on this ensemble.
	
	Axions produced by the Primakoff process can subsequently decay into photons, or into other axions. Axion decay is defined as happening when 
	$\Gamma=H(T_{\rm dec})$. The temperature of decay estimated by comparing the vacuum decay rate to the Hubble rate is close to the temperature at which decays and inverse decays re-equilibrate, which uses the full collision operator in the Boltzmann equation~\cite{2015PhRvD..92b3010M,Depta:2020wmr}. As discussed, if freeze-in, freeze-out, and misalignment production happen prior to re-equilibration via decays and inverse decays, i.e.~if $T_{\rm dec}<T_{\rm fo,fi,osc}$, then we compute the DM abundance before $T_{\rm dec}$ and scale to the effective fractional abundance today, $\xi$, as if the DM were stable. If instead $T_{\rm dec}>T_{\rm fo,fi,osc}$ we set the associated axion abundance to zero. In this section we consider only decaying axions produced by the Primakoff process from Standard Model particles, for which the only free cosmological parameter is the reheat temperature, $T_{\rm re}$. We consider three values of $T_{\rm re}$: the minimum allowed value from BBN, 5 MeV; a high reference value of $10^{10}\text{ GeV}$, and the geometric mean of the two, $T_{\rm re}\approx 10^4\text{ GeV}$. For completeness, we fix the inflationary Hubble scale to $H_I=10^8\text{ GeV}$, roughly the maximum value allowed by isocurvature constraints on the QCD axion in the pre-inflationary symmetry breaking scenario for $f_a\lesssim 10^{16}\text{ GeV}$~\cite{2009ApJS..180..330K,hertzberg2008}.
	
	We consider decays in four different cosmic epochs:
	\begin{itemize}
		\item Before BBN.
		\item Between BBN and recombination.
		\item Between recombination and today.
		\item Lifetime longer than the age of the Universe.
	\end{itemize}
	
	Decays in the first epoch, before BBN, are more strongly model-dependent than decays in later epochs: for example, problematic decay products could in principle be eliminated by a late, secondary stage of inflation.  We therefore do not consider constraints arising from such early decays in this section.
	
	Decays in the second epoch, between BBN and recombination, were studied in detail in Ref.~\cite{Balazs:2022tjl} by the \textsc{gambit} collaboration. Data products including Bayesian scan results in the parameter space $(\xi,\tau)$ can be used to transfer the limits to our setting. Axion decays during this epoch affect the light element abundances by photodissociation of nuclei, CMB spectral distortions through injection of photons that can no longe thermalize ($y$-type distortions) and through changing the Compton scattering rate ($\mu$-type distortions), and CMB anisotropies via changes to the effective number of neutrinos and the baryon to photon ratio. All these constraints (and more) are included in the posterior samples available from Ref.~\cite{Balazs:2022tjl}. The samples are restricted by prior to only cover this epoch of decays, and probe a minimum value $\xi\approx 10^{-10}$.
	
	Decays in the third and fourth epochs were studied in Ref.~\cite{Langhoff:2022bij}, and we adopt an approximate method to obtain limits in the $(\xi,\tau)$ plane presented there.
	
	Dark matter decays in the third epoch inject high energy photons into the intergalactic medium (IGM), which can lead to its heating and ionization. A hot, ionized IGM at high redshifts $z\gtrsim 10$ is excluded by the \emph{Planck} measurement of the CMB optical depth (see Refs.~\cite{Slatyer:2016qyl,Poulin:2016anj}). An approximation to the limit is given by~\cite{Langhoff:2022bij}:
	\be
	\xi<\frac{\tau}{\tau_{\rm min}}(\exp [t_{\rm CMB}/\tau])^{2/3}\, ,
	\ee
	where $t_{\rm CMB}$ is the characteristic timescale of recombination and $\tau_{\rm min}$ is the minimum allowed DM lifetime derived from the CMB anisotropies. The value of $t_{\rm CMB}=8.7\times 10^{13}\text{ s}$ adopted by Ref.~\cite{Langhoff:2022bij} arises from fitting the conservative limits of Ref.~\cite{Poulin:2016anj}. This value for $t_{\rm CMB}$ corresponds to a redshift $z\approx 320<z_{\rm rec}\approx1100$ below that of recombination itself, and a timescale approximately one order of magnitude larger than the lower prior limit of the \textsc{gambit} analysis described above. This gap between the timescales leads to a small gap in the constraint contours presented shortly. We adopt $\tau_{\rm min}=10^{25}\text{ s}$ consistent with Refs.~\cite{Poulin:2016anj,Cang:2020exa}.
	
	While the fourth category may seem to only imply that the corresponding axion is a stable DM candidate, in fact DM with lifetimes far exceeding the age of the Universe can nonetheless be probed by decays in regions where the DM number density is large, and so rare processes occur. This leads to limits from decays of DM in the Milky Way itself from X-ray observations, and from the cosmic ray background. We approximate the Milky Way X-ray limits as~\cite{Langhoff:2022bij}:
	\be
	\xi<\frac{\tau}{\tau_{\rm min}}\exp [t_{\rm U}/\tau]\, ,
	\ee
	where $t_{\rm U}\approx 1/H_0$ is the age of the Universe and we take $\tau_{\rm min}=10^{29}\text{ s}$. For simplicity we do not modify the limits accounting for the possibly warm axion DM, since we are not setting precision statistical limits (see Refs.~\cite{Langhoff:2022bij,Holm:2022eqq}).
	
	\begin{table}
		\centering
		\begin{tabular}{c| c| c| c| c}
			{$\log_{10} (T_{\rm re}/\text{eV})$} & {$h^{1,1}=50$} & {$h^{1,1}=100$} &  {$h^{1,1}=200$} &  {$h^{1,1}=491$} \\ \hline 
			6.7    & 0.01 & 0.004 & 0.003 & 0.0 \\
			12.9 & 0.87 & 0.32 & 0.17 &0.28 \\
			19.0  & 1.66 & 0.69 & 0.43 & 0.50\\ 
		\end{tabular}
		\caption{Average number of thermally produced decaying DM axions (defined as an axion flavor with $T_{\rm dec}>T_0$ and a non-zero thermal abundance) per model for different values of the reheat temperature and $h^{1,1}$.}\label{tab:num_decay}
	\end{table}
	At $h^{1,1}=50$, as is evident in Fig.~\ref{fig:mAll_gAll} and Fig.~\ref{fig:mass_coupling_main}, the invisible and visible axion populations are not widely separated at large mass, $m>1\text{ eV}$. This leads to the possibility that a large number of models will produce a significant population of thermal axions that subsequently decay. The number of decaying DM axions per model  is listed in Table~\ref{tab:num_decay} for different $T_{\rm re}$ and $h^{1,1}$. The number of thermally produced decaying axions increases with the reheat temperature, since the decay width goes like $m_a^3$, and a higher $T_{\rm re}$ allows heavier axions in the distribution to be thermally produced. However, the number also decreases with $h^{1,1}$, because the overall mass scale of the axions decreases.

	\begin{figure}
		\centering	
		\includegraphics[width=0.49\textwidth]{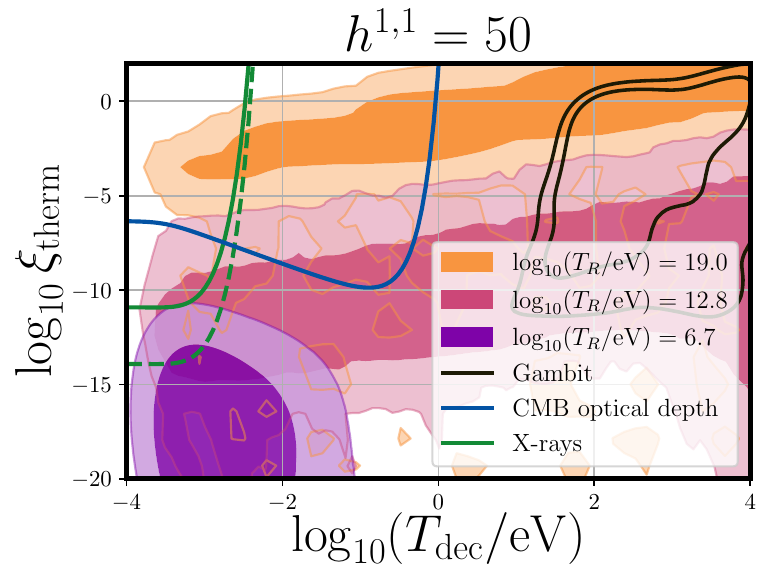}
		\includegraphics[width=0.49\textwidth]{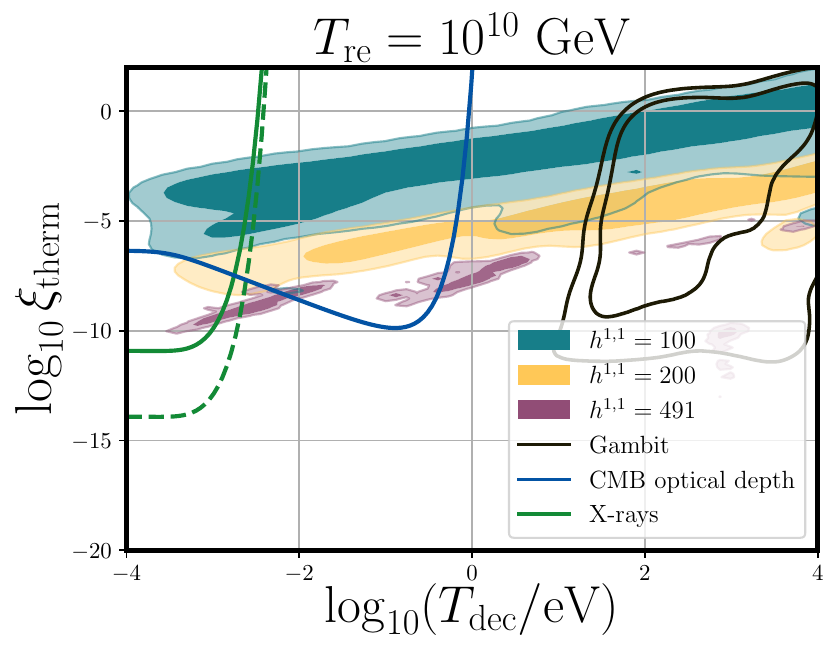}
		\caption{Filled contours give the distribution of effective relic abundance of decaying DM versus temperature of decay. In order to be allowed, a model must be below the exclusion contours shown as solid lines. The \textsc{gambit} contours are a preferred region that models should lie inside of or below. \textsc{gambit} contours are not defined for $T_{\rm dec}<10\text{ eV}$. The dashed line shows a forecast for future X-ray limits probing $\tau\approx 10^{32}\text{ s}$~\cite{Krnjaic:2023odw}. \emph{Left Panel:} Different values of the reheat temperature for $h^{1,1}=50$. \emph{Right Panel:} Different values of $h^{1,1}=50$ at fixed reheat temperature. Note that the number of samples in this population is small at $h^{1,1}=491$, after imposing control and cuts on the QED divisor volume.}
		\label{fig:decaying_DM_dists}
	\end{figure}
	Fig.~\ref{fig:decaying_DM_dists} shows the distribution of decaying axion DM relative abundances, $\xi_{\rm dec}$, versus temperature of decay for $h^{1,1}=50$ and for different values of the reheat temperature $T_{\rm re} \approx [5\text{ MeV}, 10^4\text{ GeV}, 10^{10}\text{ GeV}]$. For larger values of $h^{1,1}$ the distributions have largely the same shape, but are more poorly sampled since the number of axions per model in this population becomes smaller (see Table~\ref{tab:num_decay}). Models include all possible realizations of EM described in \S\ref{sec:distributions}. The decaying DM phenomenology is largely unaffected by how EM is realised in the UV: this can be seen in Fig.~\ref{fig:mass_coupling_main}, where we observe axions in the region corresponding to decaying DM  limits (blue constraints) in both GUT and non-GUT models.
	
	Fig.~\ref{fig:decaying_DM_dists} also displays exclusion contours, which models should lie below in order to be allowed by the observations (the \textsc{gambit} contours are 68\% and 95\% favored regions and are not defined for $\xi<10^{-12}$ and $T_{\rm dec}< 10\text{ eV}$). The exclusions assume that the axions dominantly decay to photons, rather than dark decays to other axions. We checked that decays to photons dominate the branching ratio in most models, particularly in the mass range of interest, and we leave a full study of dark decays to a future work. We observe that for the lowest possible reheat temperature, $T_{\rm re}= 5 \text{ MeV}$ (which gives the ``irreducible axion background''~\cite{Langhoff:2022bij}), the distribution lies well below all of the observational constraints on $\xi$. Furthermore, at the lowest $T_{\rm re}$, the average number of axions per model in the decaying DM population is less than one. At high $T_{\rm re}$ there is a significant amount of decaying DM, and large portions of the distribution are excluded, with the average number of axions per model also larger than at low $T_{\rm re}$.
	
	We conclude that \emph{thermal Primakoff production and subsequent decay of axions in the mass range $1\text{ keV}\lesssim m_a\lesssim 1 \text{ GeV}$ broadly disfavors $T_{\rm re}\gtrsim 10^{10}\text{ GeV}$ independent of $h^{1,1}$}. For such reheat temperatures the number of decaying axions per model becomes of order unity for all $h^{1,1}$, and the distribution of $\xi$ and $T_{\rm dec}$ lies largely in the observationally excluded region. However, particularly at large $h^{1,1}$, better statistics and a thorough dedicated study are required to reach a definite conclusion regarding the reheat temperature.
	
	\section{Summary and conclusions}\label{sec:conclusions}
	
	The goal of this paper was to systematically investigate axion-photon couplings in compactifications of type IIB string theory.

	The picture that emerged in early studies of the string axiverse \cite{axiverse} was of many-axion theories with masses $m_a$ spanning a vast range, but with decay constants $f_a$ clustered around a very high scale.
	Later work revealed that in weakly-coupled ultraviolet completions of gravity resulting from explicit Calabi-Yau compactifications of type IIB string theory --- the `Kreuzer-Skarke axiverse' --- the decay constants \emph{also} range over many orders of magnitude, with typical $f_a$ as small as $\mathcal{O}(10^9)~\text{GeV}$ in theories of $491$ axions \cite{Demirtas:2018akl,Mehta:2021pwf}.
	In single-axion theories one expects axion-photon couplings $g_{a\gamma\gamma} \sim \alpha_{\text{EM}}/(2\pi f_a)$, so if the same relation were to hold in the Kreuzer-Skarke axiverse, astrophysical limits would be severely constraining.
	
	One of our key results is that for the great majority of axions in this class of compactifications, $g_{a\gamma\gamma}$ is in fact hierarchically \emph{smaller} than $\alpha_{\text{EM}}/(2\pi f_a)$.  Thus, most axions are invisible, even though their decay constants are very small compared to the Planck scale and the GUT scale.  Just a handful of axions in each model have less-suppressed couplings to the photon, and might be detectable through electromagnetic processes.
	
	The causes of the suppression of axion-photon couplings can be described in EFT, but rest on structures dictated in the ultraviolet.  Euclidean D3-branes wrapping the four-cycle that supports QED generate an axion mass term involving a very small scale $m_{\rm QED}$, which we termed the \emph{light threshold}.  We showed that axion mass eigenstates with masses $m \ll m_{\rm QED}$ suffer hierarchical suppression of their couplings to the photon.  This is a stringy generalization of an effect studied in the context of GUTs in \cite{Agrawal:2022lsp}.
	Moreover, as stressed in \cite{Halverson:2019cmy}, the sparseness of the kinetic matrix leads to only a small amount of kinetic mixing between the electroweak axions and the other axions, which in turn leads to suppressed axion-photon couplings for these other axions.  We referred to this effect as \emph{kinetic isolation}.
	In summary, when the kinetic matrix is sparse and the instanton scales are hierarchical, axion-photon couplings are hierarchically suppressed. 
	
	In the present work we have systematically investigated the axion-photon coupling and axion mass distributions, and the resulting axion phenomenology, in explicit Calabi-Yau compactifications of type IIB string theory.  We considered the closed string axions arising from the Ramond-Ramond four-form.  We constructed Calabi-Yau threefolds from  triangulations of polytopes in the Kreuzer-Skarke database at four different values of the Hodge number, $h^{1,1}=50,100,200,491$, and at different points in K\"{a}hler moduli space, giving a total of $2\times 10^5$ models.  The axion-photon coupling is determined by a choice of divisor hosting the D-brane gauge group that contains QED, and by a choice of point in K\"{a}hler moduli space, which determines all the divisor volumes. We selected models as explained in \S\ref{ss:models}.
	
	We found that both the light threshold and kinetic isolation effects are prevalent in string theory, and only a few percent of the axions in a typical compactification have axion-photon couplings large enough to be relevant for phenomenology.  This is particularly true at high $h^{1,1}$, where kinetic matrices are extremely sparse.
	Nonetheless, the few well-coupled axions provide a target for ongoing searches.  Our results accord with those of  \cite{Halverson:2019cmy}, which pointed out and analyzed the consequences of what we here termed kinetic isolation. 
	We identified a further significant effect: the light threshold (\S\ref{ss:lightthreshold}).
	Combining these two phenomena, we found hierarchical suppression of axion-photon couplings.
	Comparing Figure \ref{fig:g_effective} to Figure 5 of \cite{Halverson:2019cmy}, we see that incorporating the light threshold in models with $h^{1,1} \sim 491$ makes these as hard to observe as models with $h^{1,1} \sim 200$ in which only kinetic isolation is included.  For the same reason, we find weaker couplings to the photon than seen in \cite{Demirtas:2021gsq}.
	
	With these models in hand, we investigated the resulting phenomenology. Our study was not intended to place rigorous statistical constraints on the axiverse (as in Ref.~\cite{Mehta:2021pwf}), but rather as an exploration of the possibilities. We have identified what appear to be some of the most promising probes to study the axiverse, and trends relating the low energy phenomenology to aspects of UV physics.
	
	We investigated how the mass of the QCD axion is correlated with $h^{1,1}$ in our models, finding it is possible to associate a given QCD axion mass to a range of values of $h^{1,1}$. Due to the observed correlation, direct detection of the QCD axion in future could narrow down the Calabi-Yau topology. Furthermore, at $h^{1,1}=491$ the predicted mass, $\mathcal{O}(1-10\text{ meV})$, is right up against existing astrophysical limits, and could lead to an observable signature in neutrino emission from a future galactic supernova.
	
	Continuing with astrophysical limits, we found that in non-GUT cases with QED realised on a divisor with volume $\gtrsim 30$, models with $h^{1,1}=491$ are also right up against existing constraints from X-ray spectrum constraints from the Chandra satellite. There is also the possibility to probe the same region of model space with the future IAXO helioscope.
	
	Moving to cosmology, we found that in a wide range of models it is possible to explain the observed birefringence of the CMB, supposing this is of cosmic origin. Explaining birefringence requires that QED is not unified with QCD, and is realized on a cycle with volume around 40. We also investigated axion dark matter production by the Primakoff process. If the reheat temperature is high, around $10^{10}\text{ GeV}$, then a large number of models produce heavy and decaying axion dark matter, which is excluded by a range of cosmological observations. This phenomenon could be used to exclude certain models and place upper limits on the reheat temperature.
	
	We have prepared a foundation for systematic studies of the visible axiverse, but much work remains.  The modeling of the visible sector here was highly simplified, and one should work with explicit orientifolds and D-brane configurations realizing visible and dark sectors, and also more thoroughly explore moduli space.
	We have only outlined the possibilities for phenomenology: given what we have found, quantifying the statistical significance of X-ray spectrum limits, and those from decaying DM at large $h^{1,1}$, could place strong limits on this part of the landscape.
	Continuing advances in theory and observations may reveal glimmers of light from the otherwise invisible axiverse.
	
	\section*{Acknowledgements}
	
	We acknowledge the following open source code packages: \textsc{numpy}~\cite{numpy}, \textsc{scipy}~\cite{scipy}, \textsc{matplotlib}~\cite{matplotlib}, \textsc{getdist}~\cite{Lewis:2019xzd}, \textsc{AxionLimits}~\cite{AxionLimits}, \textsc{jupyter}~\cite{Kluyver2016jupyter}. We used \textsc{CYTools}~\cite{CYTools} (see the documentation for the packages it relies on). We acknowledge discussions with Wenyuan Ai, Prateek Agrawal, Pierluca Carenza, Francesca Chadha-Day, Jim Halverson, Sebastian Hoof, Marco Hufnagel, Joerg Jaeckel, Andrew Long, Nate MacFadden, James Matthews, John March-Russell, Viraf Mehta, Fuminobu Takahashi, and Wen Yin. DJEM acknowledges Marco Hufnagel for providing the tabulated values of $g_{\star,Q}$ used to compute the Primakoff rate. DJEM is supported by an Ernest Rutherford Fellowship from the Science and Technologies Facilities Council (ST/T004037/1). 
	The research of NG, LM, and JM was supported in part by
	NSF grant PHY–2014071. 
	The research of NG was also supported in part by a grant from the Simons Foundation (602883,CV), the DellaPietra Foundation, and by the NSF grant PHY-2013858. LM and DJEM were supported in part by a Leverhulme Trust Reserach Project (RPG-2022-145). This work was supported by the 2022 Cornell--King’s Global Strategic Collaboration Awards program.
	
	\appendix
	
	\section{Additional phenomenology}\label{appendix:additonal_pheno}
	
	\subsection{Dark matter composition}\label{sec:warm_fuzzy}
	
	The Kreuzer-Skarke axiverse leads to axion DM with a wide spectrum of masses. As discussed in the main text, the DM relic density can be a mix of particles produced by the Primakoff process, as well as condensate produced by realignment.  We explore this briefly here. 
	
	The DM relic density depends on the cosmological model primarily through the reheat temperature. Minimal models, i.e.~those with the lowest relic abundances, occur when the reheat temperature is as low as possible, with $T_{\rm re}\approx 5\text{ MeV}$ and $H(T_{\rm re})\approx 10^{-15}\text{ eV}$. In this case, thermal production is strongly suppressed for $m_a>T_{\rm re}$, and we conservatively assume that the realignment population with $m_a>3H(T_{\rm re})$ has been diluted by entropy production.
	
\begin{figure}
		\includegraphics[width=0.5\textwidth]{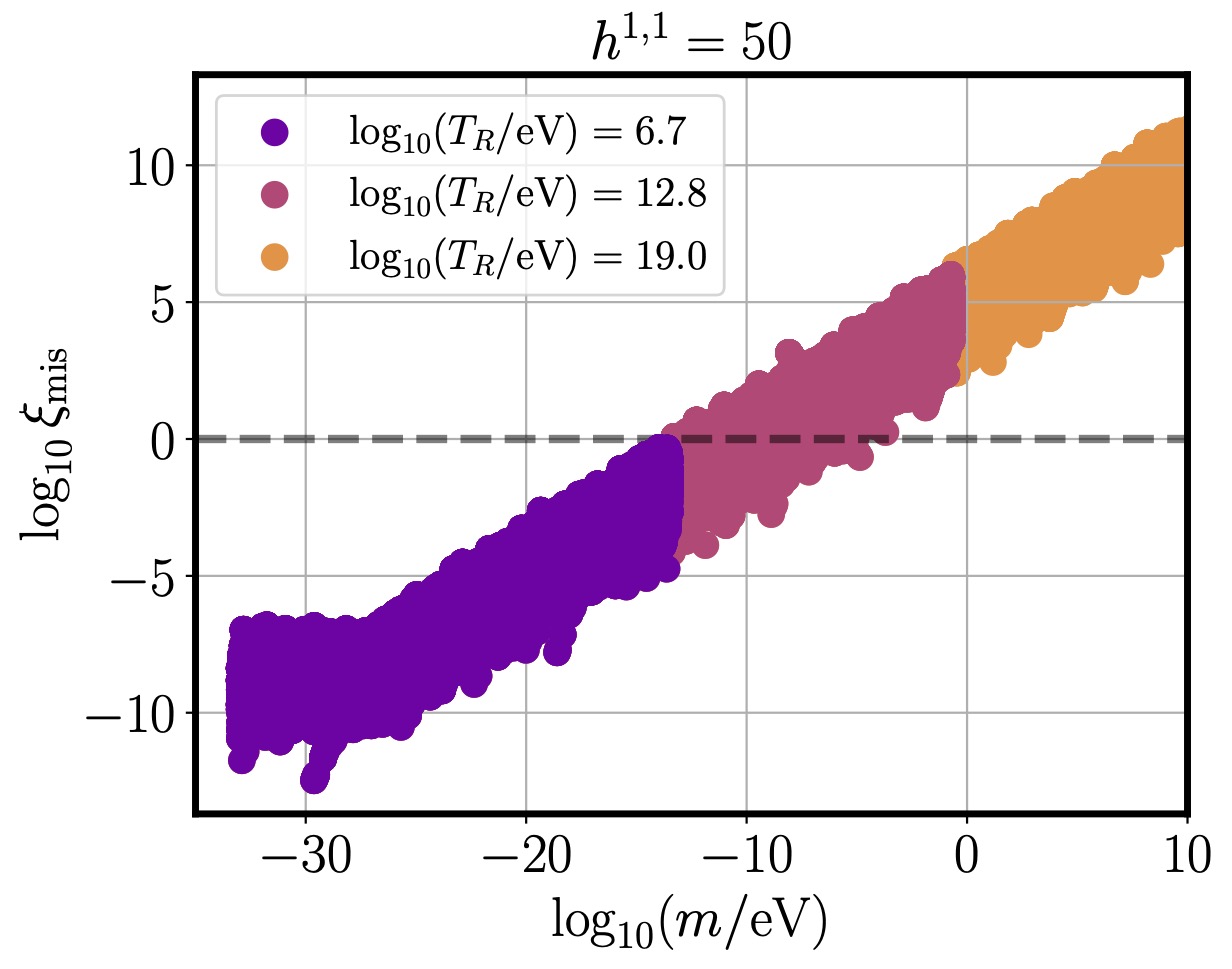}
		\includegraphics[width=0.5\textwidth]{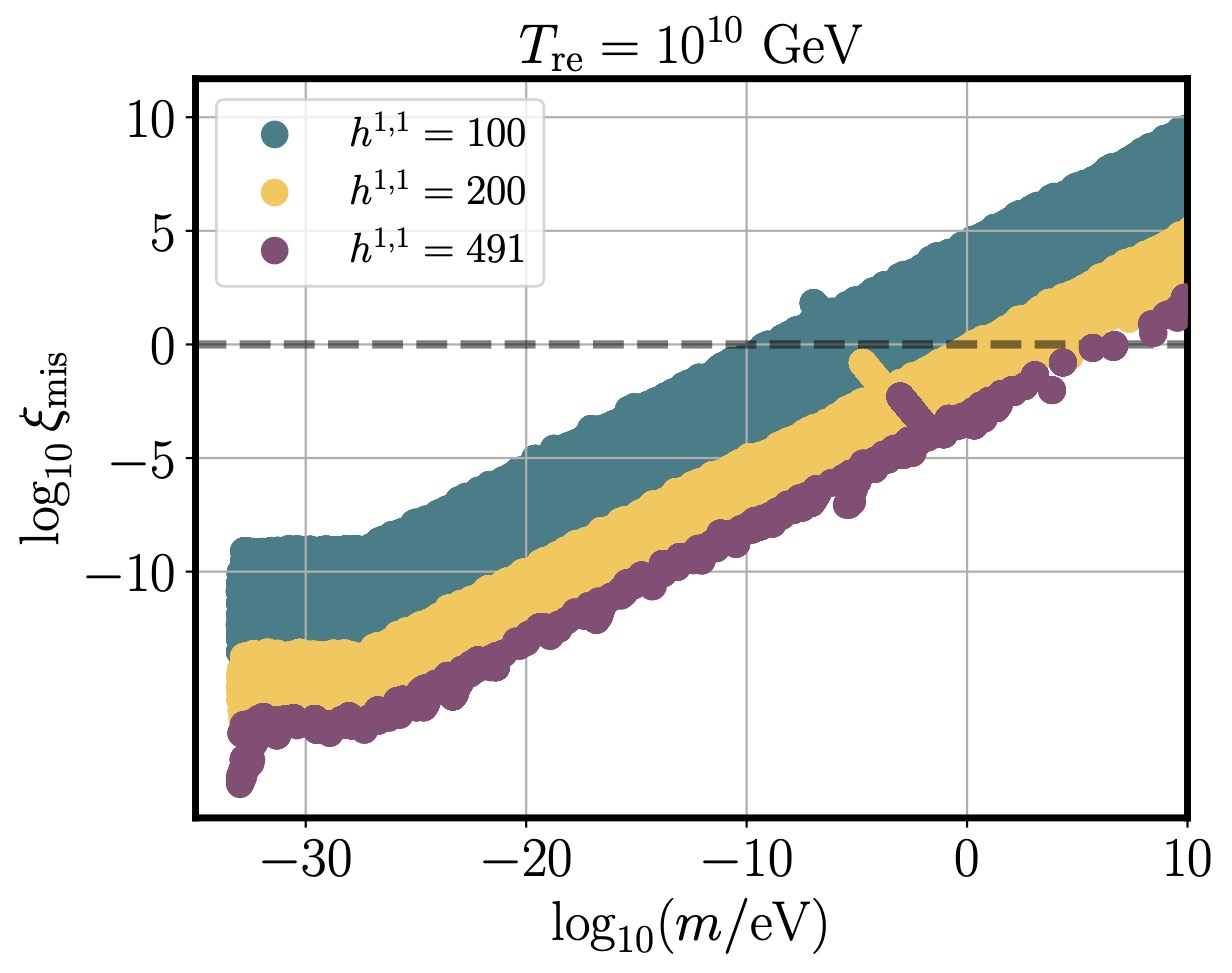}
		\caption{Dependence of the misalignment fractional relic density, $\xi_{\text{mis}}$, on the axion mass $m$. Horizontal lines indicate $\xi=1$, i.e.~$\Omega_a h^2=0.12$, the correct relic density. \emph{Left Panel:} Different values of the reheat temperature for $h^{1,1}=50$. \emph{Right Panel:} Different values of $h^{1,1}$ at fixed reheat temperature. We see that overproduction of misalignment relic DM is a risk at a wide range of reheat temperatures.}
		\label{fig:dm_composition}
	\end{figure}

	Fig.~\ref{fig:dm_composition} shows the relic abundance of misalignment axions for $h^{1,1}=50$ for different values of the reheat temperature (left), and for various $h^{1,1}$ at fixed reheat temperature (right), with order-unity misalignment angle.
	The reheat temperature affects the maximum mass available in the misalignment population, leading to too much DM for high $T_{\rm re}$. Overproduction of misalignment DM becomes less severe at larger $h^{1,1}$, due to the decreasing K\"{a}hler metric eigenvalues, and consequently decreasing decay constants. At $h^{1,1}=491$, the misalignment population gives the correct relic abundance around $m=100\text{ keV}$, which can only be produced under our assumptions for $T_{\rm re}\gtrsim 10^{10}\text{ GeV}$. 
	
	We also investigated Primakoff production of stable axions, and found that even at $T_{\rm re}=10^{19}\text{ eV}$ over production is not an issue. 
	As noted in the main text, decaying thermal axions in general disfavor $T_{\rm re}\gg 10^{19}\text{ eV}$. This leads us to conclude that the DM is more likely to be dominated by mislaignment produced light axions than Primakoff produced heavy axions.

	Finally, we consider DM produced by the \emph{collective Primakoff process} (see the discussion in \S\ref{sec:hokey_kokey}), defined as production of the components of the EM basis where $\Gamma_{\rm Prim}(T_f)>\Gamma_{\rm mix}$ (where $T_f$ is the lower of the freeze-out or freeze-in temperatures). Such production is only significant if the Primakoff freeze-out temperature is below the reheat temperature: freeze-in of relativistic particles leads to negligible abundances. The minimum freeze-out temperature for each $h^{1,1}$ in our sample is shown in Table~\ref{tab:freeze_out_temp}.
	
\begin{figure}
		\centering	
		\includegraphics[width=0.7\textwidth]{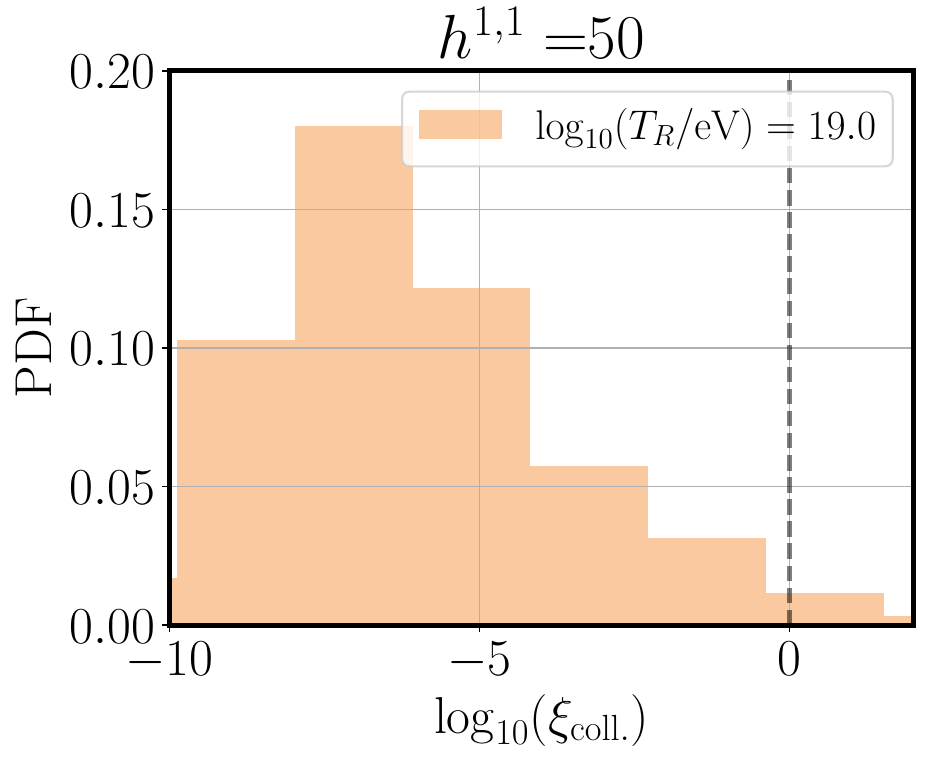}
		\caption{Dark matter produced via the `collective Primakoff' process in the EM basis, for $h^{1,1}=50$ and for reheat temperature $T_{\rm re}=10^{19}\text{ eV}$. In our ensemble the abundance in this channel is identically zero for $T_{\rm re}<10^{18}\text{ eV}$. The vertical dashed line indicates $\xi=1$, the observed cold DM abundance. DM in this channel is hot, and so must have abundance less than a few percent of the cold DM limit.}
		\label{fig:DM_coll_hist}
	\end{figure}

	The collective Primakoff process produces DM for those axions where $T_a>T_0$, so the axions are non-relativistic today. The resulting DM abundance is shown in Fig.~\ref{fig:DM_coll_hist} at $T_{\rm re}=10^{19}\text{ eV}$ for $h^{1,1}=50$.
	
	Since production in this channel requires $\Gamma(T_{\rm fo})=H(T_{\rm fo})>\Gamma_{\rm mix}$ and $T\propto \sqrt{H M_{\text{pl}}}$, this also implies $T>m_i$. Therefore, DM produced in this manner is \emph{hot} DM and in order not to be excluded must have a low abundance, much smaller than the observed cold DM abundance. From Fig.~\ref{fig:DM_coll_hist} we see that at high reheat temperatures, the collective Primakoff process overproduces hot DM.

	The observations on the total relic density in this Appendix favor low reheat temperatures, strengthening the assertions made in the main text. When the reheat temperature is low enough to not overproduce heavy, decaying axions, or hot DM, then axions produced by misalignment will be the dominant component of the axion DM. Depending on the details of the model, misalignment axions can provide a sizeable fraction of the DM for $m_a\gtrsim 10^{-15}\text{ eV}$, and an admixture of all axions, including the QCD axion, should be expected in general.
	
	In a non-GUT, if the dominant axion in the DM is the EM axion, or is well mixed with it, then the dominant axion will lie above the QCD line and be accessible to future axion searches~\cite{Adams:2022pbo}. However, depending on the realization of EM in the UV, the DM axions other than the QCD axion could all be invisible. We recall that the QCD axion mass correlates strongly with $h^{1,1}$, but that its relic density and exact axion-photon coupling are subject to the usual model building and cosmological uncertainties.
	
	In the ensemble constructed in this work, we found $f_a\lesssim 10^{16}\text{ GeV}$.
	This is too small to yield
	models of ``fuzzy DM'' with $m\ll 10^{-15}\text{ eV}$ that contribute a sizeable, $\xi\gtrsim 10^{-2}$, fraction of the relic density.
	Obtaining such models likely requires additional model-building to increase $f_a$, and is left for future work.

	\subsection{Dark radiation}
	
	We parameterize the axion dark radiation (DR) contribution by the effective number of neutrino species,
	$\Delta N_{\rm eff}$,
	which is defined from the energy density as:
	\be
	\Delta N_{\rm eff} =\frac{30}{2\pi^2 T_0^4}\frac{8}{7}\left(\frac{11}{4}\right)^{4/3} \rho_a\, ,
	\ee
	where $T_0$ is the present-day photon temperature (the CMB temperature) and the numerical factors account for the energy density contained in a two-component fermion that decouples from the Standard Model before electron-positron annihilation (thus defining a standard single neutrino)~\cite{1990eaun.book.....K}. DR can be produced in our model in two ways:
	\begin{enumerate}
		\item Primakoff production of the components of the EM basis that remain relativistic today.
		\item Decay of a heavier population of axions via cubic or quartic couplings.
	\end{enumerate}
	We do not study the second case in this work.
	
	When the reheat temperature is below the freeze-out temperature, $T_{\rm re}<T_{\rm fo}$, we approximate the Primakoff contribution to $\Delta N_{\rm eff}$ in the same way as freeze-in DM, i.e.~we take the number density at reheating to be:
	\be
	n_a \sim n_{a,\rm eq}(T_{\rm re})\frac{\Gamma_{\rm Prim}(T_{\rm re})}{H(T_{\rm re})}\, .
	\ee
	Thus,  the contribution of a non-thermalized axion to $\Delta N_{\rm eff}$ is suppressed relative to that of a fully thermalized axion by the ratio of the Primakoff rate to the Hubble rate at reheating.
	\begin{figure}
		\centering	
		\includegraphics[width=0.7\textwidth]{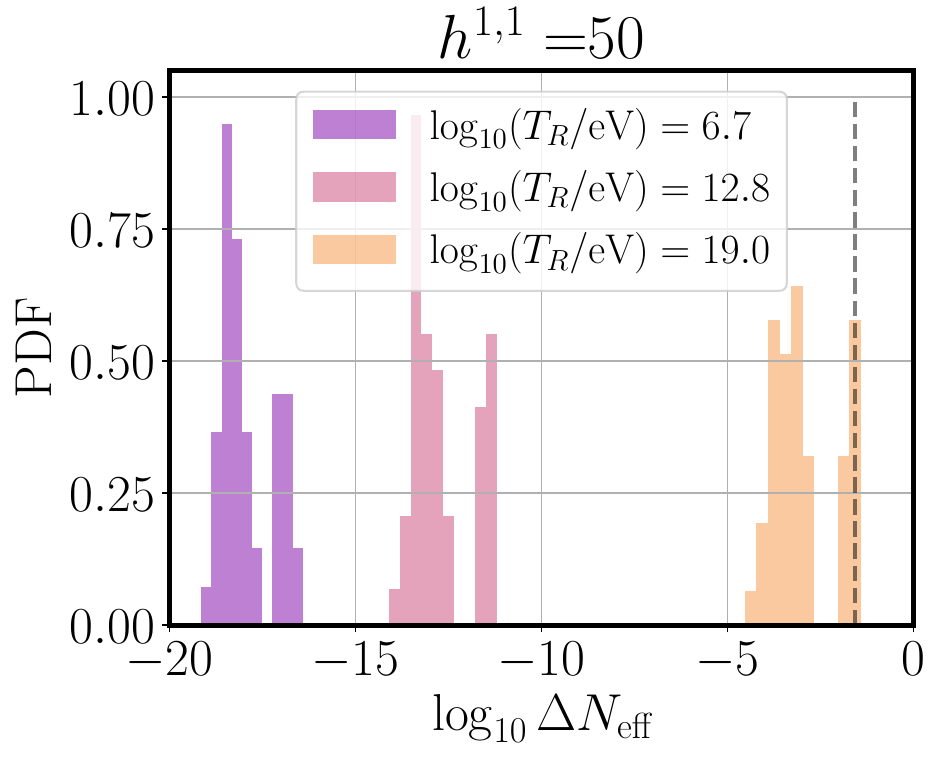}
		\caption{Dark radiation produced via the Primakoff process. The vertical dashed line indicates $\Delta N_{\rm eff}=0.027$, the expected value for a single thermalized degree of freedom in equilibrium above the top mass.  The minimum freeze-out temperature for $h^{1,1}=50$ is $2 \times 10^{18}$ (see Table~\ref{tab:freeze_out_temp}), and so the rightmost population falls in the regime where freeze-out occurs.}
		\label{fig:Neff_hist}
	\end{figure}
	
	The distribution of $\Delta N_{\rm eff}$ from the Primakoff process is shown for $h^{1,1}=50$ in Fig.~\ref{fig:Neff_hist}.
	We see that when $T_{\rm re}$ is larger than the smallest freeze-out temperature at this $h^{1,1}$ (see Table~\ref{tab:freeze_out_temp}),  
	$\Delta N_{\rm eff}$ approaches (and very occasionally exceeds) 0.027, which matches the expected value for a single bosonic degree of freedom that goes out of equilibrium above the top quark mass.
	
	Despite the large number of axions, $\Delta N_{\rm eff}$ is never much larger than 0.027. This limited production of DR is a consequence of two effects: the suppression of axion-photon couplings of all but a handful of axions, and the fact that for axions to be relativistic requires $T>\Gamma_{\rm Prim}>m_i$ at freeze-out, and only the single linear combination in the effectively massless part of the EM basis is produced as DR, i.e.~DR is necessarily produced in the collective Primakoff channel.
	
	Looking back to Appendix~\ref{sec:warm_fuzzy}, we see that when the reheat temperature is high enough to bring the Primakoff interaction into thermal equilibrium and produce $\Delta N_{\rm eff}=0.027$ (which may be accessible to future CMB observations), the Primakoff process also overproduces hot DM and decaying DM. Thus once we account for the DM abundance favoring low reheat temperatures this in turn disfavors an observably large value of thermally produced $\Delta N_{\rm eff}$.
	
	\begin{table}
		\centering
		\begin{tabular}{c |c}
			Hodge Number& {$T_{\rm fo}^{\rm min}$ [eV]}   \\ \hline 
			$h^{1,1}=50$    &  $2\times 10^{18}$ \\
			$h^{1,1}=100$  & $4\times 10^{17}$  \\
			$h^{1,1}=200$  & $1\times 10^{16}$ \\
			$h^{1,1}=491$   & $7\times 10^{14}$ \\ 
		\end{tabular}
		\caption{Minimum freeze-out temperature for different values of the Hodge number. }\label{tab:freeze_out_temp}
	\end{table}

	\subsection{Resonances and redistribution of the relic density}\label{sec:redistribution}
	\begin{figure}
		\centering	
		\includegraphics[width=0.7\textwidth]{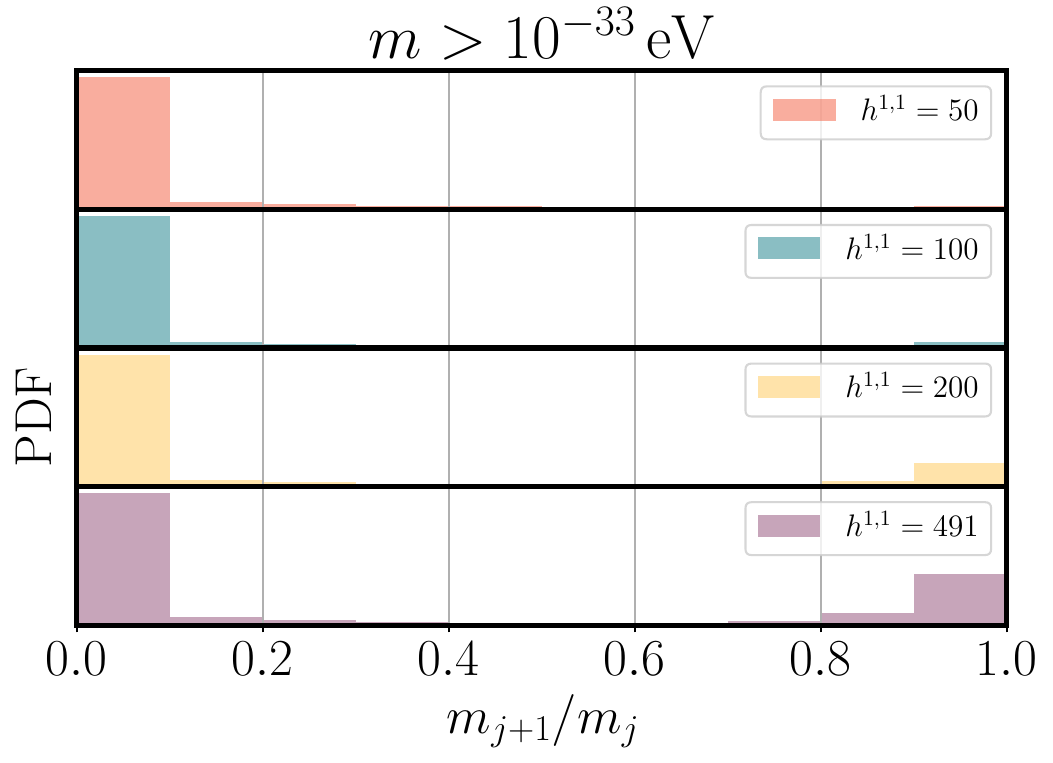}
		\caption{Distribution of the friendship parameter $\mu=m_{j+1}/m_j$, which is of interest since resonant energy transfer can occur for $\mu \gtrsim 0.75$.}
		\label{fig:Resonances_Mu}
	\end{figure}
	
	\begin{figure}
		\centering	
		\includegraphics[width=0.7\textwidth]{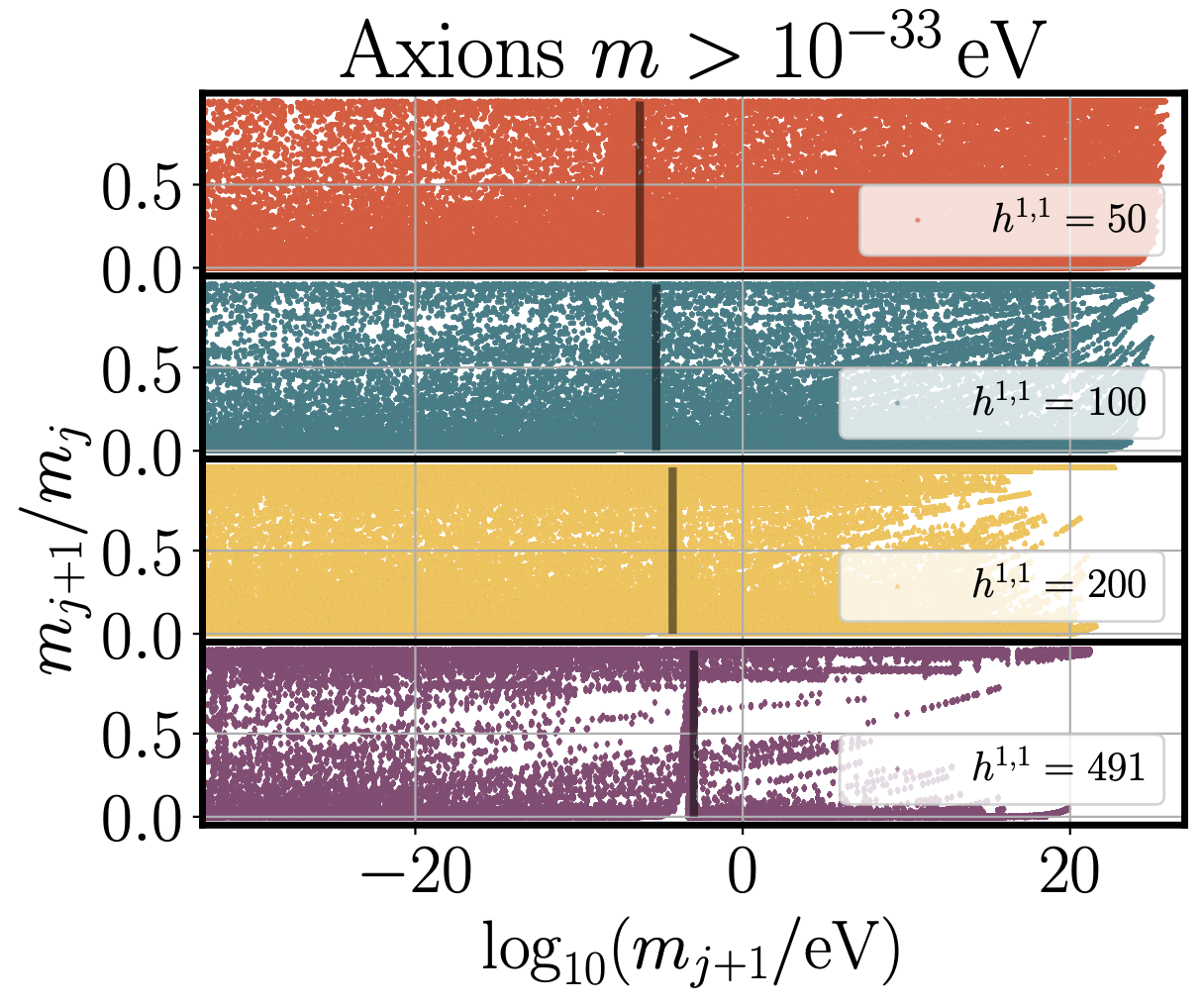}
		\caption{Distribution of the friendship parameter $\mu=m_{j+1}/m_j$ versus mass $m_{j+1}$. The vertical line in each figure indicates the mean QCD axion mass from Fig.~\ref{fig:mQCD_Hodge}.}
		\label{fig:mAll_muAll}
	\end{figure}
	
	The non-linear nature of the axion potential allows, in addition to the perturbative decays already discussed, for various processes that can redistribute the relic density created via vacuum realignment or the Primakoff process among the different axion species. We briefly discuss some of these processes here.
	
	Two-to-two annihilation of axion $i$ to either two or one axion $(i+1)$ can occur via the quartic couplings $\lambda_{ii(i+1)(i+1)}$ and $\lambda_{iii(i+1)}$. If these interactions are able to come into equilibrium, then the energy density created in species $i$ would redistribute into a hotter population in axion $(i+1)$. If the thermally averaged cross section for the interaction is $\langle \sigma v\rangle$ then equilibrium is achieved when $n_i\langle \sigma v\rangle = H$, becoming less important at low temperatures due to the decrease in $n_i$ particles in the initial state. Approximating $n_i$ by the freeze-in or freeze-out population of Primakoff-created particles, and estimating $\sigma$ from a typical value of $\lambda\sim m^2/f^2$, we find that such an interaction will not come into equilibrium at temperatures below the Planck scale for $m\approx 1\text{ GeV}$ and $f\approx 10^{12}\text{ GeV}$.  
	
	Thus, we estimate that axion-to-axion annihilations are unlikely, for typical parameters and hierarchical instanton scales, to significantly alter the relic density. Nonetheless, such annihilations could lead to indirect detection channels for axion DM. Light, relativistic axions produced by annihilation of heavy axions in the centre of the galaxy could convert to photons in the galactic magnetic field and lead to a gamma ray or X-ray signal similar to annihilating WIMPs. Such a scenario requires detailed study beyond the scope of the present work.
	
	Re-distribution of the energy density in the condensate can also occur via a resonant phenomenon explored in Refs.~\cite{Cyncynates:2021xzw,Cyncynates:2022wlq} (dubbed ``friendship''), leading to energy transfer between axions of different $f$ and comparable masses, $m$. Friendship requires large displacements of the field from the vacuum, and is not captured by our approximation of vacuum realignment which works only with the mass eigenstates. It will also be suppressed for the case of hierarchically separated masses. We will not assess the presence of friendship directly, but report on the presence or absence of close-to-resonant pairs (which may have consequences also for other observables that we outline). It should be noted that friendship only redistributes energy from the initially misaligned fields, and does not affect the overall relic density (thus our estimates of this will be unaffected), although non-linear structure formation can be significantly altered~\cite{Cyncynates:2022wlq}.
	
	Another instance of redistribution of the relic density relies specifically on the temperature dependence of the QCD axion mass relative to that of the other axions. When the QCD axion mass crosses the mass of another axion that it is well mixed with, then an avoided level crossing occurs, transferring energy between the two eigenstates, as shown in Refs.~\cite{Murai:2023xjn,Cyncynates:2023esj}. Depending on the hierarchy of masses and decay constants between the two fields, the level mixing can either enhance or deplete the energy density of the QCD axion. However, as shown in Ref.~\cite{Murai:2023xjn} (for the case where the two decay constants are not hierarchically separated, as we expect in our constructions), the total density usually remains unchanged from the naive sum of the independent axions. An overall enhancement of the total relic density requires that both masses become relevant at the same time, i.e.~that there is some $T$ such that $m_{\rm QCD}(T)\approx H(T) \approx m_X$, where $X$ labels the axion with mixing to the QCD axion. We ignore the effects of such near resonances in computing the relic density.
	
	The possible importance of resonances in the Kreuzer-Skarke axiverse is assessed in Fig.~\ref{fig:Resonances_Mu}, where we show the quantity $\mu = m_{j+1}/m_j$, i.e.~the ratio of mass of an axion to its next heaviest neighbour in the spectrum. The `friendship' phenomenon is relevant for $\mu\gtrsim 0.75$, and for masses $m>10^{-33}\text{ eV}$ (such that the field is oscillating). We notice that resonances are rare, but become less rare at large $h^{1,1}$.  Fig.~\ref{fig:mAll_muAll} shows the distribution of the resonance parameter versus mass. Resonances are slightly more common at higher mass, and we notice a slight clumping of the resonance parameter near the mass of the QCD axion (average value indicated with the vertical line). Thus, the energy transfer property caused by avoided level crossing discussed in Refs.~\cite{Murai:2023xjn,Cyncynates:2023esj} may be significant in some models.
	
	Finally, as mentioned in \S\ref{sec:cosmology}, we neglect redistribution of the relic density between the particles and the condensate.

	\renewcommand{\baselinestretch}{0.25}\normalfont
	\bibliographystyle{JHEP}
	\bibliography{bibliography}
	
\end{document}